\documentclass[reqno]{amsart}
\usepackage{makecell}
\usepackage{hyperref}
\usepackage{amsmath,amssymb} 
\pdfoutput = 1 
\usepackage{amsmath}

\usepackage{euscript}
\usepackage[T2A]{fontenc}
\usepackage{textcomp}
\usepackage{mathtools}

\usepackage{amsthm}
\theoremstyle{plain}

\newtheorem*{sle*}{Consequence}

\theoremstyle{Definition}

\theoremstyle{Remark} 

\usepackage{float}

\allowdisplaybreaks

\RequirePackage{caption}
\usepackage{graphicx}
\graphicspath{}
\DeclareGraphicsExtensions{.pdf,.png,.jpg}

\theoremstyle{plain}
\newtheorem{theorem}{Theorem}[section]

\theoremstyle{definition}

\theoremstyle{remark}

\numberwithin{equation}{section}

\begin{document}
	
	\title[ Exact solution of generalized triple Ising chains with multi-spin interactions
	]
	{ Exact solution of generalized triple Ising chains with multi-spin interactions
	}
	\author[P. Khrapov]{Pavel Khrapov}
	\address{Pavel Khrapov \\ Department of Mathematics
		\\ Bauman Moscow State Technical University \\  ul. Baumanskaya 2-ya, 5/1, Moscow \\ 105005, Moscow,  Russian Federation}  
	\email{khrapov@bmstu.ru , pvkhrapov@gmail.com }
	
	\author[N. Volkov]{Nikita Volkov}
	\address{Nikita Volkov \\ Department of Mathematics
		\\ Bauman Moscow State Technical University \\  ul. Baumanskaya 2-ya, 5/1, Moscow \\ 105005, Moscow,  Russian Federation}  
	\email{volkovns@student.bmstu.ru , nikita.volkov01@mail.ru}

	\subjclass[2010]{82B20, 82B23}
	\keywords{Chain Ising model, transfer-matrix, partition function, free energy, specific heat, internal energy, magnetization, susceptibility, entropy, thermodynamic limit, pair correlations, correlation length, ground states, triangular model, planar gonihedric model.}
	
	\begin{abstract}
		
		In this paper, we obtain the exact physical characteristics of the triple-chain Ising model on a torus with all possible multispin interactions invariant with respect to rotation by the angle $2\pi / 3$. The exact value of the partition function in a finite cyclically closed strip of length $L$, as well as the free energy, internal energy, specific heat, magnetization, susceptibility, and entropy in the thermodynamic limit at $L \to \infty$ are found by the transfer-matrix method for the model. The spectrum of the transfer-matrix and the structure of its eigenvectors are found. For two special cases — for the model with multispin interactions of even number of spins and for the model with some interactions of two, three, four and six spins, simplified expressions of the mentioned physical characteristics are obtained; in the thermodynamic limit they are expressed through the logarithm of the root of the quadratic equation.  For the model with multispin interactions of an even number of spins, a kind of pair correlations in the thermodynamic limit is found, and it is shown that the magnetization at zero magnetic field is equal to zero; the structure of the ground states of the system is found and examples of their projections of seven-dimensional space onto three-dimensional space and examples of configurations corresponding to these ground states are given. The correlation length is shown and its graphs are given.  For percolation invariant with respect to rotation about the central axis of rotation through the angle $2 \pi/3$, a limit relation for the non-percolation probability is derived. As special cases, we consider the planar triangular model with all possible interactions, including, perhaps, different triple interactions inside neighboring triangles, and the planar model with nearest-neighbor, next nearest-neighbor, and plaquette interactions. For them the main exact physical characteristics have been found. This allowed us to obtain them for the planar gonihedric model as well.  Graphs of physical characteristics illustrating the obtained results are constructed.
		
	\end{abstract}
	
	\maketitle
	
	\tableofcontents
	
	\section{Introduction}\label{in}
	\indent The Ising model, proposed by Lenz and introduced in 1925 by Ising \cite{Ising} to describe magnetic transitions, is still the subject of intensive research because it allows us to study the characteristics of magnetic systems and a number of different related physical phenomena, such as frustration. \\
	\indent The one-dimensional model was solved by Ising himself, and the exact value of the partition function and free energy for the Ising model with nearest-neighbor interaction without an external magnetic field on a $n \times \infty$ cylinder was first obtained by Onsager \cite{Onsager}. \\
	\indent Various numerical and analytical solution methods are applied to investigate such models. Among numerical methods, variations of the Monte Carlo method or the Metropolis \cite{Binder} algorithm predominate, while the transfer-matrix method, originally proposed in \cite{Kramers1}, \cite{Kramers2}, and the combinatorial method \cite{Kac}, are used for the exact analytical solution. \\
	\indent In \cite{Yurishchev1} the exact value of the free energy for the model with one central and three side spins is found, the Hamiltonian of which includes the interactions of the central spin of a layer with the central spins on the nearest neighboring layers, with the side spins of this layer, and also the interactions of the side spins with the nearest side spins of the neighboring layers. \\
	\indent In \cite{Yurishchev4} for the triple-chain Ising model with interactions of nearest and next-nearest neighbors without an external field, the exact solution of the secular transfer-matrix equation is obtained and all its eigenvalues and the form of eigenvectors are obtained; formulas for specific heat, parallel and perpendicular susceptibilities are derived; the behavior of these values as a function of temperature for various choices of the model parameters is considered. \\
	\indent In \cite{Yokota}, a triple-chain model with different interactions of nearest neighbors and next-nearest neighbors is studied by the transfer-matrix method, for which all eigenvalues and eigenvectors of the transfer-matrix are found, the structure of ground states is shown, and expressions for pair correlations of spins in the thermodynamic limit are obtained. These expressions are calculated using the spin-layer matrices, the structure of which is constructed by analogy with the two-dimensional \cite{Kalok} and one-dimensional \cite{Baxter} cases. \\
	\indent Besides triple-chain model, other models in three-dimensional space, such as those with cubic lattices, and various planar and linear models are also being actively investigated. \\
	\indent Thus, for example, in \cite{Yurishchev2} the exact solution of two Ising models with a cyclic-closed lattice $2\times2\times \infty$, the first of which has completely anisotropic interactions, and the second consists of two different types of linear chains including non-intersecting diagonal interactions on the outer faces of the parallelepiped, is obtained using the transfer-matrix method, and the behavior of specific heat is analyzed, and in \cite{Ratner} by the transfer-matrix method, taking into account its symmetry, the eigenvalues and the structure of eigenvectors for the three-dimensional cubic model $n=m=2$ with nearest-neighbor interactions are obtained, among which the interactions between spins located vertically in one layer, horizontally in one layer, and all interactions between neighboring layers are equal. \\
	\indent In \cite{Khrapov_PV_Nikishina_VA} the generalized Ising model in the $2\times2\times\infty$ strip with a Hamiltonian invariant with respect to the central axis of rotation through the angle $\pi/2$, which includes all possible multiplicative interactions of an even number of spins in the unit cube is considered.The exact value of free energy and specific heat in the thermodynamic limit is found.  \\
	\indent A separate chapter is devoted to the gonihedric model in the $2\times2\times\infty$ strip, the exact values of the free energy and specific heat in the thermodynamic limit are found for the case of free boundary conditions and an analogue of cyclically closed boundary conditions in both directions perpendicular to the central axis of rotation. 	\\
	\indent In \cite{Dos Anjos} for a cubic lattice with nearest-neighbor and  next-nearest-neighbor interactions, a detailed study of phase diagrams is carried out. \\
	\indent In \cite{Khrapov5} we obtain formulas for finding the free energy in the thermodynamic limit on the set of exact disordered solutions for the two-dimensional generalized Ising model in an external field with the interaction of nearest neighbors, next-nearest neighbors, all possible triple interactions and the interaction of four spins for the planar model, and for the three-dimensional generalized model in an external field with all possible interactions in the tetrahedron. \\	
	\indent In \cite{Minlos_RA_Sinai_YG} the spectra of stochastic operators are investigated.\\
	\indent The technique of cluster decompositions \cite{Malyshev_VA_Minlos_RA} has been very influential in the study of lattice models. \\
	\indent  In \cite{Minlos_RA_Khrapov_PV} сluster properties and bound states of the transfer matrix of the Yang-Mills model with a compact gauge group are shown, and in \cite{Khrapov_PV} cluster expansion and spectrum of the transfer matrix of the two-dimensional ising model with strong external field are shown. \\
	\indent In addition to classical ones, alternative methods for solving these models are being developed. For example, in \cite{Andreev_AS_Khrapov_PV} on a quantum computer emulator, in particular, a triple-chain model with a Hamiltonian invariant with respect to rotation by an angle $\frac{2\pi}{3}$ and taking into account pair and plaquette interactions and interaction with the external field is investigated, - a method of searching with the help of a variational quantum algorithm for the free energy and magnetization in the thermodynamic limit for one-, double- and triple-chain Ising models, and with the help of special parameterization of the state of the qubit system and transfer-matrix decomposition, the free energy and magnetization are calculated for the triple-chain model, and in \cite{Cirillo} the study of the cubic lattice model with nearest-neighbor, next-nearest-neighbor, and plaquette interactions by the method of cluster variation is carried out. In \cite{Khrapov3} the Fourier transform of the elementary transfer-matrix of the generalized two-dimensional Ising model with special boundary conditions with a helical-type shift and a type of Hamiltonian covering the generalized Ising model with a multispin interaction, as well as models equivalent to models on a triangular lattice with a checkerboard-type Hamiltonian, is performed.\\	
	\indent In \cite{Khrapov5} we obtain formulas for finding the free energy in the thermodynamic limit on the set of exact disordered solutions for the two-dimensional generalized Ising model in an external field with the interaction of nearest neighbors, next-nearest neighbors, all possible triple interactions and the interaction of four spins for the planar model, and for the three-dimensional generalized model in an external field with all possible interactions in the tetrahedron. \\    
	\indent A meaningful review and results on percolation theory can be found in \cite{Men'shikov_M_V_Molchanov_S_A_Sidorenko_A_F}. \\
	\indent In \cite{Coniglio} percolation in the Ising model is studied, based on the generalized results of \cite{Miyamoto}.  \indent Percolation in a strip of finite width for an independent field and the Ising model was studied in \cite{Khrapov_P_V}, for continuous models in \cite{Minlos_R_A_Khrapov_P_V_}. \\
	In recent years, interest in simple systems of Ising spins has arisen as a result of considering the random surface model in the context of string theory \cite{Savvidy},\cite{Johnston}. This model was called gonihedric and its three-dimensional case was studied by approximation methods. \\
	\indent In this paper, for the generalized triple-chain Ising model on the torus with a Hamiltonian with all possible multispin interactions invariant with respect to rotation by an angle $\frac{2\pi}{3}$, we obtain the exact values of the partition function in a finite closed strip of length $L$, as well as the free energy, internal energy, specific heat, magnetization, susceptibility, and entropy for a closed chain of finite length $L$ and in the thermodynamic limit at $L \to \infty$.  \\
	\indent For a model with multispin interactions of an even number of spins, the form of pair correlations in the thermodynamic limit is found, and it is shown that the magnetization at zero external field is equal to zero; the structure of the ground states of the system is found. A formula for the correlation length is derived. The obtained results are in agreement with the results of the paper \cite{Yokota}, in which the interaction Hamiltonian containing nearest-neighbor and next-nearest neighbor interactions is considered. \\
	\indent  Limit relations are obtained for percolation invariant with respect to rotation by an angle $\frac{2\pi}{3}$. \\
	\indent As special cases, exact physical characteristics are obtained for the planar triangular model with all possible interactions, including various triple interactions inside neighboring triangles, and for the planar model with nearest-neighbor, next-nearestneighbor and cladding interactions. For specific parameters of interactions this gives exact physical characterizations for the planar gonihedric model.\\
	\indent Next, let us describe the structure of the paper. \\
	\indent In paragraph 2 we describe the model and its physical characteristics, introduce the set of elementary generating carriers of the Hamiltonian of the system and the Hamiltonian itself, formulate two theorems for the general case of the model with all possible multispin interactions in the thermodynamic limit and in the strip of finite length, which were proved: exact analytical expressions for the physical characteristics in the thermodynamic limit are given and the expression for the partition function in the strip of finite length and all eigenvalues of the transfer-matrix are written out. \\
	\indent The third paragraph presents the first special case — a triple-chain model with multispin interactions of even number of spins, for which two theorems similar to the theorems of the second paragraph are formulated and all eigenvalues of its transfer-matrix are written out. \\
	\indent For this model, exact analytical expressions of pair correlations in the thermodynamic limit are obtained using the spin-layer matrices  and the transfer-matrix, for which the transition matrix to the diagonal form is written out. With its help it is shown that the magnetization of this model in the absence of an external field is zero, which coincides with the results of \cite{Fedro}. The general expressions for pair correlations for the model with multispin interactions of an even number of spins coincide with the expressions for pair correlations of the triple-chain model with nearest- and next-nearest neighbor interactions considered in \cite{Yokota}.  \\
	\indent Besides, for this case we describe the structure of ground states \cite{Kashapov}, give examples of corresponding model configurations and two example-illustrations explicitly showing the picture of ground states in two partial projections of the seven-dimensional space of ground states onto the three-dimensional one, and show with their help the possibility of special behavior of the correlation length near the boundary points of the regions of ground states. \\
	\indent In paragraph 4 we present the second special case — a model with some interactions of two, three, four and six spins. For it two theorems similar to the theorems of the second paragraph are formulated and all eigenvalues are written out, two of which have multiplicity 1 and two have multiplicity 3. \\
	\indent In the fifth paragraph, we formulate two theorems on the non-percolation probability, invariant with respect to rotation by an angle $\frac{2\pi}{3}$, in a closed strip of finite length $L$, and in the thermodynamic limit.  \\
	\indent In the sixth paragraph we consider a model with nearest neighbor, next-nearest neighbor and plaquette interactions. This is an important special case of the more general model of paragraph 3, and all the results of the third paragraph are true for it. The results for the gonihedric model are formulated separately. \\
	\indent In the seventh paragraph we consider the planar triangular model, the validity of the theorems \ref{th1} and \ref{th2} is shown for it. \\
	\indent In the eighth paragraph we consider an example illustrating all physical characteristics given in the main theorem of the second paragraph in the thermodynamic limit for the general case of the model with all possible multispin interactions at two variable parameters corresponding to the next-nearest neighbor interactions and equal to each other, and a variable temperature — three-dimensional plots of these quantities and cross sections of some of them at fixed temperatures are shown. \\
	\indent In paragraph 9 we prove six theorems (1 - 4, 6, 7), formulated in paragraphs 2 — 4: with the help of the theorem on preservation of eigenspaces by commuting matrices \cite{Horn} we obtained the structures of eigenvectors of transfer-matrices for each case, due to which the problem of finding eigenvalues of initial transfer-matrices of size $8\times8$ was reduced to the search of roots of characteristic equations of matrices of smaller dimensions — for the general case of the model with all possible multispin interactions, 4 eigenvalues, including the largest one, were found as roots of the equation of the fourth degree, the solution of which was obtained by the Ferrari method, and the other four as roots of two quadratic trinomials. For the first special case, four eigenvalues, including the largest one, are found as roots of two square trinomials, and the remaining four are found as roots of four linear equations. In the second special case, all eigenvalues are found as roots of two quadratic trinomials — two roots of multiplicity 1, and two roots of multiplicity 3. \\
	\indent In the appendix, we give formulas for the Ferrari method of solving the fourth degree equation, and a scheme for computing the partial derivatives on the external field $H$ of the coefficients of the equation \text{(\ref{ny1})}.
	
	\section{Model description and main result}  \label{Model_description}
	\begin{figure}[H]
		\center{\includegraphics[width=1\linewidth]{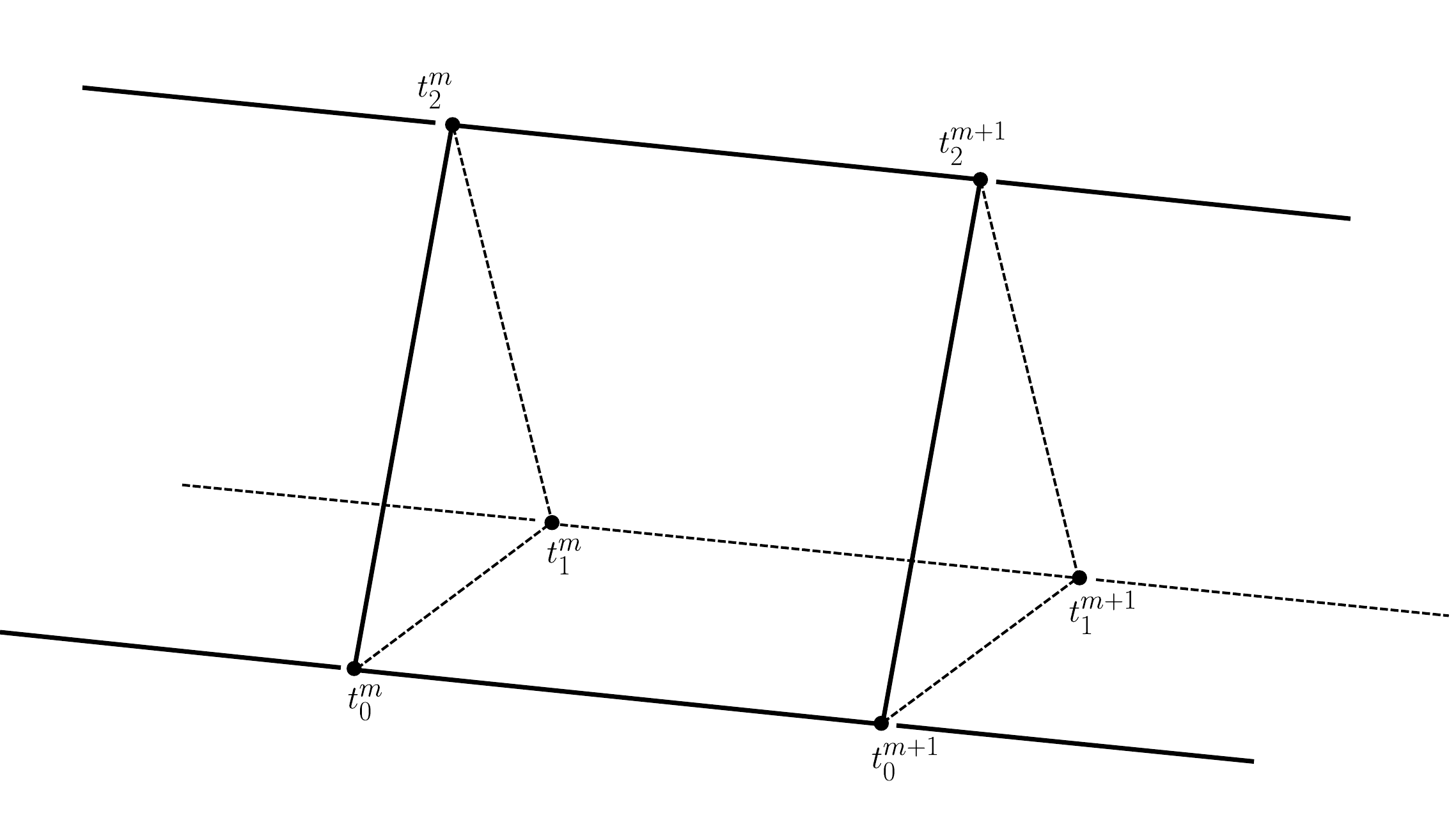}}
		\caption{Triple-chain lattice of the Ising model}
		\label{p1}
	\end{figure}
	\indent Consider a three-dimensional triple-chain cyclically closed lattice model of size $3 \times L$ (Fig. \ref{p1}) with a total number of nodes in the lattice $N = 3L$, where $L$ is the number of layers in the model:
	\begin{equation} \label{1}
		\mathcal{L}_{L} = \{t_i^m; m = 0, 1, \ldots, L-1; t_i^L = t_i^0; i = 0 , 1, 2\}
	\end{equation}
	\indent Let us consider that in each node $t_i^m$ there is a particle, the state of which is determined by the value of spin $\sigma_i^m \equiv \sigma_{t_i^m}$, $m = 0, 1, \ldots, L-1$; $i = 0, 1, 2$.\\
	\indent Let $\Omega ^m_P = \{t_i^m, t_i^{m+1}, i = 0, 1, 2\}$ be a triangular prism (Fig. \ref{p1}). Let us introduce the set of elementary generating carriers of the Hamiltonian:
	\begin{equation} \label{2}
		\begin{gathered}
			\Phi^m_1 = \{\{t^m_0\},\{t^{m+1}_0\}\}, \\
			\Phi^m_2 = \{\{t_0^m, t_1^m\}, \{t_0^m, t_0^{m+1}\}, \{t_0^{m}, t_1^{m+1}\}, \{t_0^{m}, t_2^{m+1}\}, \{t_0^{m+1}, t_1^{m+1}\}\}, \\
			\Phi^m_3 = \{\{t^m_0, t^m_1, t^m_2\}, \{t^m_0, t^m_1, t^{m+1}_0\},\{t^m_0, t^m_1, t^{m+1}_1\},\{t^m_0, t^m_1, t^{m+1}_2\},\{t^m_0, t^{m+1}_0, t^{m+1}_2\}, \\
			\{t^m_0, t^{m+1}_0, t^{m+1}_1\}, \{t^m_0, t^{m+1}_1, t^{m+1}_2\}, \{t^{m+1}_0, t^{m+1}_1, t^{m+1}_2\}\}, \\
			\Phi_4^m = \{ \{t_0^m,t_1^m,t_2^{m},t_0^{m+1}\}, \{t_0^m,t_1^m,t_0^{m+1},t_1^{m+1}\}, \{t_0^m,t_1^{m},t_0^{m+1},t_2^{m+1}\}, \\ \{t_0^m,t_1^m,t_1^{m+1},t_2^{m+1}\}, \{t_0^m,t_0^{m+1},t_1^{m+1},t_2^{m+1}\} \}, \\
			\Phi^m_5 = \{ \{t_0^m, t_1^m, t_2^m, t_1^{m+1}, t_2^{m+1}\}, \{t_1^m, t_2^m, t_0^{m+1}, t_1^{m+1}, t_2^{m+1}\} \}, \\
			\Phi_6^m = \{ \{t_0^m, t_1^m, t_2^m, t_0^{m+1}, t_1^{m+1}, t_2^{m+1}\} \}.
		\end{gathered}
	\end{equation}
	\indent Let us introduce an elementary generating Hamiltonian for all interactions:
	\begin{equation} \label{3}
		\hat{H}^m = \hat{H}^m_1 + \hat{H}^m_2 + \hat{H}^m_3 + \hat{H}^m_4 + \hat{H}^m_5 + \hat{H}^m_6,
	\end{equation}
	for which 
	\[
	\hat{H}^m_r = - \sum_{\alpha_m \in \Phi_r^m} J_{\alpha_m} \sigma_{\alpha_m},
	\]
	\[
	r = 1, 2, \ldots, 6
	\]
	— components of the Hamiltonian responsible for the interaction of $r$ spins in the $\Omega_P^m$ prism, $J_{\alpha_m}$ are the corresponding coefficients of interspin interactions. \\
	\indent For the function $\varphi(\sigma_0^m, \sigma_1^m, \sigma_2^m, \sigma_0^{m+1}, \sigma_1^{m+1}, \sigma_2^{m+1})$ that depends on $\sigma_t$, $t \in \Omega ^m_P$, we introduce the "rotation" operator $R^m_{\frac{2\pi}{3}}$ by the angle $\frac{2\pi}{3}$:
	\begin{equation} \label{4}
		R^m_{\frac{2\pi}{3}}(\varphi(\sigma_0^m, \sigma_1^m, \sigma_2^m, \sigma_0^{m+1}, \sigma_1^{m+1}, \sigma_2^{m+1})) = 
		\varphi(\sigma_1^m, \sigma_2^m, \sigma_0^m, \sigma_1^{m+1}, \sigma_2^{m+1}, \sigma_0^{m+1}), 
	\end{equation}
	then the Hamiltonian of the model has the form 
	\begin{equation} \label{5}
		\mathcal{H}_L(\sigma) = \sum_{m = 0}^{L-1}H^m,
	\end{equation}
	where
	\begin{equation} \label{nnn5}
		H^m = \hat{H}^m + R^m_{\frac{2\pi}{3}}\hat{H}^m + \big(R^m_{\frac{2\pi}{3}}\big)^2\hat{H}^m,
	\end{equation}
	that is, all carriers of the Hamiltonian in the prism $\Omega ^m_P$ are generated by elementary generating carriers (\ref{2}) at the rotation of the model by the angle $\frac{2\pi}{3}$ and the angle $\frac{4\pi}{3}$. \\
	\indent In this formulation of the Hamiltonian for the interaction $J_{\{t_0^m,t_1^m,t_2^m\}}$ and for $\hat{H}^m_6$ one must take into account the $\frac{1}{3}$ multiplier compensating the rotations.\\
	\indent In addition, when going from the prism $\Omega_P^m$ to the prism $\Omega_P^{m+1}$, the elementary generating carriers $\{t_0^{m}\}, \{t_0^{m}, t_1^{m}\}, \{t^m_0, t^{m}_1, t^{m}_2\}$ in the prism $\Omega_P^m$ generate, respectively, elementary generating carriers $\{t_0^{m+1}\}, \{t_0^{m+1}, \{t_0^{m+1}\}, \{t^{m+1}\}, \{t^{m+1}_0, t^{m+1}_1, t^{m+1}_2\}$, so that the latter can be omitted explicitly in (\ref{2}), but we include them to emphasize the following clearly:
	\begin{equation}\label{nnn7}
		\begin{gathered}
			\Phi_4^m = \{\Omega ^m_P \setminus \{t_0^m, t_1^m\}, \Omega ^m_P \setminus \{t_0^m, t_0^{m+1}\}, \Omega ^m_P \setminus \{t_0^{m}, t_1^{m+1}\}, \\
			\Omega ^m_P \setminus \{t_0^{m}, t_2^{m+1}\}, \Omega ^m_P \setminus \{t_0^{m+1}, t_1^{m+1}\}\}, \\
			\Phi^m_5 = \{\Omega ^m_P \setminus \{t^m_0\}, \Omega ^m_P \setminus \{t^{m+1}_0\}\}.
		\end{gathered}
	\end{equation}
	\indent Note also that this transition must be taken into account when composing the Hamiltonian of the model, indicating in front of its components, containing the interactions $\{t_0^{m}\}, \{t_0^{m}, t_1^{m}\}, \{t^m_0, t^{m}_1, t^{m}_2\}, \{t_0^{m+1}\}, \{t_0^{m+1}, t_1^{m+1}\}, \{t^{m+1}_0, t^{m+1}_1, t^{m+1}_2\}$ and generated by them, the multiplier $\frac{1}{2}$. \\
	\indent The partition function of the model will be written in the form:
	\begin{equation} \label{6}
		Z_{L} = \sum_{\sigma} \exp\bigg\{-\frac{\mathcal{H}_L(\sigma)}{k_BT}\bigg\},
	\end{equation}
	where $k_B$ is the Boltzmann constant (for simplification in further calculations we take $k_B=1$), and the summation is taken over all configurations of spins. \\
	\indent To find the partition function, we introduce a transfer-matrix of size $8 \times 8$. Its nonzero elements are defined as follows:
	\begin{equation} \label{12}
		\theta_{k,l} = \theta_{(\sigma_{0}^m, \sigma_{1}^m, \sigma_{2}^m), (\sigma_0^{m+1}, \sigma_1^{m+1}, \sigma_2^{m+1})} = \exp \bigg\{- \frac{H^m}{T}\bigg\},
	\end{equation}
	\[
	\begin{gathered}
		k = \frac{1 - \sigma_0^m}{2} + (1 - \sigma_1^m) + 2(1 - \sigma_2^m), \\
		l = \frac{1 - \sigma^{m+1}_0}{2} + (1 - \sigma^{m+1}_1) + 2(1 - \sigma^{m+1}_2).
	\end{gathered}
	\]
	\indent Then the partition function of the model is expressed as follows:
	\begin{equation} \label{13}
		Z_{L} = \operatorname{tr}(\theta^L).
	\end{equation}
	\indent The free energy of the system per lattice node is defined in the standard way \cite{Baxter}:
	\begin{equation} \label{nn1}
		f(H, T) = - \frac{T}{N} \ln Z_{L}(H, T),
	\end{equation}
	where $N=3L$ is the number of nodes of the considered lattice.\\
	\indent Internal energy per lattice node is equal to \cite{Baxter}:
	\begin{equation} \label{nnn2}
		u(H, T) = -T^2 \frac{\partial}{ \partial T}\biggl(\frac{f(H, T)}{T}\biggr). 
	\end{equation}
	\indent The specific heat per node is defined as \cite{Baxter}:
	\begin{equation} \label{nnn3}
		c(H, T) = \frac{\partial}{\partial T}u(H,T) = -2T\frac{\partial}{\partial T}\biggl(\frac{f(H, T)}{T}\biggr) - T^2 \frac{\partial^2}{\partial T^2}\biggl(\frac{f(H, T)}{T}\biggr).
	\end{equation}
	\indent Magnetization and susceptibility, respectively, are equal to \cite{Baxter}:
	\begin{equation} \label{last1}
		M(H,T)=-\frac{\partial}{\partial H} f(H,T),
	\end{equation}
	\begin{equation} \label{last2}
		\chi(H,T) = \frac{\partial}{\partial H} M(H,T),
	\end{equation}
	and entropy:
	\begin{equation} \label{last9}
		S(H,T)=-\frac{\partial}{\partial T} f(H,T).
	\end{equation}
	\indent To formulate the following theorems, we introduce matrices: 
	\begin{equation} \label{19}
		\tau^1 = 
		\begin{pmatrix}
			\tau_{11}^1& \tau_{12}^1& \tau_{13}^1& \tau_{14}^1\\
			\tau_{21}^1& \tau_{22}^1& \tau_{23}^1& \tau_{24}^1\\
			\tau_{31}^1& \tau_{32}^1& \tau_{33}^1& \tau_{34}^1\\
			\tau_{41}^1& \tau_{42}^1& \tau_{43}^1& \tau_{44}^1
		\end{pmatrix} =
		\begin{pmatrix}
			\theta_{11}& \theta_{12} + \theta_{13} + \theta_{15}& \theta_{14} + \theta_{16} + \theta_{17}& \theta_{18}\\
			\theta_{21}& \theta_{22} + \theta_{23} + \theta_{25}& \theta_{24} + \theta_{26} + \theta_{27}& \theta_{28}\\
			\theta_{41}& \theta_{42} + \theta_{43} + \theta_{45}& \theta_{44} + \theta_{46} + \theta_{47}& \theta_{48}\\
			\theta_{81}& \theta_{82} + \theta_{83} + \theta_{85}& \theta_{84} + \theta_{86} + \theta_{87}& \theta_{88}\\
		\end{pmatrix},
	\end{equation}
	\begin{equation} \label{n10}
		\tau^2 = 
		\begin{pmatrix}
			\tau_{11}^2& \tau_{12}^2\\
			\tau_{21}^2& \tau_{22}^2
		\end{pmatrix} =
		\begin{pmatrix}
			e^{\frac{2 \pi}{3} i }  \theta_{52} + e^{-\frac{2 \pi}{3} i }  \theta_{53} + \theta_{55}& e^{-\frac{2 \pi}{3} i }  \theta_{54} + e^{\frac{2 \pi}{3} i }  \theta_{56} + \theta_{57} \\
			e^{\frac{2 \pi}{3} i }  \theta_{72} + e^{-\frac{2 \pi}{3} i }  \theta_{73} + \theta_{75}& e^{-\frac{2 \pi}{3} i }  \theta_{74} + e^{\frac{2 \pi}{3} i }  \theta_{76} + \theta_{77}
		\end{pmatrix},
	\end{equation}
	\begin{equation} \label{n14}
		\tau^3 = 
		\begin{pmatrix}
			\tau_{11}^3& \tau_{12}^3\\
			\tau_{21}^3& \tau_{22}^3
		\end{pmatrix} =
		\begin{pmatrix}
			e^{-\frac{2 \pi}{3} i } \theta_{52} + e^{\frac{2 \pi}{3} i }  \theta_{53} + \theta_{55}& e^{\frac{2 \pi}{3} i }  \theta_{54} + e^{-\frac{2 \pi}{3} i }  \theta_{56} + \theta_{57} \\
			e^{-\frac{2 \pi}{3} i }  \theta_{72} + e^{\frac{2 \pi}{3} i } \theta_{73} + \theta_{75}& e^{\frac{2 \pi}{3} i }  \theta_{74} + e^{-\frac{2 \pi}{3} i }  \theta_{76} + \theta_{77} \\
		\end{pmatrix},
	\end{equation}
	and also expressions:
	\begin{equation} \label{ny1.1}
		\begin{gathered}
			b_0 = \frac{a}{2} - \sqrt{\frac{a^2}{4} -b + y_1},\\
			c_0 = \frac{y_1}{2} - \operatorname{sign}\biggl(\frac{a}{2}y_1 - c\biggr)\sqrt{\frac{y_1^2}{4} - d},\\
			b_1 = \frac{a}{2} + \sqrt{\frac{a^2}{4} -b + y_1},\\
			c_1 = \frac{y_1}{2} + \operatorname{sign}\biggl(\frac{a}{2}y_1 - c\biggr)\sqrt{\frac{y_1^2}{4} - d},
		\end{gathered}
	\end{equation}
	where $a, b, c, d$ are the coefficients of the characteristic equation of the fourth degree of the matrix (\ref{19}):
	\begin{equation} \label{ny1}
		\lambda^4 + a\lambda^3 + b\lambda^2 + c\lambda + d = 0,
	\end{equation}
	and $y_1$ is one of the roots of the cubic equation obtained for equation (\ref{ny1}) according to Ferrari's method described in the appendix.\\
	\begin{theorem} \label{th1}
		\textbf{\textit{Main theorem}}\\
		\indent\textit{In the thermodynamic limit, the free energy, internal energy, specific heat, magnetization, susceptibility per lattice node and entropy, respectively, are calculated as follows:}
		\begin{equation} \label{16}
			f(H, T) = -T\frac{\ln(\lambda_{max}(H,T))}{3},
		\end{equation}
		\begin{equation} \label{17}
			u(H, T) =  T^2 \frac{\partial}{\partial T} \frac{\ln(\lambda_{max}(H, T))}{3},
		\end{equation}
		\begin{equation} \label{18}
			c(H, T) = 2T \frac{\partial}{\partial T} \frac{\ln (\lambda_{max}(H,T))}{3} + T^2 \frac{\partial^2}{\partial T^2}\frac{\ln (\lambda_{max}(H, T))}{3},
		\end{equation}
		\begin{equation} \label{last3}
			M(H,T) = \frac{T}{3} \frac{1}{\lambda_{max}(H,T)} \frac{\partial \lambda_{max}(H,T)}{\partial H},
		\end{equation}
		\begin{equation} \label{last4}
			\chi(H,T) = \frac{T}{3}\biggl(\frac{1}{\lambda_{max}(H,T)}\frac{\partial^2 \lambda_{max}(H,T)}{\partial H^2}    +     \biggl(\frac{1}{\lambda_{max}(H,T)} \frac{\partial \lambda_{max}(H,T)}{\partial H}\biggr)^2 \biggr),
		\end{equation}
		\begin{equation} \label{last10}
			S(H,T) = \frac{1}{3} \biggl(\ln (\lambda_{max}(H,T)) + \frac{T}{\lambda_{max}(H,T)} \frac{\partial \lambda_{max}(H,T)}{\partial T}\biggr),
		\end{equation}
		\textit{where the largest eigenvalue $\lambda_{\max}(H, T)$ of the transfer-matrix $\theta$ has the form:}
		\begin{equation} \label{n5}
			\lambda_{max}(H,T) = \frac{-b_0 + \sqrt{b_0^2 - 4c_0}}{2}
		\end{equation}
		\textit{and its partial derivatives:}
		\begin{equation} \label{last11}
			\frac{\partial \lambda_{max}(H,T)}{\partial T} = -\frac{a'_T\lambda_{max}^3+b'_T\lambda_{max}^2 +c'_T\lambda_{max} + d'_T}{4\lambda_{max}^3+3a\lambda_{max}^2+2b\lambda_{max} + c},
		\end{equation}
		\begin{equation} \label{last5}
			\frac{\partial \lambda_{max}(H,T)}{\partial H} = -\frac{a'_H\lambda_{max}^3+b'_H\lambda_{max}^2 +c'_H\lambda_{max}}{4\lambda_{max}^3+3a\lambda_{max}^2+2b\lambda_{max} + c},
		\end{equation}
		\begin{equation} \label{last6}
			\begin{gathered}
				\frac{\partial^2 \lambda_{max}(H,T)}{\partial H^2} = -\frac{1}{4\lambda_{max}^3 + 3a\lambda_{max}^2 + 2b\lambda_{max} + c}  \biggl(12\lambda_{max}^2\biggl(\frac{\partial \lambda_{max}(H,T)}{\partial H}\biggr)^2  +\\ 
				+ a''_{HH}\lambda_{max}^3 +  6a'_H\lambda_{max}^2 \frac{\partial \lambda_{max}(H,T)}{\partial H}  + 6a\lambda_{max}\biggl(\frac{\partial \lambda_{max}(H,T)}{\partial H}\biggr)^2 + b''_{HH}\lambda_{max}^2 + \\
				+ 4b'_H\lambda_{max}\frac{\partial \lambda_{max}(H,T)}{\partial H}  + 2b\biggl(\frac{\partial \lambda_{max}(H,T)}{\partial H} \biggr)^2 + c''_{HH}\lambda_{max} + 2c'_H \frac{\partial \lambda_{max}(H,T)}{\partial H} \biggr).
			\end{gathered}
		\end{equation}
	\end{theorem}
	\begin{theorem} \label{th2}
		\textit{The partition function in a finite closed strip of length $L$ can be written as:}
		\begin{equation} \label{nnn1}
			Z_{L} = \sum_{k = 1}^{8}\lambda_k^L,
		\end{equation}
		\textit{where $\lambda_k$, $k=1,\ldots,8$ are the roots of the characteristic polynomial of the transfer matrix $\theta$, for which:}
		\begin{equation} \label{ny2}
			\lambda_{1,2} = \frac{-b_0 \pm \sqrt{b_0^2 - 4c_0}}{2},
		\end{equation}
		\begin{equation} \label{ny3}
			\lambda_{3,4} = \frac{-b_1 \pm \sqrt{b_1^2 - 4c_1}}{2}
		\end{equation}
		\textit{— are the roots of the characteristic equation} (\ref{ny1}) \textit{ of the matrix} (\ref{19})\textit{, found by the Ferrari method, and the eigenvalues:}
		\begin{equation} \label{ny4}
			\lambda_{5,6} = \frac{-b_2 \pm \sqrt{b_2^2 - 4c_2}}{2},
		\end{equation}
		\begin{equation} \label{ny4.1}
			\lambda_{7,8} = \frac{-b_3 \pm \sqrt{b_3^2 - 4c_3}}{2}
		\end{equation}
		\textit{- are respectively the roots of the characteristic square polynomials of the matrices} (\ref{n10}) \textit{and} (\ref{n14})\textit{, for which:}
		\begin{equation} \label{ny1.2}
			\begin{gathered}
				b_j = -(\tau_{11}^j + \tau_{22}^j),\\
				c_j = \tau_{11}^j\tau_{22}^j - \tau_{12}^j\tau_{21}^j, \\
				j = 2, 3.
			\end{gathered}
		\end{equation}
	\end{theorem}
	
	\newpage
	\section{The main special case: a model with multispin interactions of even number of spins}
	\subsection{Model description and results}
	\indent \\
	\indent In this case, the components $\hat{H}^m_1, \hat{H}^m_3, \hat{H}^m_5$ of the elementary generating Hamiltonian (\ref{3}) are zero, and the transfer-matrix $\theta$ has the center-symmetric form.\\
	\indent To formulate the following theorems, we introduce matrices:
	\begin{equation} \label{45}
		\tau^4 = 
		\begin{pmatrix}
			\tau_{11}^4& \tau_{12}^4\\
			\tau_{21}^4& \tau_{22}^4
		\end{pmatrix} =
		\begin{pmatrix}
			\theta_{11} + \theta_{18}& \theta_{12} + \theta_{13}+\theta_{14} + \theta_{15}+\theta_{16} + \theta_{17}\\
			\theta_{21} + \theta_{28}& \theta_{22} + \theta_{23}+\theta_{24} + \theta_{25}+\theta_{26} + \theta_{27}
		\end{pmatrix},
	\end{equation}
	\begin{equation} \label{54}
		\tau^5 = 
		\begin{pmatrix}
			\tau_{11}^5& \tau_{12}^5\\
			\tau_{21}^5& \tau_{22}^5
		\end{pmatrix} =
		\begin{pmatrix}
			\theta_{11} - \theta_{18}& \theta_{12} + \theta_{13} - \theta_{14} + \theta_{15} - \theta_{16} - \theta_{17}\\
			\theta_{21} - \theta_{28}& \theta_{22} + \theta_{23} - \theta_{24} + \theta_{25} - \theta_{26} - \theta_{27}
		\end{pmatrix}.
	\end{equation}
	\begin{theorem} \label{th3}
		\textit{In the thermodynamic limit, the free energy, internal energy, specific heat, and entropy per lattice node are respectively of the form} (\ref{16})\textit{,} (\ref{17})\textit{,} (\ref{18})\textit{,} (\ref{last10})\textit{, where the largest eigenvalue $\lambda_{max}$ of the transfer-matrix $\theta$ is of the form:} 
		\begin{equation} \label{ny5}
			\lambda_{max} = \frac{-b_4 + \sqrt{b_4^2-4c_4}}{2}
		\end{equation}
	\end{theorem}
	\begin{theorem} \label{th4}
		\textit{The partition function in a finite closed band of length $L$ can be written in the form} (\ref{nnn1})\textit{, where $\lambda_k, k = 1, \ldots, 8$ are the roots of the characteristic polynomial of the transfer-matrix $\theta$, for which:}
		\begin{equation} \label{ny6}
			\lambda_{1,2} = \frac{-b_4 \pm \sqrt{b_4^2 - 4c_4}}{2},
		\end{equation}
		\begin{equation} \label{ny7}
			\lambda_{3,4} = \frac{-b_5 \pm \sqrt{b_5^2 - 4c_5}}{2}
		\end{equation}
		\textit{— are respectively the roots of the characteristic square polynomials of matrices} (\ref{45}) \textit{and} (\ref{54})\textit{, for which:}
		\[
		\begin{gathered}
			b_j = -(\tau_{11}^j + \tau_{22}^j),\\
			c_j = \tau_{11}^j\tau_{22}^j - \tau_{12}^j\tau_{21}^j, \\
			j = 4, 5,
		\end{gathered}
		\]
		\textit{and the remaining eigenvalues are of the form:}
		\begin{equation} \label{61}
			\lambda_5 = \big(\theta_{22} + \theta_{27}\big) + e^{\frac{2\pi}{3}i}\big(\theta_{23} + \theta_{26}\big) + e^{-\frac{2\pi}{3}i}\big(\theta_{24} + \theta_{25}\big),
		\end{equation}
		\begin{equation} \label{62}
			\lambda_6 = \big(\theta_{22} - \theta_{27}\big) + e^{\frac{2\pi}{3}i}\big(\theta_{23} - \theta_{26}\big) - e^{-\frac{2\pi}{3}i}\big(\theta_{24} - \theta_{25}\big),
		\end{equation}
		\begin{equation} \label{63}
			\lambda_7 = \big(\theta_{22} + \theta_{27}\big) + e^{-\frac{2\pi}{3}i}\big(\theta_{23} + \theta_{26}\big) + e^{\frac{2\pi}{3}i}\big(\theta_{24} + \theta_{25}\big),
		\end{equation}
		\begin{equation} \label{64}
			\lambda_8 = \big(\theta_{22} - \theta_{27}\big) + e^{-\frac{2\pi}{3}i}\big(\theta_{23} - \theta_{26}\big) - 	e^{\frac{2\pi}{3}i}\big(\theta_{24} - \theta_{25}\big),
		\end{equation}
		\textit{where $\theta_{2l}$ are the elements of the transfer-matrix $\theta$, $l = 2, \ldots, 7$.}
	\end{theorem}
	\subsection{Pair correlations of the model with multispin interactions of even number of spins}
	\indent \\
	\indent The Hamiltonian of this model coincides with (\ref{5}), for which: 
	\begin{equation} \label{29.1}
		\begin{gathered}
			H^m = -\frac{J_1}{2}\bigg(\sigma_{0}^m\sigma_{1}^m + \sigma_{0}^m\sigma_{2}^m + \sigma_{1}^m\sigma_{2}^m + \sigma_{0}^{m+1}\sigma_{1}^{m+1} + \sigma_{0}^{m+1}\sigma_{2}^{m+1} + \sigma_{1}^{m+1}\sigma_{2}^{m+1}\bigg) - \\ -J_2\bigg(\sigma_{0}^m\sigma_{0}^{m+1} 
			+ \sigma_{1}^m\sigma_{1}^{m+1} + \sigma_{2}^m\sigma_{2}^{m+1}\bigg) - \\
			-J_3\bigg(\sigma_{0}^m\sigma_{1}^{m+1} + \sigma_{1}^m\sigma_{0}^{m+1} + \sigma_{2}^m\sigma_{0}^{m+1} + \sigma_{0}^m\sigma_{2}^{m+1} + \sigma_{1}^m\sigma_{2}^{m+1} + \sigma_{2}^m\sigma_{1}^{m+1}\bigg) - \\ -J_4\bigg(\sigma_{0}^m\sigma_{1}^m\sigma_{2}^m\sigma_{0}^{m+1} + \sigma_{0}^m\sigma_{1}^m\sigma_{2}^m\sigma_{1}^{m+1} + \sigma_{0}^m\sigma_{1}^m\sigma_{2}^m\sigma_{2}^{m+1} + \\
			+\sigma_{0}^m\sigma_{0}^{m+1}\sigma_{1}^{m+1}\sigma_{2}^{m+1} + \sigma_{1}^m\sigma_{0}^{m+1}\sigma_{1}^{m+1}\sigma_{2}^{m+1} + \sigma_{2}^m\sigma_{0}^{m+1}\sigma_{1}^{m+1}\sigma_{2}^{m+1}\bigg) - \\ -J_5\bigg(\sigma_{0}^m\sigma_{1}^m\sigma_{0}^{m+1}\sigma_{1}^{m+1} + \sigma_{0}^m\sigma_{2}^m\sigma_{0}^{m+1}\sigma_{2}^{m+1} + \sigma_{1}^m\sigma_{2}^m\sigma_{1}^{m+1}\sigma_{2}^{m+1}\bigg) - \\
			-J_6\bigg(\sigma_{0}^m\sigma_{1}^m\sigma_{0}^{m+1}\sigma_{2}^{m+1} + \sigma_{0}^m\sigma_{2}^m\sigma_{1}^{m+1}\sigma_{2}^{m+1} + \sigma_{1}^m\sigma_{2}^m\sigma_{0}^{m+1}\sigma_{1}^{m+1} + \sigma_{0}^m\sigma_{1}^m\sigma_{1}^{m+1}\sigma_{2}^{m+1} + \\ +\sigma_{0}^m\sigma_{2}^m\sigma_{0}^{m+1}\sigma_{1}^{m+1} + \sigma_{1}^m\sigma_{2}^m\sigma_{0}^{m+1}\sigma_{2}^{m+1}\bigg) - \\
			-J_7\sigma_{0}^m\sigma_{1}^m\sigma_{2}^m\sigma_{0}^{m+1}\sigma_{1}^{m+1}\sigma_{2}^{m+1},
		\end{gathered}
	\end{equation}
	and the transfer-matrix (\ref{12}) has the structure of the form
	\begin{equation} \label{29.2}
		\theta = 
		\begin{pmatrix}
			a&b&b&c&b&c&c&d\\
			b&e&f&g&f&g&h&c\\
			b&f&e&g&f&h&g&c\\
			c&g&g&e&h&f&f&b\\
			b&f&f&h&e&g&g&c\\
			c&g&h&f&g&e&f&b\\
			c&h&g&f&g&f&e&b\\
			d&c&c&b&c&b&b&a
		\end{pmatrix},
	\end{equation}
	for which
	\begin{equation} \label{29.3}
		\begin{gathered}
			a = \exp\bigg\{\frac{3J_1 + 3J_2 + 6J_3 + 6J_4 + 3J_5 + 6J_6 + J_7}{T}\bigg\},\\
			b = \exp\bigg\{\frac{J_1 + J_2 + 2J_3 - 2J_4 - J_5 - 2J_6 - J_7}{T}\bigg\},\\
			c = \exp\bigg\{\frac{J_1 - J_2 - 2J_3 + 2J_4 - J_5 - 2J_6 + J_7}{T}\bigg\},\\
			d = \exp\bigg\{\frac{3J_1 - 3J_2 - 6J_3 - 6J_4 + 3J_5 + 6J_6 - J_7}{T}\bigg\},\\
			e = \exp\bigg\{\frac{- J_1 + 3J_2 - 2J_3 - 2J_4 + 3J_5 - 2J_6 + J_7}{T}\bigg\},\\
			f = \exp\bigg\{\frac{- J_1 - J_2 + 2J_3 - 2J_4 - J_5 + 2J_6 + J_7}{T}\bigg\},\\
			g = \exp\bigg\{\frac{-J_1 + J_2 - 2J_3 + 2J_4 - J_5 + 2J_6 - J_7}{T}\bigg\},\\
			h = \exp\bigg\{\frac{-J_1 - 3J_2 + 2J_3 + 2J_4 + 3J_5 - 2J_6 - J_7}{T}\bigg\}.
		\end{gathered}
	\end{equation}
	and its eigenvalues (according to \ref{ny6} - \ref{64}):
	\begin{equation} \label{29.22} 
		\begin{gathered}
			\lambda_{1,2} = \frac{a + d + e + 2f + 2g + h \pm \sqrt{C_1}}{2},\\
			\lambda_{3,4} = \frac{a - d + e + 2f - 2g - h \pm \sqrt{C_2}}{2},\\
			\lambda_{5,6} = e + h - g - f,\\
			\lambda_{7,8} = e - h + g - f,
		\end{gathered}
	\end{equation}
	for which:
	\begin{equation} \label{29.23} 
		\begin{gathered}
			C_1 = \big(a+d-e-2f-2g-h\big)^2 +12\big(b+c\big)^2,\\
			C_2 = \big(-a+d+e+2f-2g-h\big)^2 + 12\big(b-c\big)^2.
		\end{gathered}
	\end{equation}
	\begin{theorem} \label{th7}
		\textit{In the thermodynamic limit the correlations of two spins for the model with Hamiltonian} (\ref{29.1}) \textit{are written as}:
		\begin{equation} \label{29.18} 
			\begin{gathered}
				G_{(i,0),(j,0)} = G_{(i,1),(j,1)} = G_{(i,2),(j,2)} = \\
				=\biggl(A_1B_2 + \frac{1}{3}B_1A_2\biggr)^2\biggl(\frac{\lambda_3}{\lambda_1}\biggr)^{j-i} + \biggl(A_1A_2 - \frac{1}{3}B_1B_2\biggr)^2\biggl(\frac{\lambda_4}{\lambda_1}\biggr)^{j-i} +\frac{8}{9}B_1^2\biggl(\frac{\lambda_7}{\lambda_1}\biggr)^{j-i},
			\end{gathered}
		\end{equation}
		\begin{equation} \label{29.19} 
			\begin{gathered}
				G_{(i,0),(j,1)} = G_{(i,0),(j,2)} = G_{(i,1),(j,2)} = \\
				=\biggl(A_1B_2 + \frac{1}{3}B_1A_2\biggr)^2\biggl(\frac{\lambda_3}{\lambda_1}\biggr)^{j-i} + \biggl(A_1A_2 - \frac{1}{3}B_1B_2\biggr)^2\biggl(\frac{\lambda_4}{\lambda_1}\biggr)^{j-i} - \\ - \frac{4}{9}B_1^2\biggl(\frac{\lambda_7}{\lambda_1}\biggr)^{j-i},
			\end{gathered}
		\end{equation}
		\textit{where}
		\begin{equation}
			\begin{gathered}
				A_1 = \sqrt{\frac{\sqrt{C_1} + a + d - e - 2f -2g - h}{2\sqrt{C1}}},\\
				B_1 = \frac{\sqrt{6}(b+c)}{\sqrt{(\sqrt{C_1} + a + d - e -2f - 2g - h)\sqrt{C_1}}}, \\
				A_2 = \sqrt{\frac{\sqrt{C_2} - a + d + e +2f - 2g - h}{2\sqrt{C_2}}},\\
				B_2 = \frac{\sqrt{6}(b-c)}{\sqrt{(\sqrt{C_2} - a + d + e + 2f - 2g - h)\sqrt{C_2}}}
			\end{gathered}
		\end{equation}
		\textit{and the average value of one spin is zero:}
		\begin{equation}  \label{29.20} 
			\big\langle\sigma\big\rangle  = 0
		\end{equation}
	\end{theorem}
	\indent \\
	\indent The correlations of two spins \cite{Baxter}, \cite{Yokota} are given by the expression:
	\begin{equation}  \label{29.4}
		G_{(i,\alpha),(j,\beta)} = \big\langle\sigma_{\alpha}^i \sigma_{ \beta}^j\big\rangle - \big\langle\sigma_{ \alpha}^i\big\rangle\big\langle\sigma_{\beta}^j\big\rangle,
	\end{equation}
	where the average values \cite{Binder} are calculated as follows \cite{Yokota}, \cite{Baxter}, \cite{Fisher}:
	\begin{equation}  \label{29.5}
		\big\langle\sigma_{\alpha}^i \sigma_{\beta}^j\big\rangle = \frac{\operatorname{Tr}\big(S_\alpha\theta^{j-i}S_\beta\theta^{L+i-j}\big)}{Z_L},
	\end{equation}
	\begin{equation}  \label{29.6}
		\big\langle\sigma_\gamma^k\big\rangle = \frac{\operatorname{Tr}\big(S_\gamma\theta^{L}\big)}{Z_L},
	\end{equation}
	where $S_l$ are the diagonal spin-layer matrices, $l = 0,1,2$, which for this model, according to the kind of transfer-matrix, are written as follows:
	\begin{equation} \label{29.7}
		S_0 = 
		\begin{pmatrix}
			1&0&0&0&0&0&0&0  \\
			0&-1&0&0&0&0&0&0 \\
			0&0&1&0&0&0&0&0   \\
			0&0&0&-1&0&0&0&0  \\
			0&0&0&0&1&0&0&0    \\
			0&0&0&0&0&-1&0&0   \\
			0&0&0&0&0&0&1&0     \\
			0&0&0&0&0&0&0&-1
		\end{pmatrix},
	\end{equation}
	\begin{equation} \label{29.8}
		S_1 = 
		\begin{pmatrix}
			1&0&0&0&0&0&0&0\\
			0&1&0&0&0&0&0&0\\
			0&0&-1&0&0&0&0&0\\
			0&0&0&-1&0&0&0&0\\
			0&0&0&0&1&0&0&0\\
			0&0&0&0&0&1&0&0\\
			0&0&0&0&0&0&-1&0\\
			0&0&0&0&0&0&0&-1
		\end{pmatrix},
	\end{equation}
	\begin{equation} \label{29.9}
		S_2 = 
		\begin{pmatrix}
			1&0&0&0&0&0&0&0\\
			0&1&0&0&0&0&0&0\\
			0&0&1&0&0&0&0&0\\
			0&0&0&1&0&0&0&0\\
			0&0&0&0&-1&0&0&0\\
			0&0&0&0&0&-1&0&0\\
			0&0&0&0&0&0&-1&0\\
			0&0&0&0&0&0&0&-1
		\end{pmatrix}.
	\end{equation}
	\indent The expression (\ref{29.5}) can be rewritten as:
	\begin{equation} \label{29.10}
		\big\langle\sigma_{\alpha}^i \sigma_{\beta}^j\big\rangle = \frac{\operatorname{Tr}\big(\theta^{i-1}S_{\alpha}\theta^{j-i}S_{\beta}\theta^{L-j+1}\big)}{\operatorname{Tr}\theta^L} = \frac{\operatorname{Tr}\big(\Lambda^{i-1}S'_{\alpha}\Lambda^{j-i}S'_{\beta}\Lambda^{L-j+1}\big)}{\operatorname{Tr}\Lambda^L}.
	\end{equation}
	\indent Similarly, the expression (\ref{29.6}) is written in the form
	\begin{equation}  \label{29.11}
		\big\langle\sigma_\gamma^k\big\rangle = \frac{\operatorname{Tr}\big(S'_\gamma\Lambda^{L}\big)}{\operatorname{Tr}\Lambda^L},
	\end{equation}
	for which $\Lambda$ is the diagonal matrix obtained from the transfer-matrix by diagonalization transformation, and 
	\begin{equation}  \label{29.12}
		S'_l = P^{-1}S_lP,
	\end{equation} 
	where $P$ is the transition matrix of the transfer-matrix to the diagonal form:
	\begin{equation} \label{29.13}
		P = \frac{1}{\sqrt{6}}
		\begin{pmatrix}
			\sqrt{3}A_1&\sqrt{3}B_1&0&0&\sqrt{3}B_2&-\sqrt{3}A_2&0&0\\
			B_1&-A_1&\omega&1&A_2&B_2&-\omega&-1\\
			B_1&-A_1&\omega^2&\omega^2&A_2&B_2&-\omega^2&-\omega^2\\
			B_1&-A_1&1&\omega&-A_2&-B_2&1&\omega\\
			B_1&-A_1&1&\omega&A_2&B_2&-1&-\omega\\
			B_1&-A_1&\omega^2&\omega^2&-A_2&-B_2&\omega^2&\omega^2\\
			B_1&-A_1&\omega&1&-A_2&-B_2&\omega&1\\
			\sqrt{3}A_1&\sqrt{3}B_1&0&0&-\sqrt{3}B_2&\sqrt{3}A_2&0&0
		\end{pmatrix},
	\end{equation}
	\begin{equation} \label{29.14}
		P^{-1} = \frac{1}{\sqrt{6}}
		\begin{pmatrix}
			\sqrt{3}A_1&B_1&B_1&B_1&B_1&B_1&B_1&\sqrt{3}A_1\\
			\sqrt{3}B_1&-A_1&-A_1&-A_1&-A_1&-A_1&-A_1&\sqrt{3}B_1\\
			0&\bar{\omega}&\bar{\omega}^2&1&1&\bar{\omega}^2&\bar{\omega}&0\\
			0&1&\bar{\omega}^2&\bar{\omega}&\bar{\omega}&\bar{\omega}^2&1&0\\
			\sqrt{3}B_2&A_2&A_2&-A_2&A_2&-A_2&-A_2&-\sqrt{3}B_2\\
			-\sqrt{3}A_2&B_2&B_2&-B_2&B_2&-B_2&-B_2&\sqrt{3}A_2\\
			0&-\bar{\omega}&-\bar{\omega}^2&1&-1&\bar{\omega}^2&\bar{\omega}&0\\
			0&-1&-\bar{\omega}^2&\bar{\omega}&-\bar{\omega}&\bar{\omega}^2&1&0
		\end{pmatrix},
	\end{equation}
	\begin{equation} \label{29.15}
		\begin{gathered}
			\omega = e^{\frac{2\pi}{3}i}
		\end{gathered}
	\end{equation}
	\indent From the expressions (\ref{29.12}) — (\ref{29.14}) we obtain the matrix $S'_0$:
	\begin{equation} \label{29.16} 
		\begin{psmallmatrix}
			0&0&0&0&A_1B_2+\frac{B_1A_2}{3}&\frac{B_1B_2}{3}-A_1A_2&-\frac{2}{3}e^{-\frac{\pi i}{3}}B_1&\frac{2}{3}B_1\\
			0&0&0&0&B_1B_2-\frac{A_1A_2}{3}&-\frac{A_1B_2}{3}-B_1A_2&\frac{2}{3}e^{-\frac{\pi i}{3}}A_1&-\frac{2}{3}A_1\\
			0&0&0&0&\frac{2}{3}e^{\frac{\pi i}{3}}A_2&\frac{2}{3}e^{\frac{\pi i}{3}}B_2&-\frac{1}{3}&-\frac{2}{3}e^{\frac{\pi i}{3}}\\
			0&0&0&0&-\frac{2}{3}A_2&-\frac{2}{3}B_2&-\frac{2}{3}e^{-\frac{\pi i}{3}}&-\frac{1}{3}\\
			A_1B_2+\frac{B_1A_2}{3}&B_1B_2-\frac{A_1A_2}{3}&\frac{2}{3}e^{-\frac{\pi i}{3}}A_2&-\frac{2}{3}A_2&0&0&0&0\\
			\frac{B_1B_2}{3}-A_1A_2&-B_1A_2-\frac{A_1B_2}{3}&\frac{2}{3}e^{-\frac{\pi i}{3}}B_2&-\frac{2}{3}B_2&0&0&0&0\\
			-\frac{2}{3}e^{\frac{\pi i}{3}}B_1&\frac{2}{3}e^{\frac{\pi i}{3}}A_1&-\frac{1}{3}&-\frac{2}{3}e^{\frac{\pi i}{3}}&0&0&0&0\\
			\frac{2}{3}B_1&-\frac{2}{3}A_1&-\frac{2}{3}e^{-\frac{\pi i}{3}}&-\frac{1}{3}&0&0&0&0
		\end{psmallmatrix},
	\end{equation}
	and the matrix $S'_1$:
	\begin{equation} \label{29.17} 
		\begin{psmallmatrix}
			0&0&0&0&A_1B_2+\frac{B_1A_2}{3}&\frac{B_1B_2}{3}-A_1A_2&-\frac{2}{3}e^{\frac{\pi i}{3}}B_1&-\frac{2}{3}e^{\frac{\pi i}{3}}B_1\\
			0&0&0&0&B_1B_2-\frac{A_1A_2}{3}&-\frac{A_1B_2}{3}-B_1A_2&\frac{2}{3}e^{\frac{\pi i}{3}}A_1&\frac{2}{3}e^{\frac{\pi i}{3}}A_1\\
			0&0&0&0&\frac{2}{3}e^{-\frac{\pi i}{3}}A_2&\frac{2}{3}e^{-\frac{\pi i}{3}}B_2&-\frac{1}{3}&\frac{2}{3}\\
			0&0&0&0&\frac{2}{3}e^{-\frac{\pi i}{3}}A_2&\frac{2}{3}e^{-\frac{\pi i}{3}}B_2&\frac{2}{3}&-\frac{1}{3}\\
			A_1B_2+\frac{B_1A_2}{3}&B_1B_2-\frac{A_1A_2}{3}&\frac{2}{3}e^{\frac{\pi i}{3}}A_2&\frac{2}{3}e^{\frac{\pi i}{3}}A_2&0&0&0&0\\
			\frac{B_1B_2}{3}-A_1A_2&-B_1A_2-\frac{A_1B_2}{3}&\frac{2}{3}e^{\frac{\pi i}{3}}B_2&\frac{2}{3}e^{\frac{\pi i}{3}}B_2&0&0&0&0\\
			-\frac{2}{3}e^{-\frac{\pi i}{3}}B_1&\frac{2}{3}e^{-\frac{\pi i}{3}}A_1&-\frac{1}{3}&\frac{2}{3}&0&0&0&0\\
			-\frac{2}{3}e^{-\frac{\pi i}{3}}B_1&\frac{2}{3}e^{-\frac{\pi i}{3}}A_1&\frac{2}{3}&-\frac{1}{3}&0&0&0&0
		\end{psmallmatrix}.
	\end{equation}
	\indent From (\ref{29.10}), taking into account (\ref{29.16}) and (\ref{29.17}), we obtain the expressions for $\big\langle\sigma_{0}^i \sigma_{0}^j\big\rangle$ and $\big\langle\sigma_{0}^i \sigma_{1}^j\big\rangle$, which in the thermodynamic limit at $L \to \infty$ coincide with (\ref{29.18}) and (\ref{29.19}), and $\big\langle\sigma_{1}^i \sigma_{1}^j\big\rangle= \big\langle\sigma_{2}^i \sigma_{2}^j\big\rangle =\big\langle\sigma_{0}^i \sigma_{0}^j\big\rangle$ and $\big\langle\sigma_{0}^i \sigma_{2}^j\big\rangle =  \big\langle\sigma_{1}^i \sigma_{2}^j \big\rangle  = \big\langle\sigma_{0}^i\sigma_{1}^j\big\rangle$, given the invariance of the system with respect to rotation.\\
	\indent Thus, from (\ref{29.11}), given (\ref{29.16}), (\ref{29.17}), we have:
	\begin{equation}  \label{29.201} 
		\big\langle\sigma_{0}^{k_0}\big\rangle = \big\langle\sigma_{1}^{k_1}\big\rangle = \big\langle\sigma_{2}^{k_2}\big\rangle = 0
	\end{equation}
	\indent Note that it follows from the expression (\ref{29.201}) that the magnetization of the system is zero in this case, and the correlations of the two spins (\ref{29.4}) coincide with (\ref{29.18}) and (\ref{29.19}).
	\subsection{Ground states and correlation length of the model with multispin interactions of even number of spins}
\begin{figure}[h]
		{\includegraphics[width=1\linewidth]{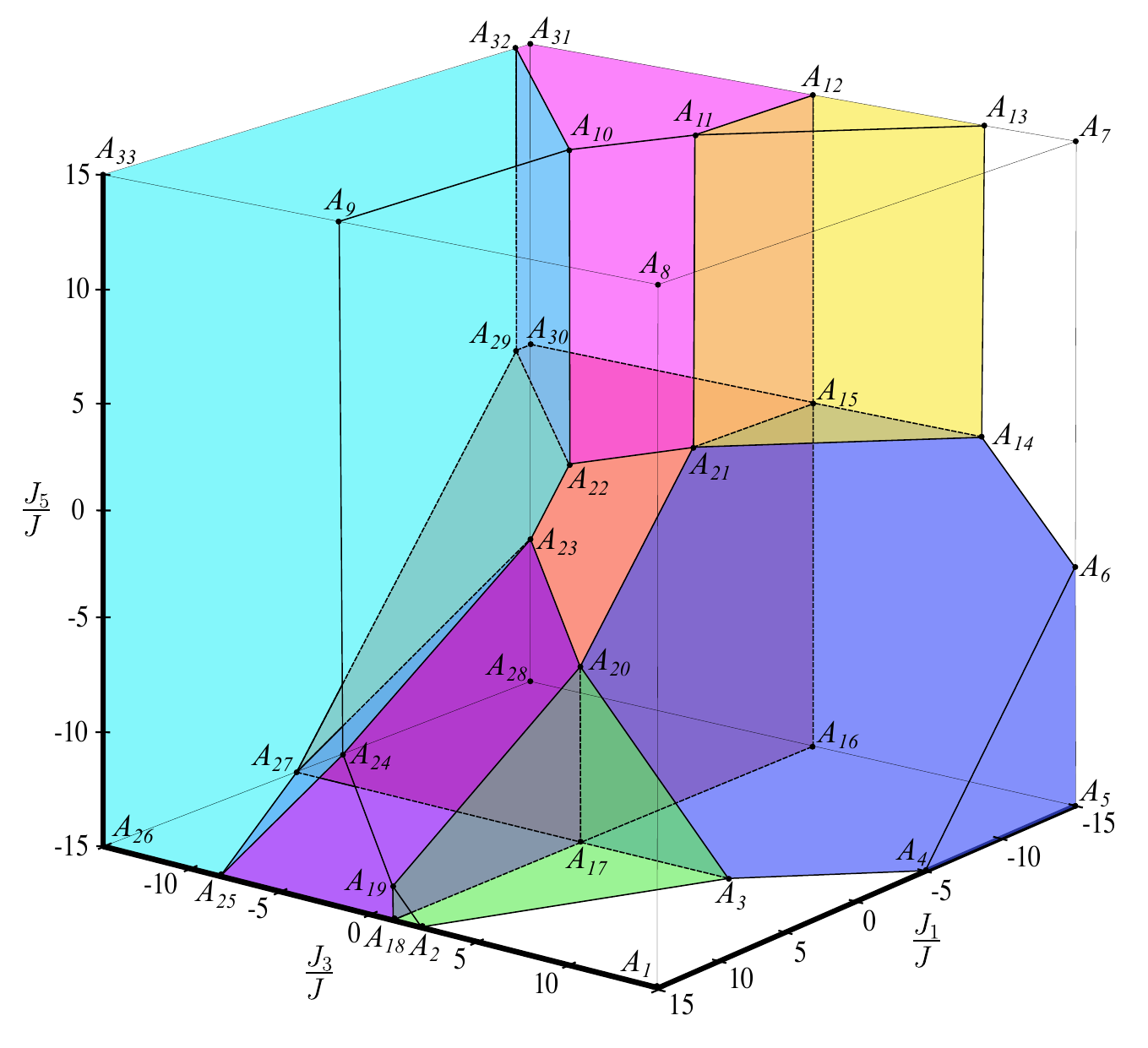}}
		\caption{The ground states of the model in coordinates $\frac{J_1}{J}, \frac{J_3}{J}, \frac{J_5}{J}$ at $J_2=J_4=J_6=J_7=J$} \label{ris1}
	\end{figure}

\begin{figure}[h]
		\begin{minipage}[h]{0.49\linewidth}
			\center{\includegraphics[width=1\linewidth]{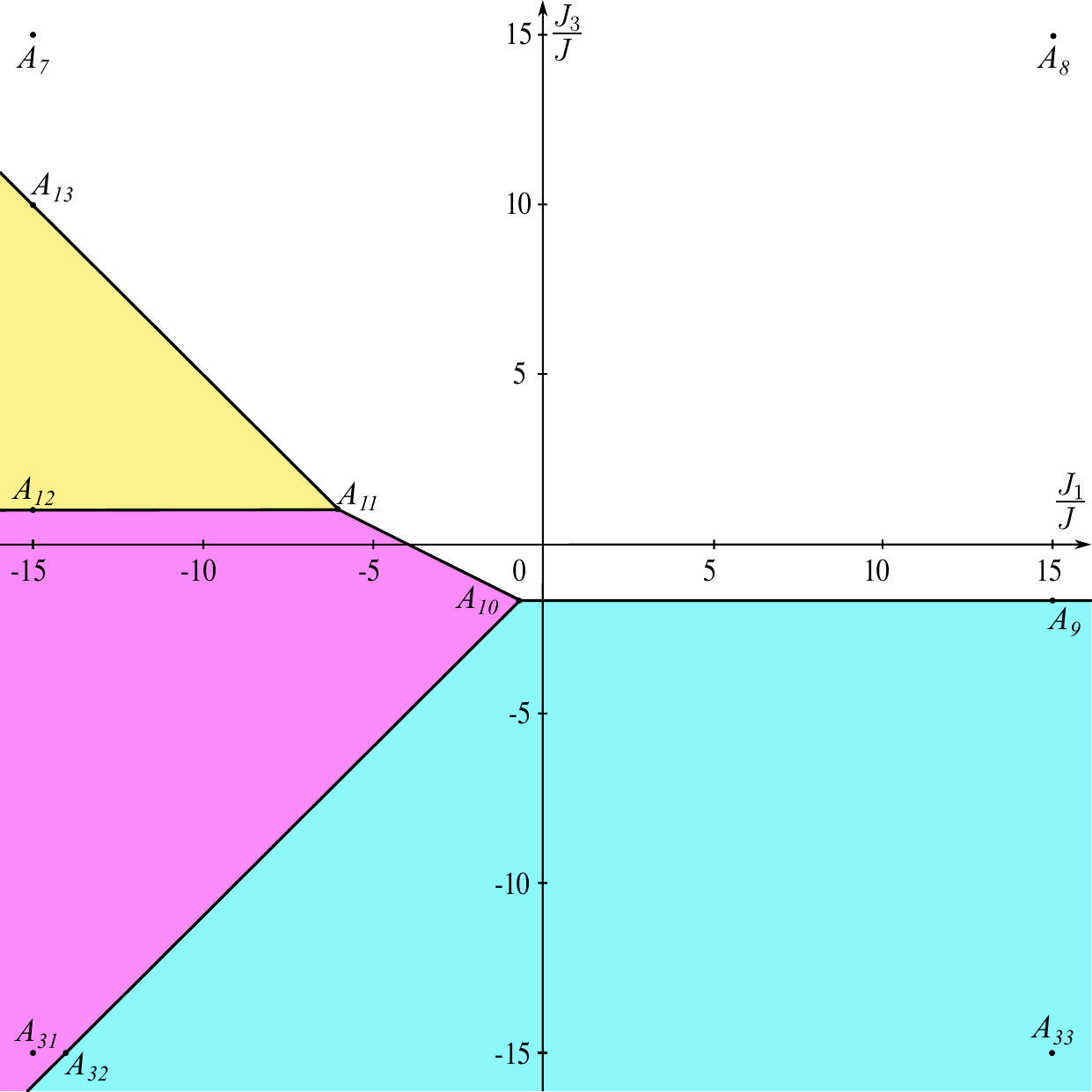} \\ a)}
		\end{minipage}
		\hfill
		\begin{minipage}[h]{0.49\linewidth}
			\center{\includegraphics[width=1\linewidth]{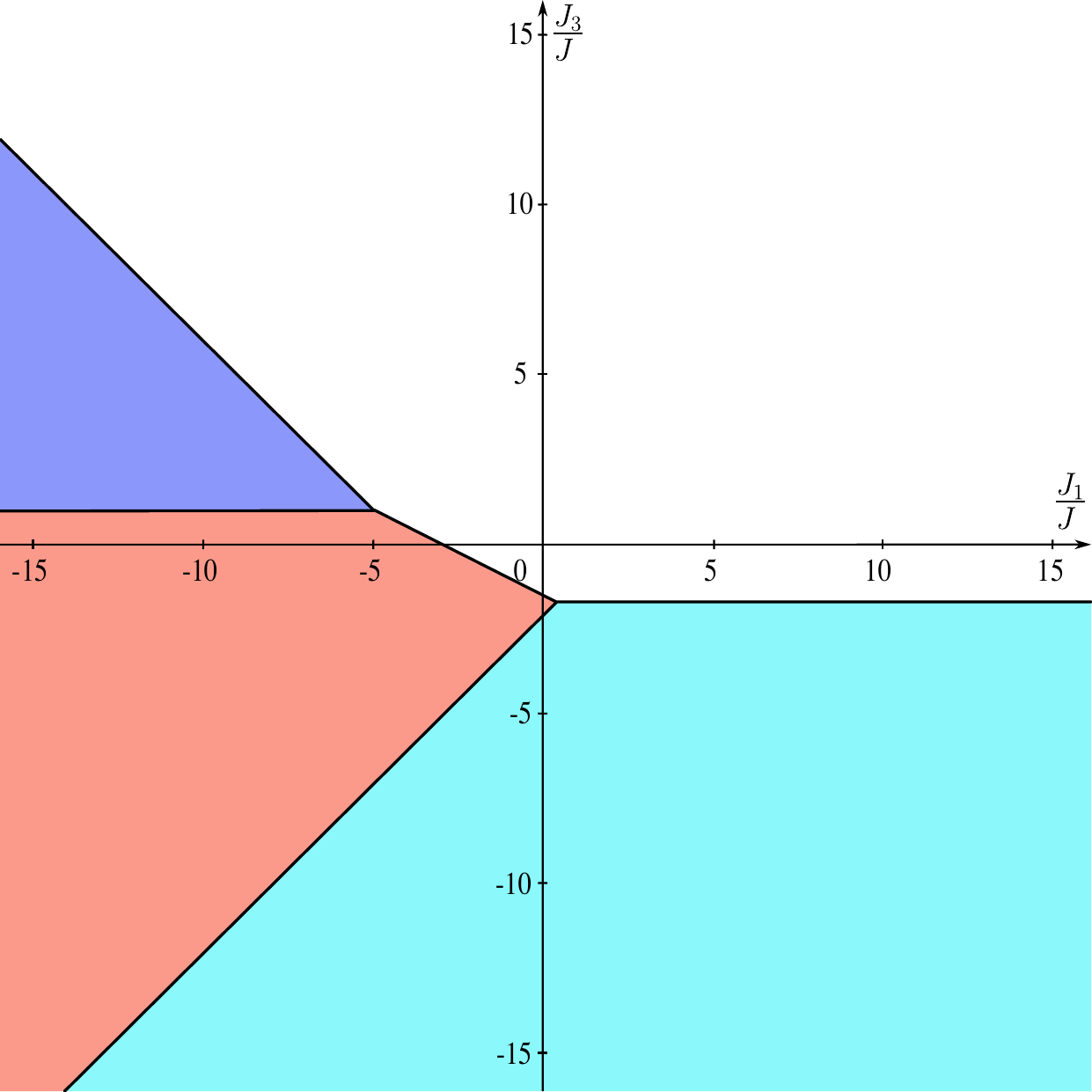} \\ b)}
		\end{minipage}
		\vfill
		\begin{minipage}[h]{0.49\linewidth}
			\center{\includegraphics[width=1\linewidth]{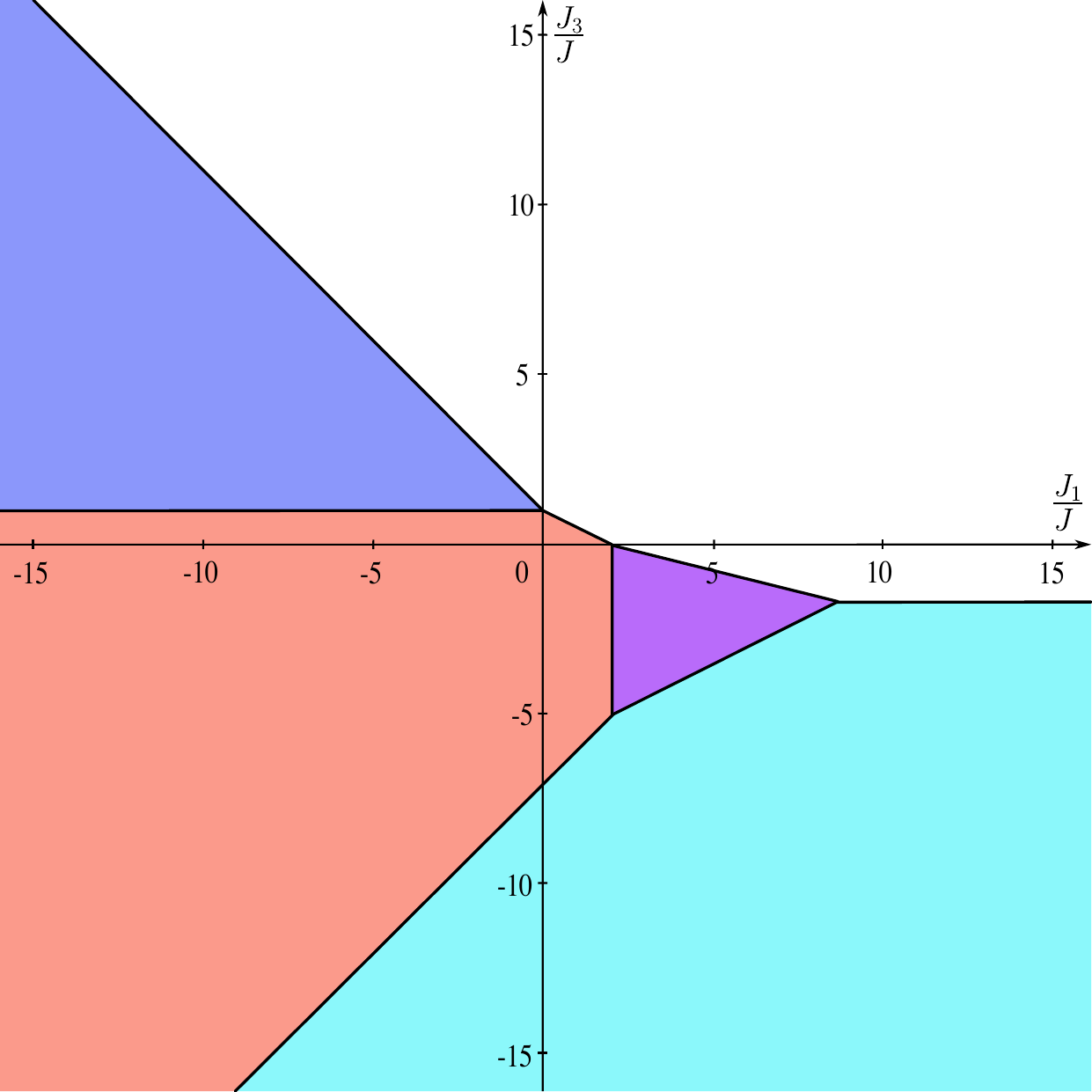} \\ c)} \\
		\end{minipage}
		\hfill
		\begin{minipage}[h]{0.49\linewidth}
			\center{\includegraphics[width=1\linewidth]{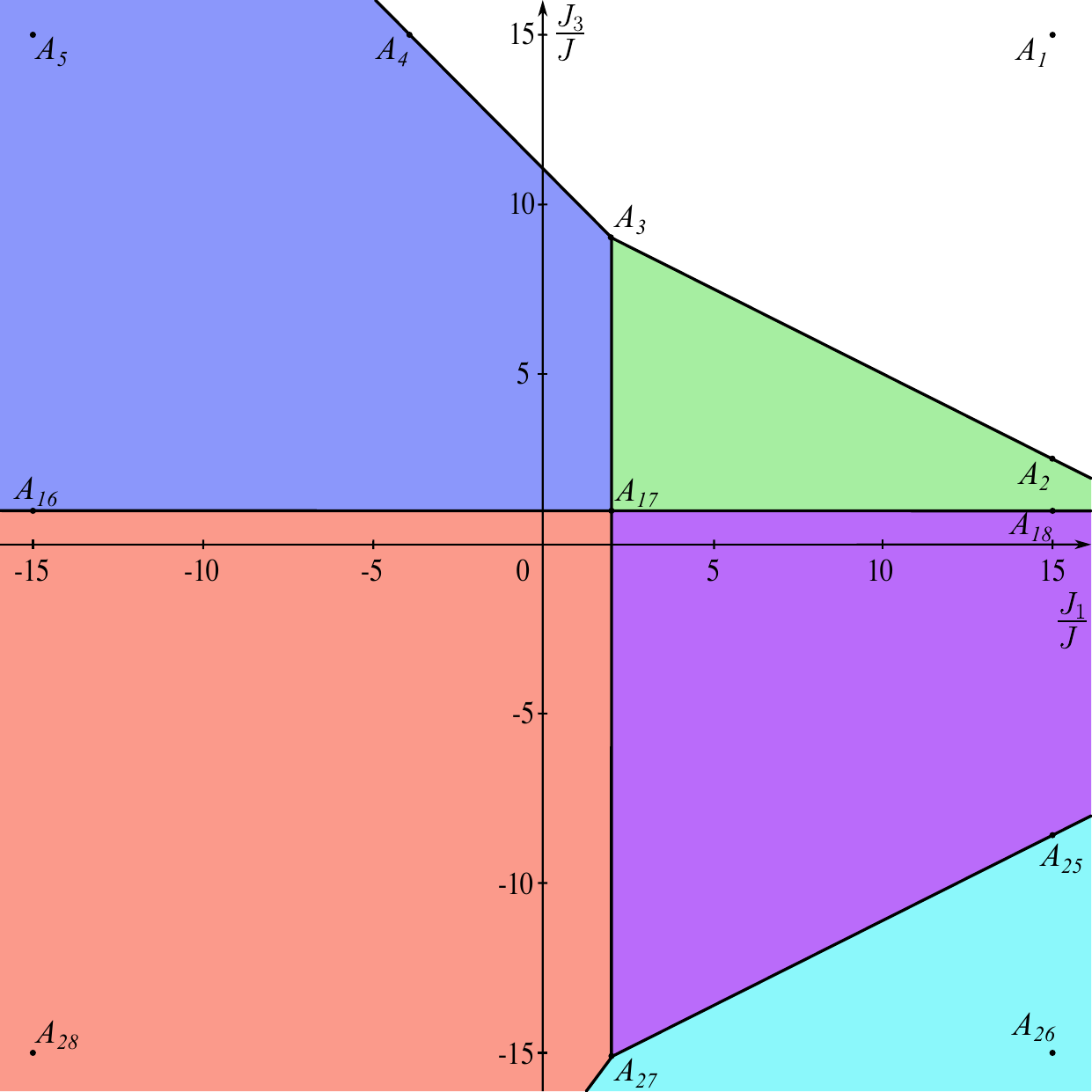} \\ d)} \\
		\end{minipage}
		\caption{Ground states of the model in coordinates $\frac{J_1}{J}, \frac{J_3}{J}$ at $J_2=J_4=J_6=J_7=J$ and a) $\frac{J_5}{J}=15$, b) $\frac{J_5}{J}=0$, c) $\frac{J_5}{J}=-5$, d) $\frac{J_5}{J}=-15$} \label{ris33}
	\end{figure}

	\indent \\
	\indent It follows from the structure of the transfer-matrix (\ref{29.2}) that the Hamiltonian (\ref{29.1}) corresponding to one prism of the $\Omega_P^m$ model can take eight different values depending on the configuration at fixed values of the interspin interactions $J_i$. Fixing the interactions $J_2=J_4=J_6=J_7=J$ and varying $\frac{J_1}{J}, \frac{J_3}{J}, \frac{J_5}{J}$ from $-15$ to $15$, we show a picture of the ground states obtained by minimizing the \cite{Kashapov} (\ref{29. 1}) in this section of the seven-dimensional space (Fig. \ref{ris1}), as well as its section at $\frac{J_5}{J} = 15, 0 , -5, -15$ (Fig. \ref{ris33}). \\
	\indent The element of the transfer-matrix corresponding to this ground state, the vertices of the bodies bounding the regions of ground states given in the figures, the color designation, and one of the configurations corresponding to this ground state are given in the table (\ref{table2}), and the coordinates of the indicated vertices are given in the table (\ref{table3}).\\
	\indent Note that all model configurations that are in the same class \cite{Kashapov} as those given in the table (\ref{table3}) also correspond to ground states. \\
	\indent Moreover, the number of configurations in the ground state to which the elements $b,c,f,g$ of the transfer-matrix correspond is infinite in the thermodynamic limit according to the structure of this transfer-matrix. 
	\clearpage
	
	\newpage
	\indent \\
	\indent \\
	\indent \\
	\indent \\
	\indent \\
	\indent \\
	\clearpage
	
\begin{figure}[h]
		\center{\includegraphics[width=1\linewidth]{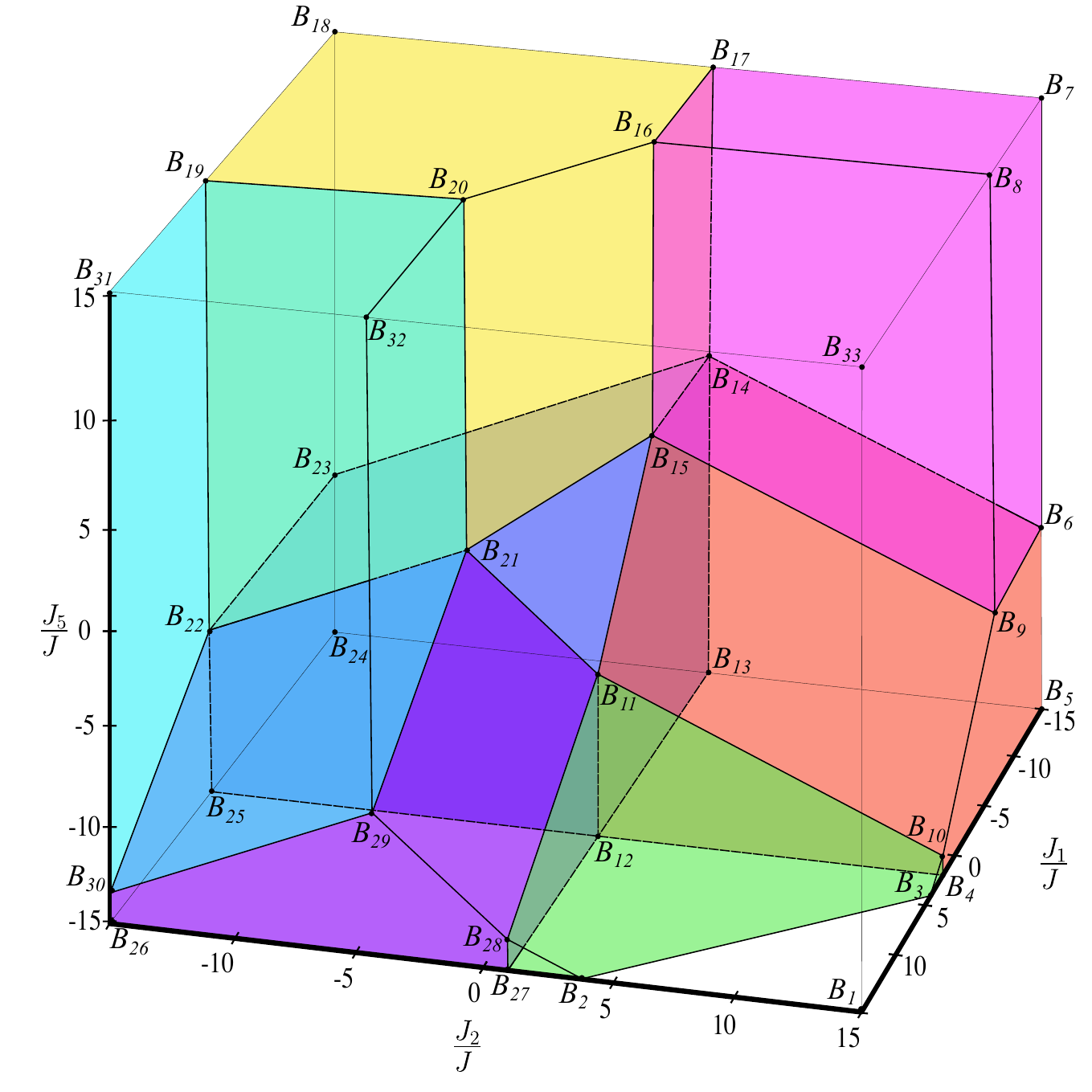}}
		\caption{Ground states of the model in coordinates $\frac{J_1}{J}, \frac{J_2}{J}, \frac{J_5}{J}$ at $J_3=J_4=J_6=J_7=J$}  \label{ris444}
	\end{figure}

\begin{figure}[h]
		\begin{minipage}[h]{0.49\linewidth}
			\center{\includegraphics[width=1\linewidth]{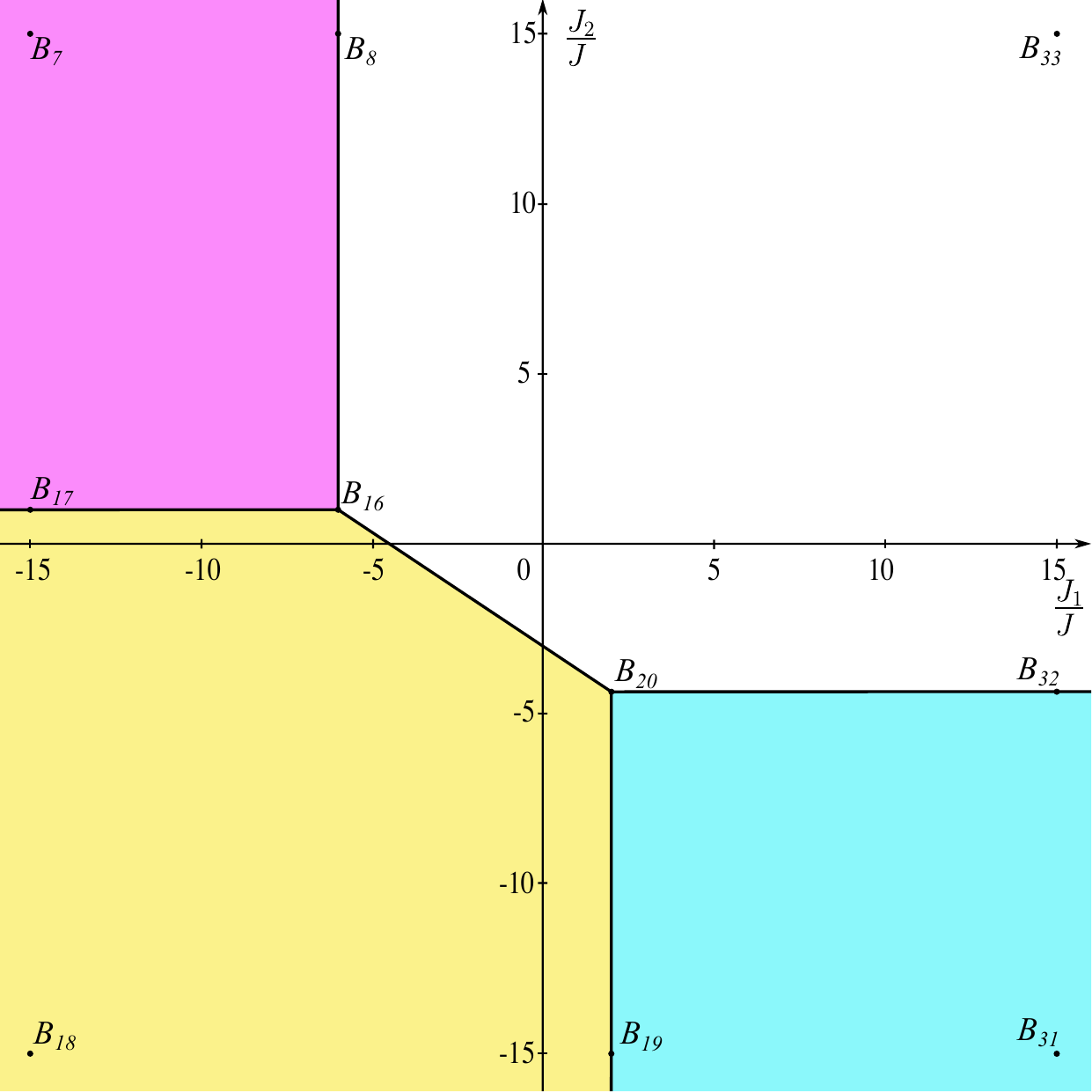} \\ a)} \\
		\end{minipage}
		\hfill
		\begin{minipage}[h]{0.49\linewidth}
			\center{\includegraphics[width=1\linewidth]{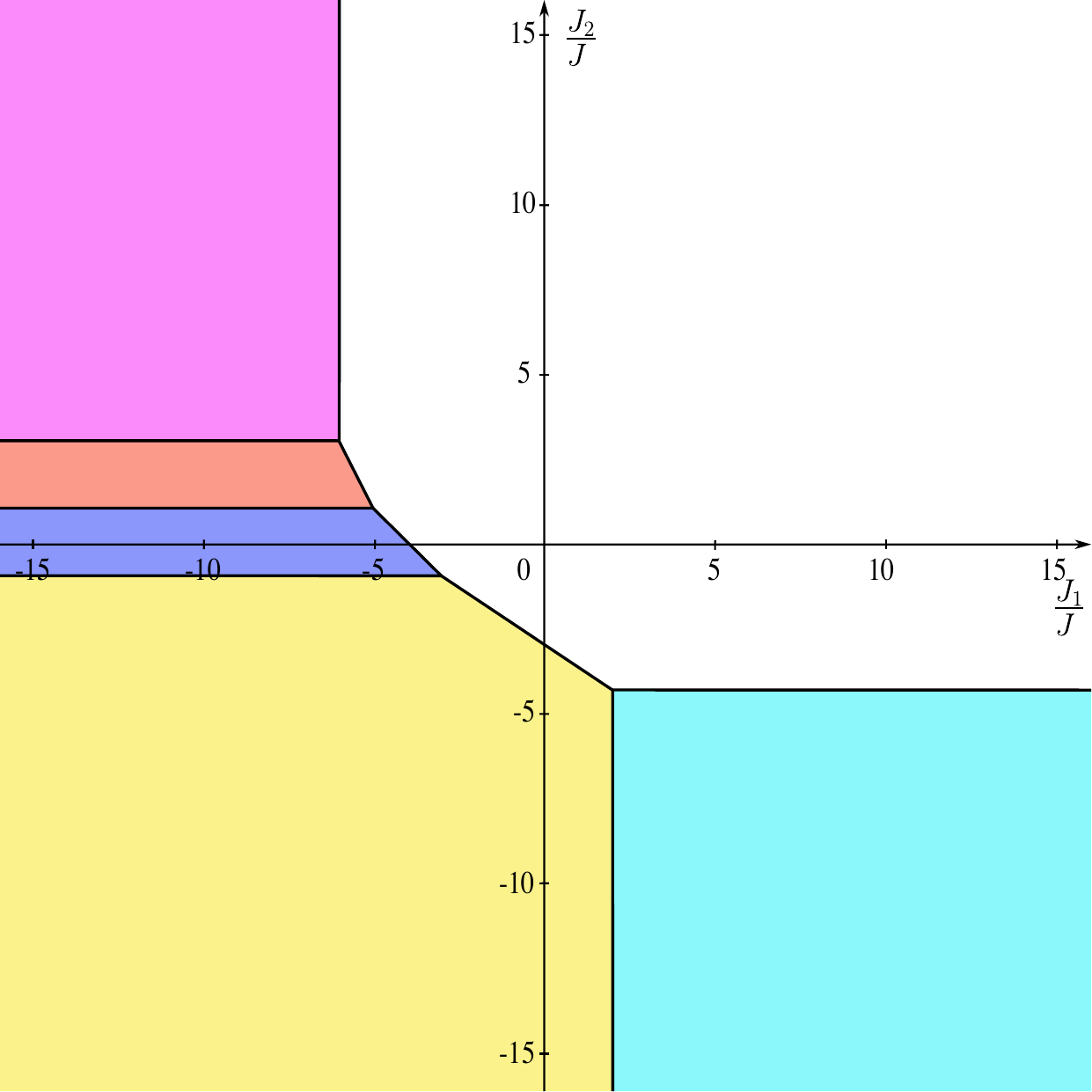} \\ b)} \\
		\end{minipage}
		\vfill
		\begin{minipage}[h]{0.49\linewidth}
			\center{\includegraphics[width=1\linewidth]{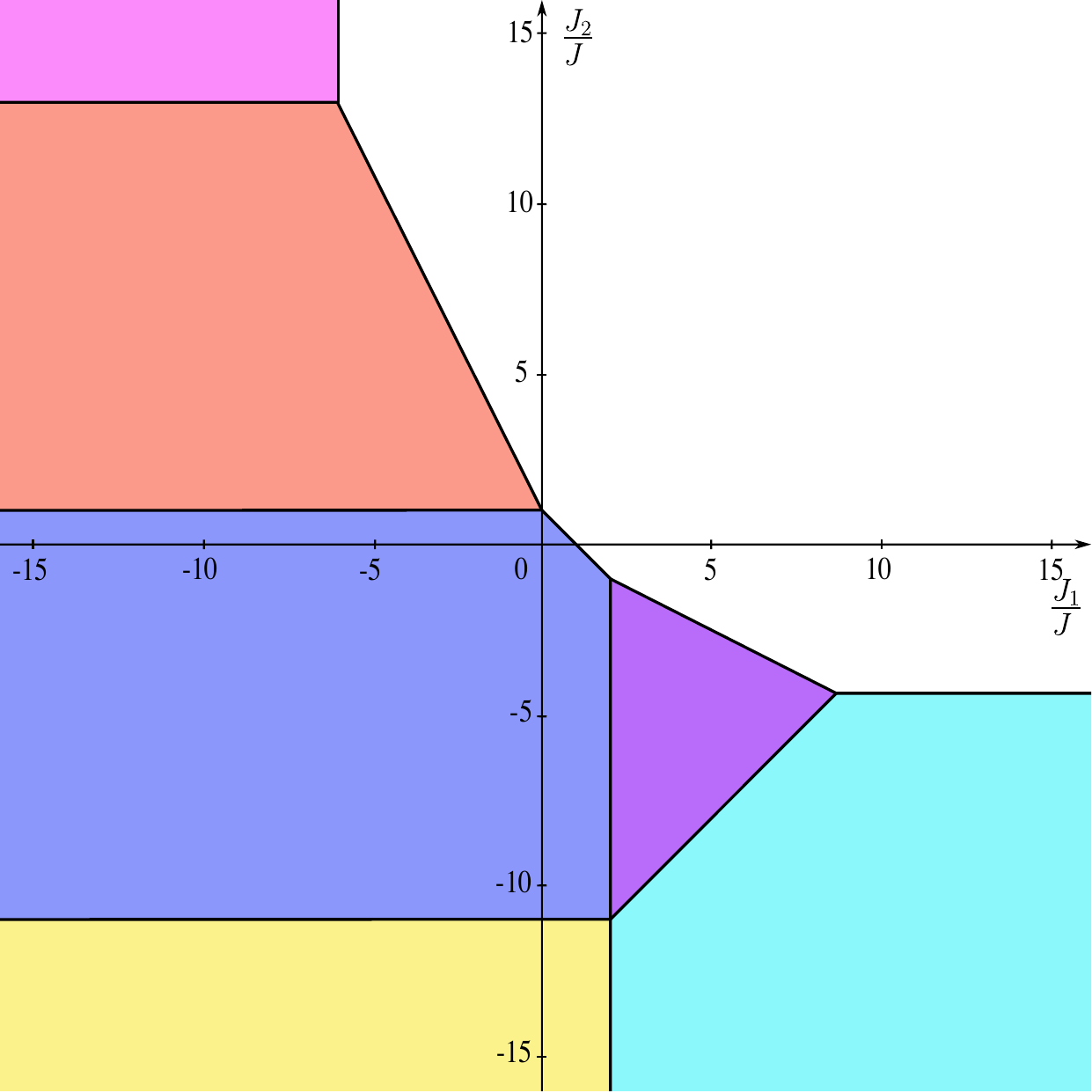} \\ c)} \\
		\end{minipage}
		\hfill
		\begin{minipage}[h]{0.49\linewidth}
			\center{\includegraphics[width=1\linewidth]{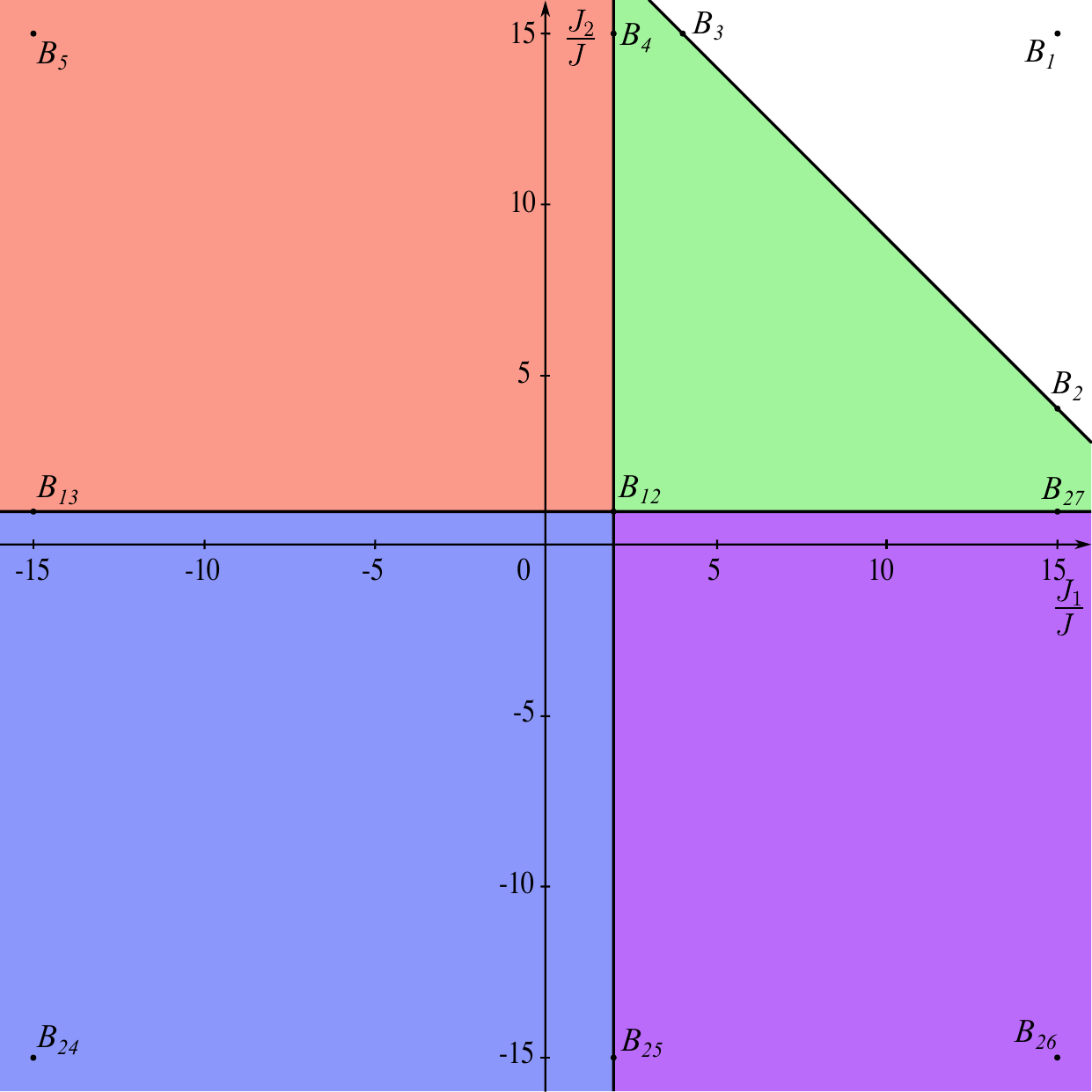} \\ d)} \\
		\end{minipage}
		\caption{Ground states of the model in coordinates $\frac{J_1}{J}, \frac{J_2}{J}$ at $J_3=J_4=J_6=J_7=J$ and a)$\frac{J_5}{J}=15$, b)$\frac{J_5}{J}=0$, c)$\frac{J_5}{J}=-5$, d)$\frac{J_5}{J}=-15$} \label{ris66}
	\end{figure}

	\newpage
	\indent Similarly, fixing the interactions $J_3=J_4=J_6=J_7 = J$ and varying $\frac{J_1}{J}, \frac{J_2}{J}, \frac{J_5}{J}$ from $-15$ to $15$, we show a picture of the ground states in a given section of seven-dimensional space (Fig. \ref{ris444}), as well as its cross-section at $\frac{J_5}{J} = 15, 0 , -5, -15$ (Fig. \ref{ris66}).
	\indent \\
	\indent \\
	\indent \\
	\indent \\
	\indent \\
	\indent \\
	\clearpage
	
	\newpage
	\indent \\
	\indent \\
	\indent \\
	\indent \\
	\indent \\
	\indent \\
	\clearpage
	
	\newpage
	\begin{table}[h]
		\centering
		\caption{Transfer-matrix elements, boundary points, color designations, and possible configurations corresponding to the ground states indicated in the figures.} 	\label{table2}
		\begin{tabular}{| c | c | c |} \hline
			\makecell{Transfer-matrix element and \\ boundary points} & \makecell{Color \\ designation} & Configuration example \\ \hline
			\makecell{$a$,\\ $A_1A_2A_3A_4A_6A_7A_8A_9A_{10}A_{11}$\\
				$A_{13}A_{14}A_{19}A_{20}A_{21}A_{22}A_{23}A_{24}$,\\
				$B_1B_2B_3B_8B_9B_{10}B_{11}$\\
				$B_{15}B_{16}B_{20}B_{21}B_{28}B_{29}B_{32}$} & \makecell{\includegraphics[width=0.2\linewidth]{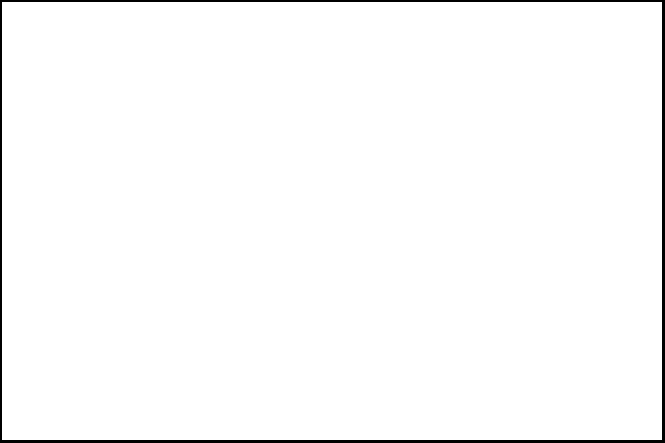}}  & \makecell{\includegraphics[width=0.4\textwidth]{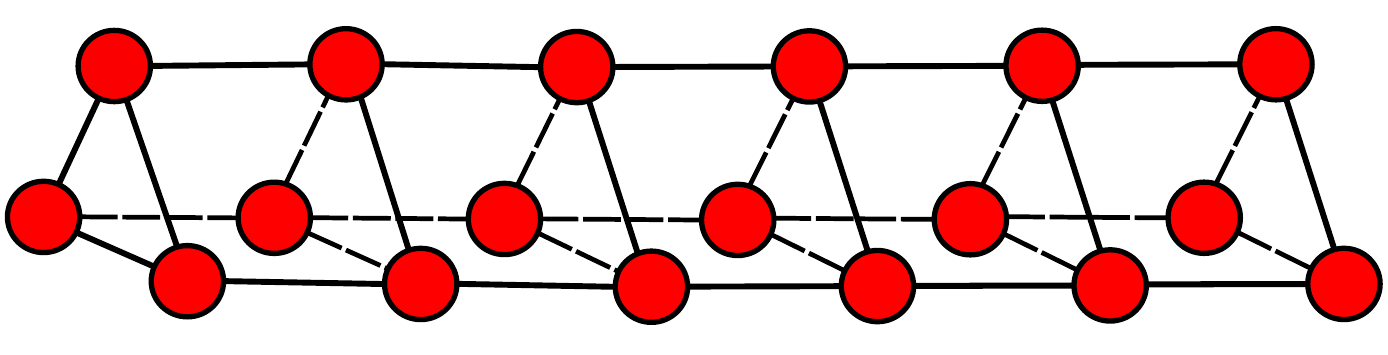}} \\ \hline
			\makecell{$b$,\\ $A_2A_3A_{17}A_{18}A_{19}A_{20}$,\\ $B_2B_3B_4B_{10}B_{11}B_{12}B_{27}B_{28}$} &	\makecell{\includegraphics[width=0.2\linewidth]{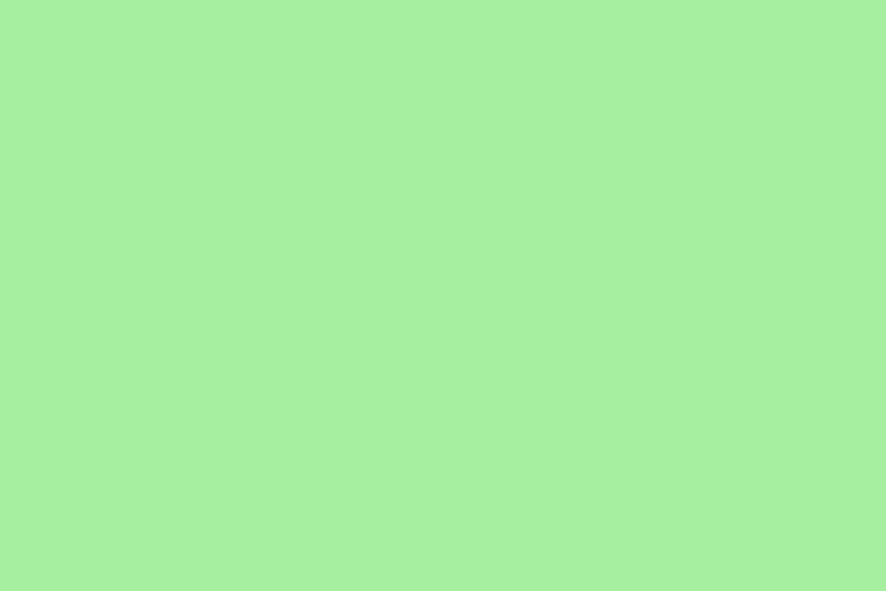} \\} & \makecell{\includegraphics[width=0.4\textwidth]{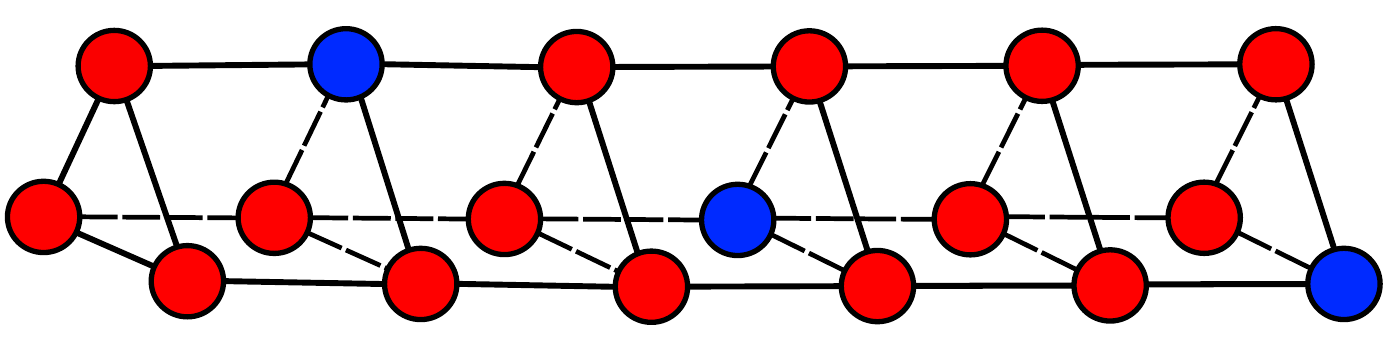}} \\ \hline
			\makecell{$c$,\\ $A_{17}A_{18}A_{19}A_{20}A_{23}A_{24}A_{25}A_{27}$,\\ $B_{11}B_{12}B_{21}B_{22}B_{25}$\\
				$B_{26}B_{27}B_{28}B_{29}B_{30}$} & \makecell{\includegraphics[width=0.2\linewidth]{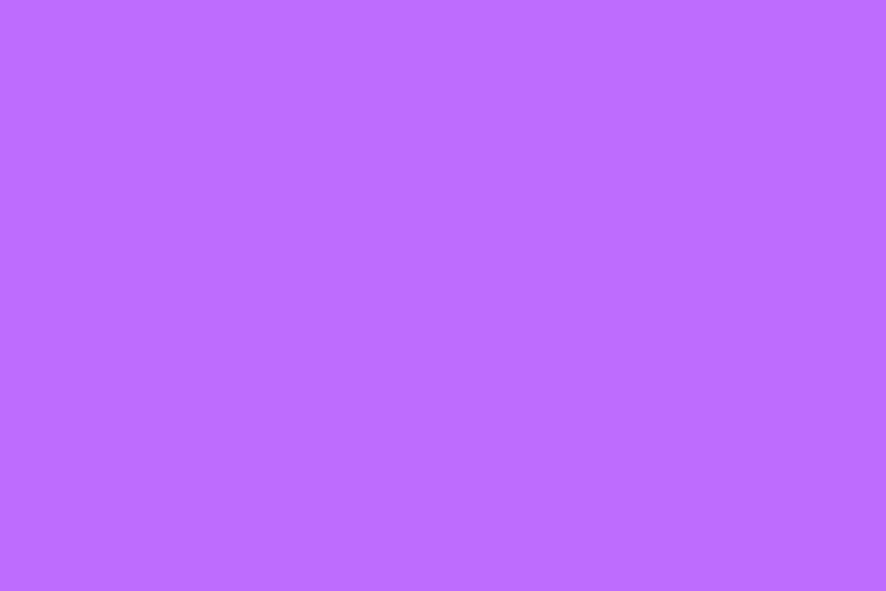}} &
			\makecell{\includegraphics[width=0.4\textwidth]{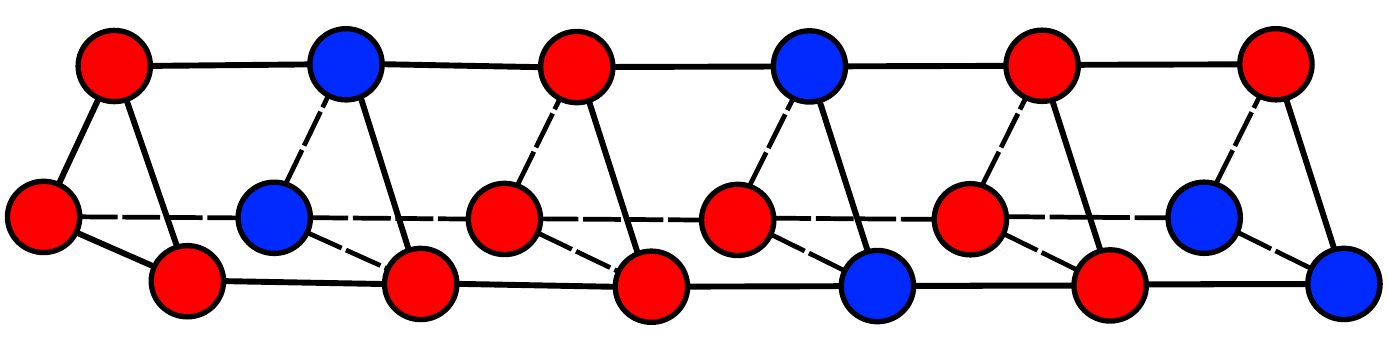}} \\ \hline
			\makecell{$d$,\\ $A_9A_{10}A_{22}A_{23}A_{24}A_{25}$\\
				$A_{26}A_{27}A_{29}A_{32}A_{33}$,\\ $B_{19}B_{20}B_{21}B_{22}B_{29}B_{30}B_{31}B_{32}$}& \makecell{\includegraphics[width=0.2\linewidth]{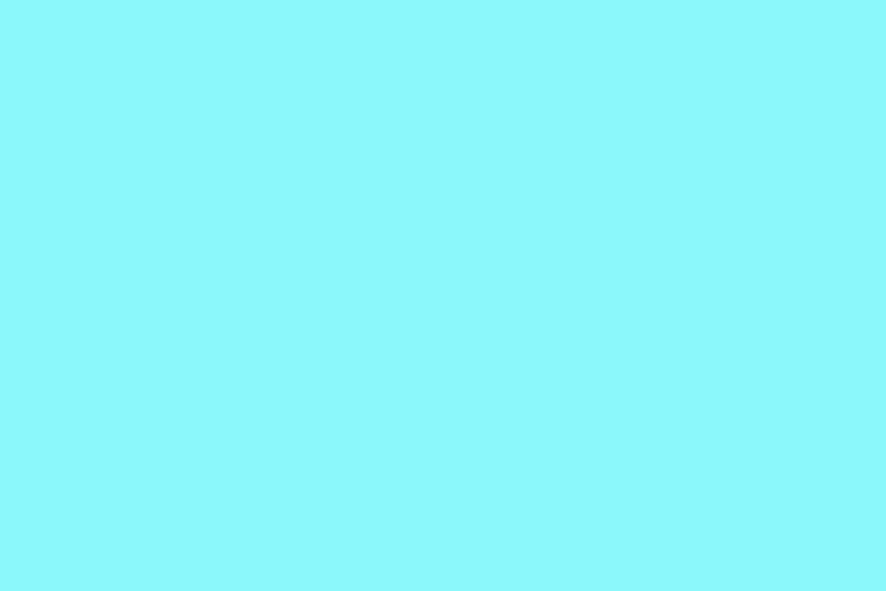}} &  
			\makecell{\includegraphics[width=0.4\textwidth]{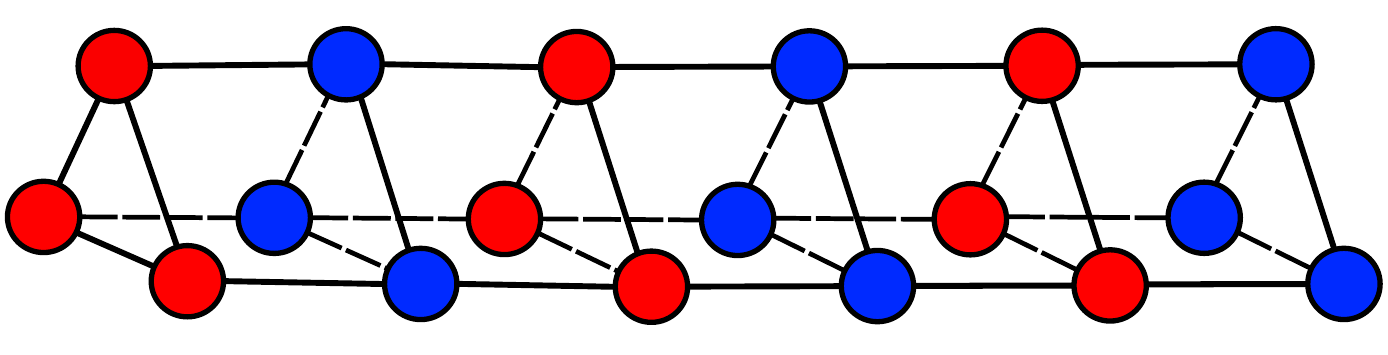}} \\ \hline
			\makecell{$e$,\\ $A_{10}A_{11}A_{12}A_{15}A_{21}$\\$A_{22}A_{29}A_{30}A_{31}A_{32}$,\\
				$B_6B_7B_8B_9B_{14}B_{15}B_{16}B_{17}$} & \makecell{\includegraphics[width=0.2\linewidth]{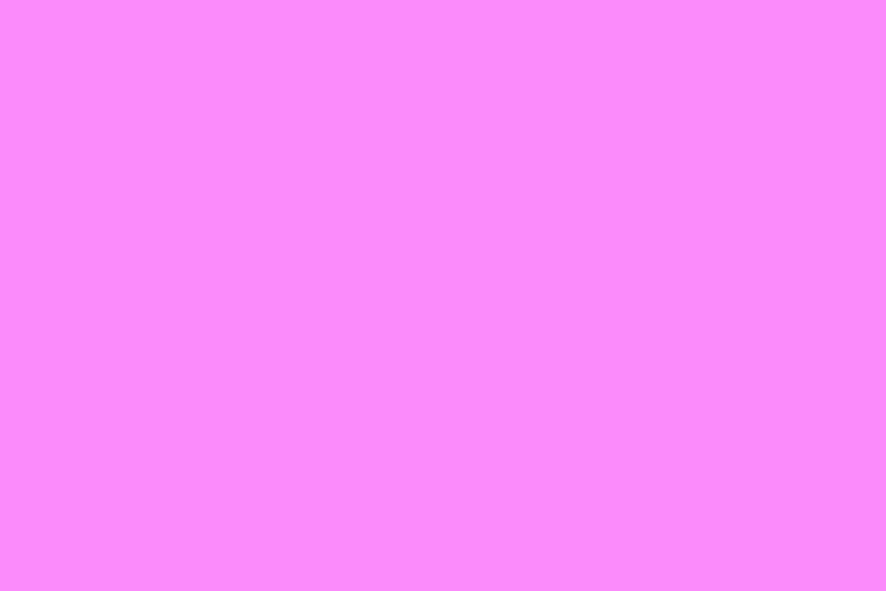}} &
			\makecell{\includegraphics[width=0.4\textwidth]{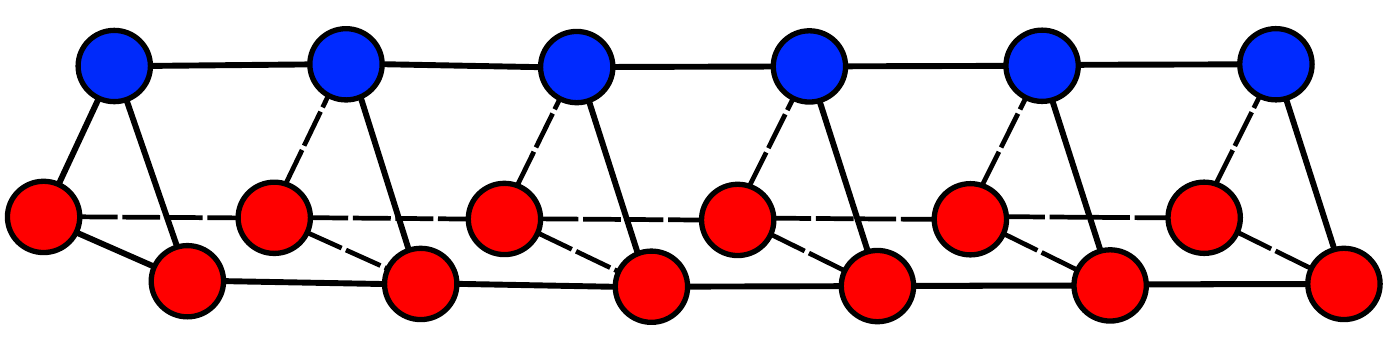}} \\ \hline
			\makecell{$f$,\\ $A_3A_4A_5A_6A_{14}$\\$A_{15}A_{16}A_{17}A_{20}A_{21}$,\\ 
				$B_{11}B_{12}B_{13}B_{14}B_{15}$\\$B_{21}B_{22}B_{23}B_{24}B_{25}$} & \makecell{\includegraphics[width=0.2\linewidth]{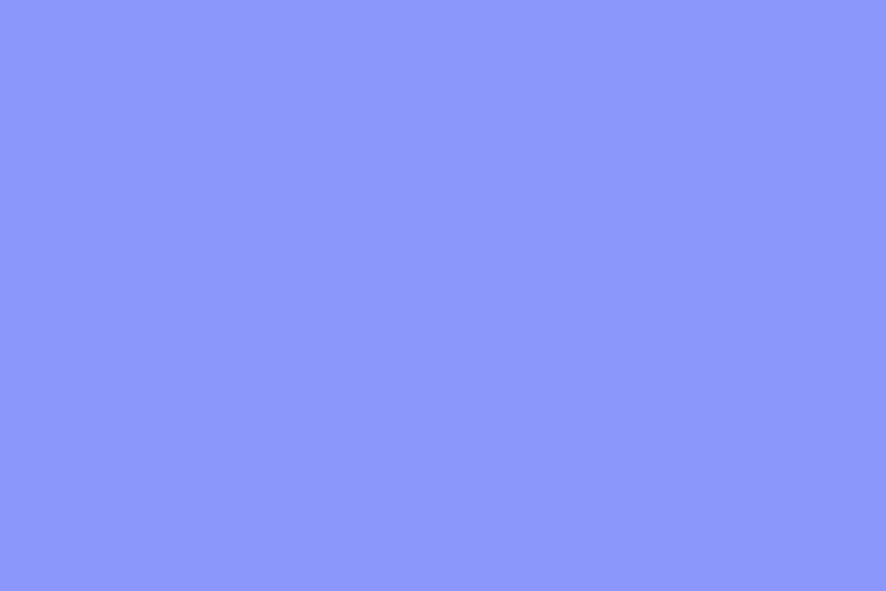}} &
			\makecell{\includegraphics[width=0.4\textwidth]{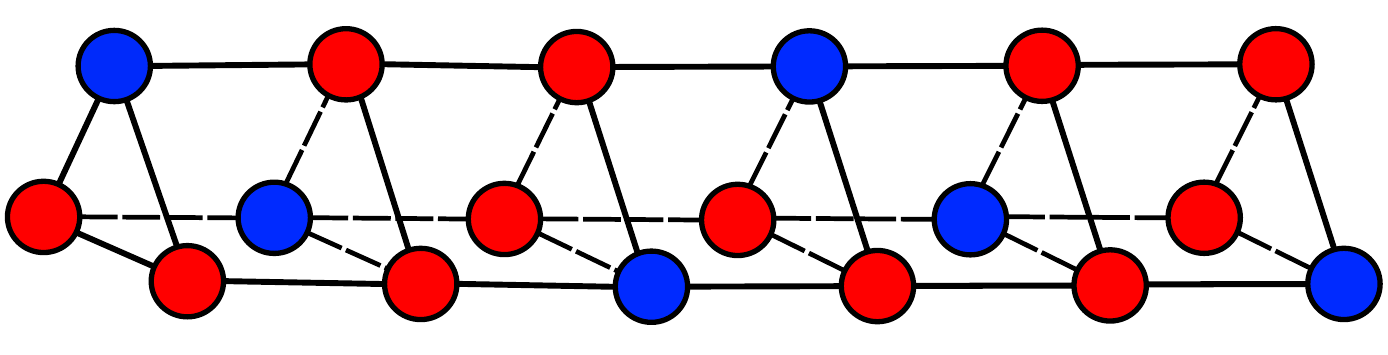}} \\ \hline
			\makecell{$g$,\\ $A_{15}A_{16}A_{17}A_{20}A_{21}A_{22}$\\$A_{23}A_{27}A_{28}A_{29}A_{30}$,\\
				$B_4B_5B_6B_9B_{10}B_{11}B_{12}B_{13}B_{14}B_{15}$} & \makecell{\includegraphics[width=0.2\linewidth]{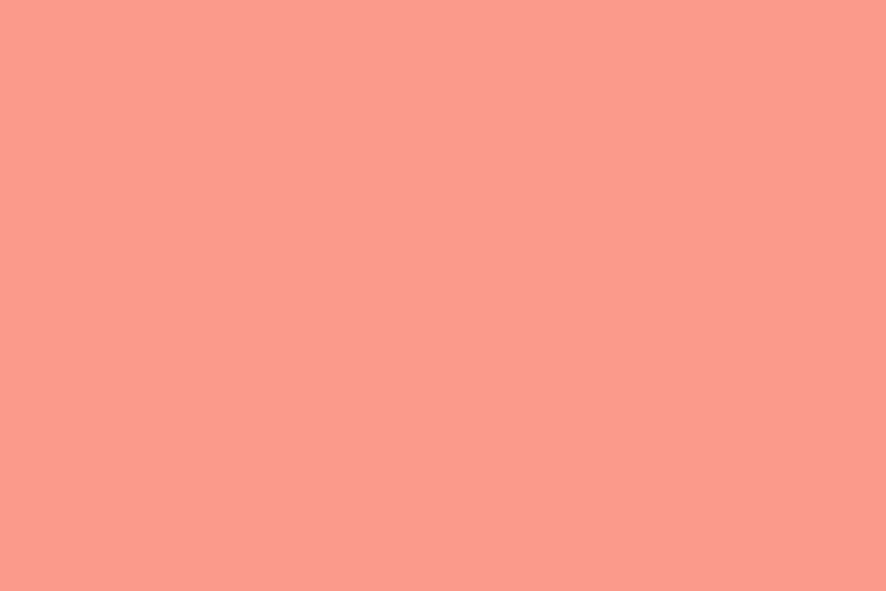}} &  
			\makecell{\includegraphics[width=0.4\textwidth]{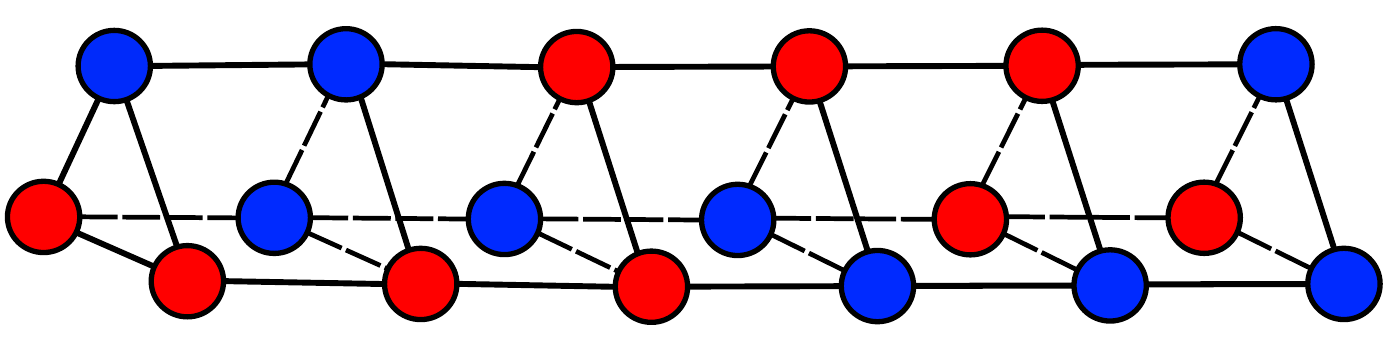}} \\ \hline
			\makecell{$h$,\\ $A_{11}A_{12}A_{13}A_{14}A_{15}A_{21}$,\\ 
				$B_{14}B_{15}B_{16}B_{17}B_{18}$\\$B_{19}B_{20}B_{21}B_{22}B_{23}$} & \makecell{\includegraphics[width=0.2\linewidth]{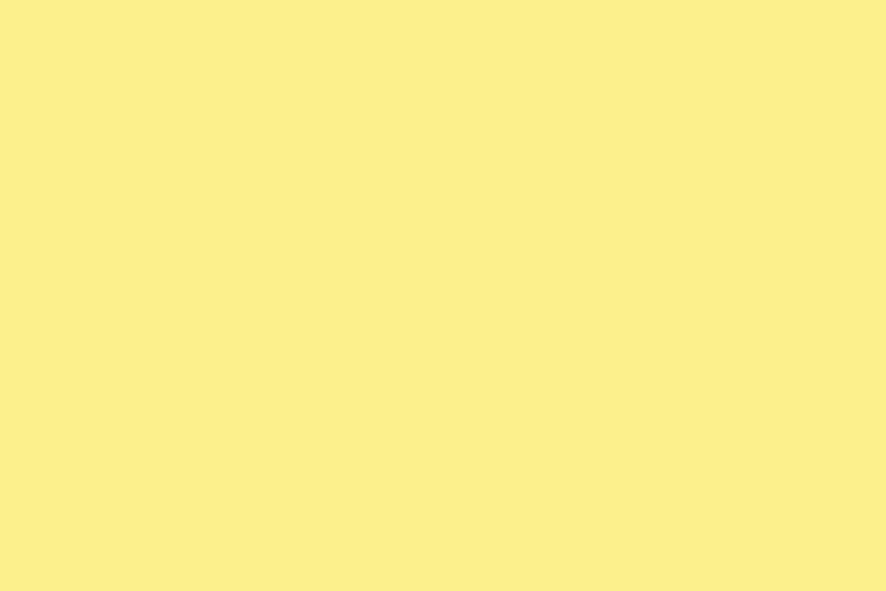}} &
			\makecell{\includegraphics[width=0.4\textwidth]{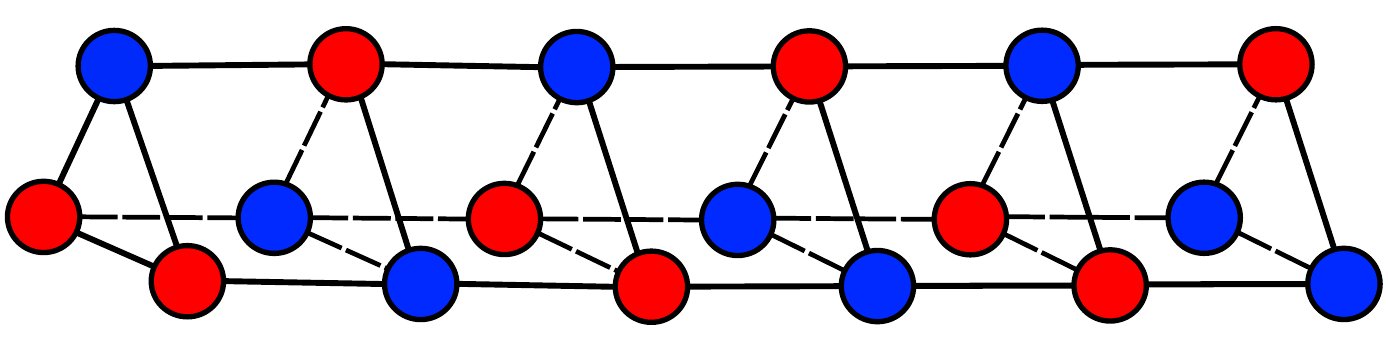}} \\ \hline
		\end{tabular}
	\end{table}
	\newpage
	
\begin{figure}[H]
		\centering{\includegraphics[width=0.65\linewidth]{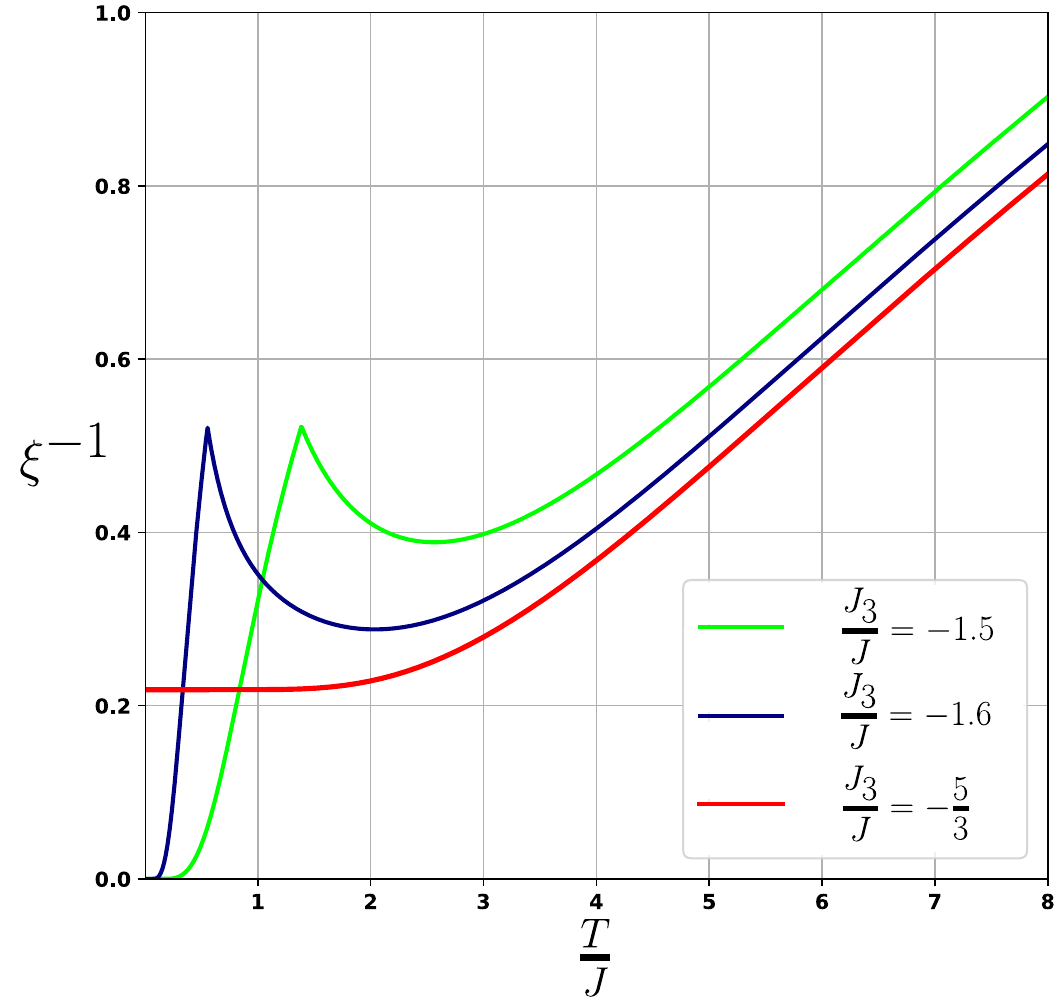}}
		\caption{Inverse correlation length plot in the low-temperature region at $\frac{J_1}{J} = 2, \frac{J_5}{J} = -\frac{5}{3}$, $J_2=J_4=J_6=J_7=J$ and $\frac{J_3}{J} = -1.5, -1.6, -\frac{5}{3}$} \label{ris5}
	\end{figure}

\begin{figure}[H]
		\centering{\includegraphics[width=0.65\linewidth]{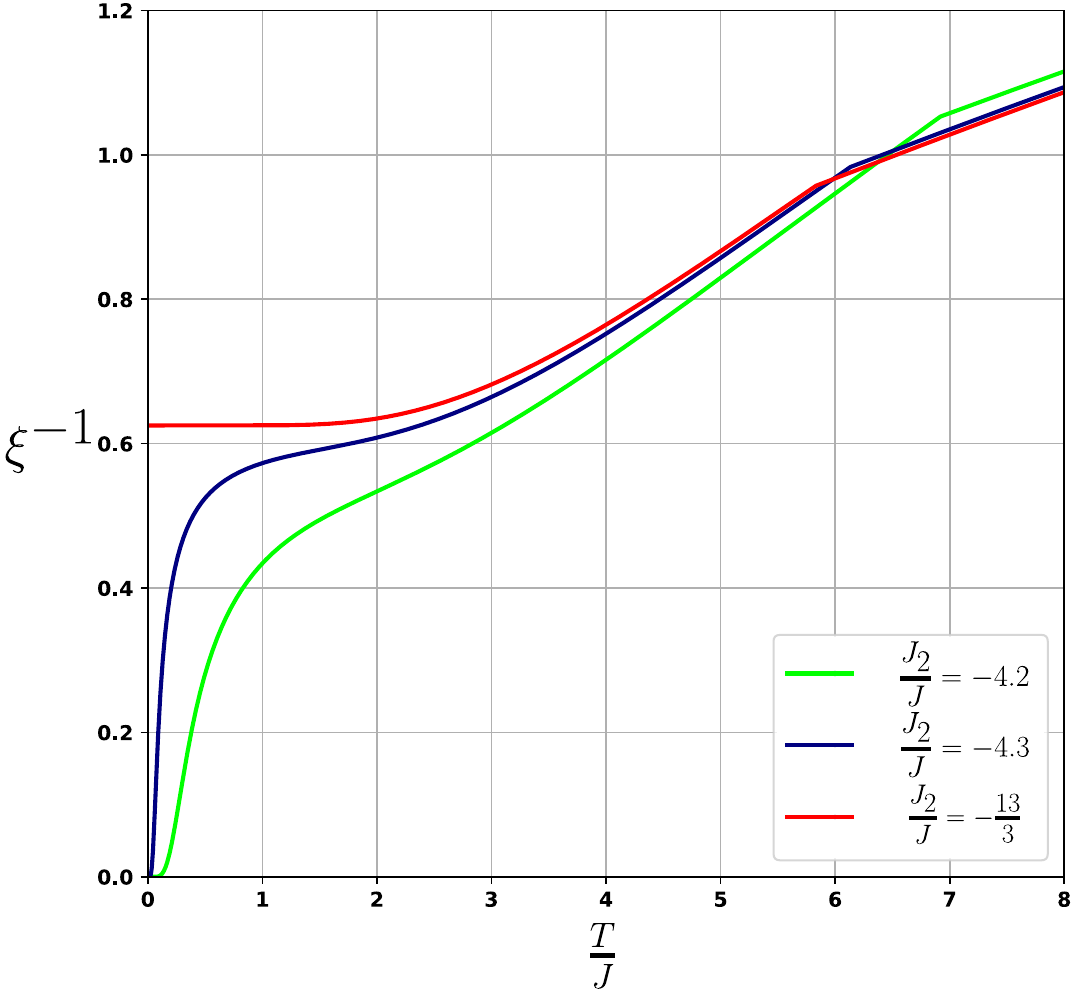} 
			\caption{Inverse correlation length plot in the low-temperature region at $\frac{J_1}{J} = 2, \frac{J_5}{J} = -\frac{5}{3}$, $J_3=J_4=J_6=J_7=J$ and $\frac{J_2}{J} = -4.2, -4.3, -\frac{13}{3}$} \label{ris7}}
	\end{figure}

	\begin{table}[h]
		\centering
		\caption{Coordinates of the boundary points of the regions of the ground states shown in the figures \ref{ris1}, \ref{ris444}.} \label{table3}
		\begin{tabular}{| c | c | c | c |} \hline
			Point & Coordinates & Point & Coordinates \\ \hline
			$A_1$& (15,15,$-15$) & $B_1$ & (15,15,$-15$) \\ \hline
			$A_2$& (15,$\frac{5}{2}$,$-15$) & $B_2$ & (15,4,$-15$) \\ \hline
			$A_3$& (2,9,$-15$) & $B_3$ & (4,15,$-15$) \\ \hline
			$A_4$& ($-4,15,-15$) & $B_4$ & (2,15,$-15$) \\ \hline
			$A_5$& ($-15,15,-15$) & $B_5$ & ($-15,15,-15$) \\ \hline
			$A_6$& ($-15,15,-4$) & $B_6$ & ($-15$,15,$-6$) \\ \hline
			$A_7$& ($-15$,15,15) & $B_7$ & ($-15$,15,15) \\ \hline
			$A_8$& (15,15,15) & $B_8$ & ($-6$,15,15) \\ \hline
			$A_9$& (15,$-\frac{5}{3}$,15) & $B_9$ & ($-6,15,-6$) \\ \hline
			$A_{10}$& ($-\frac{2}{3}$,$-\frac{5}{3}$,15) & $B_{10}$ & (2,15,$-14$) \\ \hline
			$A_{11}$& ($-6$,1,15) & $B_{11}$ & (2,1,$-7$) \\ \hline
			$A_{12}$& ($-15$,1,15) & $B_{12}$ & (2,1,$-15$) \\ \hline
			$A_{13}$& ($-15$,10,15) & $B_{13}$ & ($-15,1,-15$) \\ \hline
			$A_{14}$& ($-15$,10,1) & $B_{14}$ & ($-15$,1,1) \\ \hline
			$A_{15}$& ($-15$,1,1) & $B_{15}$ & ($-6$,1,1) \\ \hline
			$A_{16}$& ($-15$,1,$-15$) & $B_{16}$ & ($-6$,1,15) \\ \hline
			$A_{17}$& (2,1,$-15$) & $B_{17}$ & ($-15$,1,15) \\ \hline
			$A_{18}$& (15,1,$-15$) & $B_{18}$ & ($-15,-15$,15) \\ \hline
			$A_{19}$& (15,1,$-\frac{27}{2}$) & $B_{19}$ & (2,$-15$,15) \\ \hline
			$A_{20}$& (2,1,$-7$) & $B_{20}$ & (2,$-\frac{13}{3}$,15) \\ \hline
			$A_{21}$& ($-6$,1,1) & $B_{21}$ & (2,$-\frac{13}{3}$,$-\frac{5}{3}$) \\ \hline
			$A_{22}$& ($-\frac{2}{3}$,$-\frac{5}{3}$,1) & $B_{22}$ & (2,$-15$,-7) \\ \hline
			$A_{23}$& (2,$-\frac{5}{3}$,$-\frac{5}{3}$) & $B_{23}$ & ($-15,-15$,-7) \\ \hline
			$A_{24}$& (15,$-\frac{5}{3}$,$-\frac{49}{6}$) & $B_{24}$ & ($-15,-15,-15$) \\ \hline
			$A_{25}$& (15,$-\frac{17}{2},-15$) & $B_{25}$ & (2,$-15,-15$) \\ \hline
			$A_{26}$& (15,$-15,-15$) & $B_{26}$ & (15,$-15,-15$) \\ \hline
			$A_{27}$& (2,$-15,-15$) & $B_{27}$ & (15,1,$-15$) \\ \hline
			$A_{28}$& ($-15,-15,-15$) & $B_{28}$ & (15,1,$-\frac{27}{2}$) \\ \hline
			$A_{29}$& ($-14,-15$,1) & $B_{29}$ & (15,$-\frac{13}{3}$,$-\frac{49}{6}$) \\ \hline
			$A_{30}$& ($-15,-15$,1) & $B_{30}$ & (15,$-15$,$-\frac{27}{2}$) \\ \hline
			$A_{31}$& ($-15,-15$,15) & $B_{31}$ & (15,$-15$,15) \\ \hline
			$A_{32}$& ($-14,-15$,15) & $B_{32}$ & (15,$-\frac{13}{3}$,15) \\ \hline
			$A_{33}$& ($15,-15$,15) & $B_{33}$ & (15,15,15) \\ \hline
		\end{tabular}
	\end{table}
	\newpage
	\indent The boundary points of the ground states are of particular interest, since near them and in them themselves a change in the behavior of the correlation length can be observed at certain temperatures. The correlation length can be written in the form:
	\begin{equation} \label{29.21} 
		\xi^{-1} = \ln\biggl(\min_{\lambda_i \neq \lambda_{max}} \bigg|\frac{\lambda_{max}}{\lambda_i}\bigg|\biggr).
	\end{equation}
	\indent  As an example, we present plots (Figures \ref{ris5} — \ref{ris7}) of the inverse correlation lengths in the low-temperature region near some boundary points and directly at the boundary points themselves for the regions shown in Figures \ref{ris1} and \ref{ris444}.
	
	\begin{figure}[H]
		\centering{\includegraphics[width=0.65\linewidth]{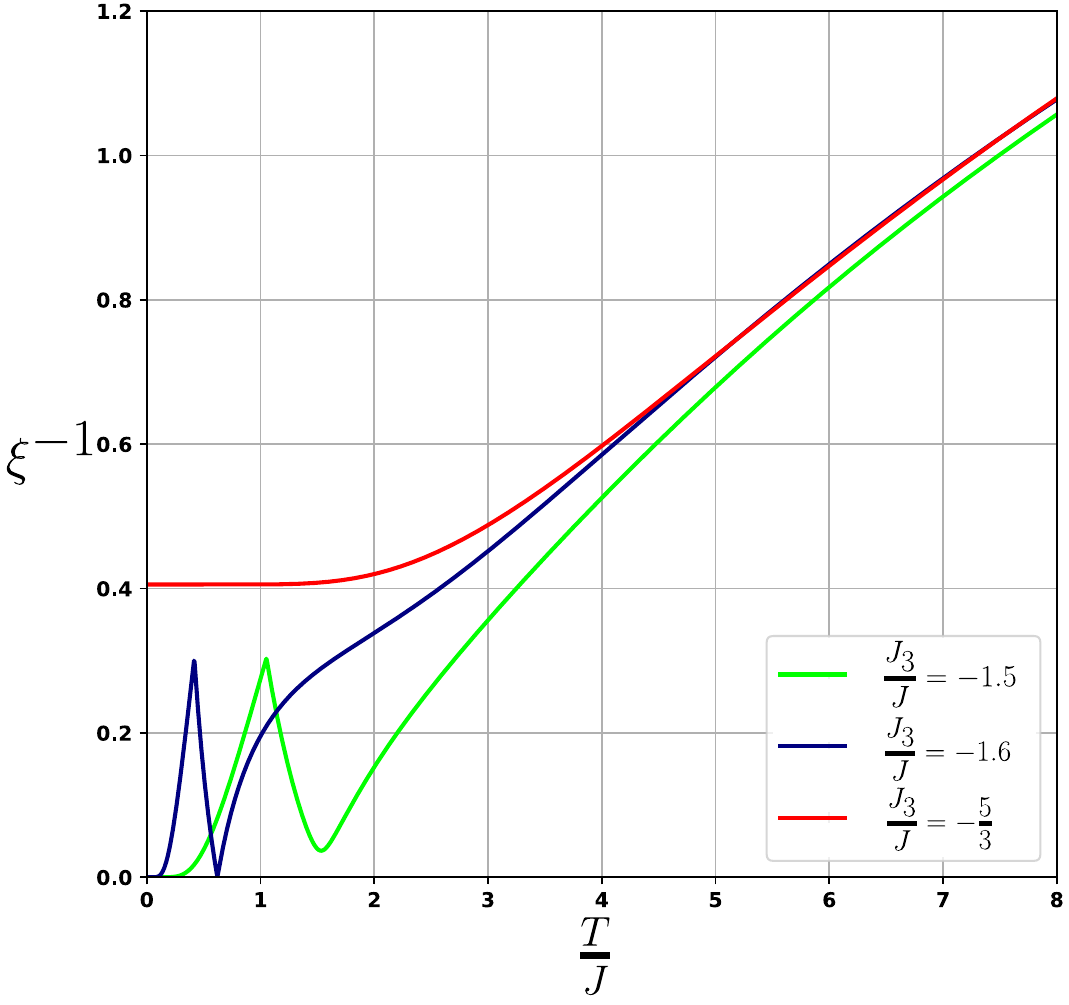} \label{ris6}}
		\caption{Inverse correlation length plot in the low-temperature region at $\frac{J_1}{J} = -\frac{2}{3}, \frac{J_5}{J} = 1$, $J_2=J_4=J_6=J_7=J$ and $\frac{J_3}{J} = -1.5, -1.6, -\frac{5}{3}$}
	\end{figure}
	
	\indent The figures show that, first of all, at the boundary points of the ground state regions, the inverse correlation length can be non-zero when the temperature tends to zero.\\
	\indent In addition, peaks can be observed at and near the boundary points at certain temperature values in the inverse correlation length plots.
	\indent \\

	\section{Model with some interactions of two, three, four and six spins}
	\indent Let us consider the second case in which only interspin interactions generated by elementary carriers $\{t_0^m, t_0^{m+1}\}$, $\{t_0^m,t_1^m,t_2^m\}$, $\{t_0^{m+1},t_1^{m+1},t_2^{m+1}\}$, $\{t_0^{m},t_1^{m},t_2^{m+1}\}$, $\{t_0^{m},t_1^{m+1},t_2^{m+1}\}$, $\{t_0^{m},t_1^{m},t_0^{m+1}, t_1^{m+1}\}$, $\{t_0^{m},t_1^{m},t_2^{m}, t_0^{m+1},t_1^{m+1},t_2^{m+1}\}$ are nonzero.\\
	\indent The component (\ref{nnn5}) of the Hamiltonian (\ref{5}) in this case has the form:
	\begin{equation} \label{ny8}
		\begin{gathered}
			H^m=-\bigg[J_1\bigg(\sigma_{0}^m \sigma_{0}^{m+1}  +  \sigma_{1}^m \sigma_{1}^{m+1} + \sigma_{2}^m \sigma_{2}^{m+1}\bigg)  + \\
			+ \frac{1}{2}J_2\bigg(  \sigma_{0}^m \sigma_{1}^{m}\sigma_{2}^m  + \sigma_{0}^{m+1} \sigma_{1}^{m+1}\sigma_{2}^{m+1}\bigg) + \\
			+ J_3\bigg( \sigma_{0}^{m} \sigma_{1}^{m}\sigma_{2}^{m+1}   +   \sigma_{0}^{m} \sigma_{2}^{m}\sigma_{1}^{m+1}  +  \sigma_{1}^{m} \sigma_{2}^{m}\sigma_{0}^{m+1}\bigg) + \\
			+ J_4\bigg(\sigma_{0}^{m} \sigma_{1}^{m+1}\sigma_{2}^{m+1} + \sigma_{1}^{m} \sigma_{0}^{m+1}\sigma_{2}^{m+1} + \sigma_{2}^{m} \sigma_{0}^{m+1}\sigma_{1}^{m+1}\bigg) + \\ 
			+J_5\bigg(\sigma_{0}^{m} \sigma_{1}^{m}\sigma_{0}^{m+1}\sigma_{1}^{m+1} + \sigma_{0}^{m} \sigma_{2}^{m}\sigma_{0}^{m+1}\sigma_{2}^{m+1} + \sigma_{1}^{m} \sigma_{2}^{m}\sigma_{1}^{m+1}\sigma_{2}^{m+1}\bigg)  + \\
			+ J_6\sigma_{0}^{m} \sigma_{1}^{m}\sigma_{2}^{m}\sigma_{0}^{m+1}\sigma_{1}^{m+1} \sigma_{2}^{m+1}\bigg],
		\end{gathered}
	\end{equation}
	where $\frac{1}{2}$ before $J_2$ is a coefficient that accounts for the transition from the $n$th to the $(n+1)$th layer of the model.\\
	\indent To simplify further formulations we introduce the notations $p=\exp\big\{\frac{J_1}{T}\big\}$, $q_1=\exp\big\{\frac{J_2}{T}\big\}$, $q_2=\exp\big\{\frac{J_3}{T}\big\}$, $q_3=\exp\big\{\frac{J_4}{T}\big\}$, $u=\exp\big\{\frac{J_5}{T}\big\}$, $r=\exp\big\{\frac{J_6}{T}\big\}$.\\
	\indent To formulate the following theorems, we introduce notation:
			\[     \begin{gathered}
				h = p^6q_1q_2^4q_3^4r^2u^2,\\
				h_1 = -p^5q_2q_3r^3u\bigg(3q_2^2q_3^2\big(q_1^2+q_2^2q_3^2\big) + p^4u^4\big(1+q_1^2q_2^6q_3^6\big)\bigg), \\
				h_2 =  q_1\bigg(9p^4q_2^4q_3^4\big(r^4-p^4\big) - 3p^4u^4\big(q_2^8+q_3^8\big) + 3p^8r^4u^4\big(1+q_2^8q_3^8\big) + q_2^4q_3^4u^8\big(p^{12}r^4-1\big)\bigg),\\
				h_3 = p^5q_2q_3r^3u\bigg(q_2^2q_3^2\big(q_1^2+q_2^2q_3^2\big) - p^4u^4\big(1+q_1^2q_2^6q_3^6\big)\bigg),\\
				h_4 = q_1\bigg(p^4q_2^4q_3^4\big(r^4-p^4\big) +p^4u^4\big(q_2^8+q_3^8\big) - p^8r^4u^4\big(1+q_2^8q_3^8\big) + q_2^4q_3^4u^8\big(p^{12}r^4-1\big) \bigg)
			\end{gathered}
			\]
			\begin{theorem} \label{th5}
				\textit{In the thermodynamic limit, the free energy, internal energy, specific heat, and entropy per lattice node are respectively} (\ref{16})\textit{,} (\ref{17})\textit{,} (\ref{18})\textit{,} (\ref{last10})\textit{, where the largest eigenvalue $\lambda_{max}$ of the transfer-matrix $\theta$} \textit{has the form:} 
				\begin{equation} \label{ny10}
					\lambda_{max} = \frac{-h_1 + \sqrt{h_1^2 - 4hh_2}}{2h}
				\end{equation}
			\end{theorem}
			\begin{theorem} \label{th6}
				\textit{The partition function in a finite closed strip of length $L$ can be written in the form} (\ref{nnn1})\textit{, where $\lambda_k, k = 1, \ldots, 8$ are the roots of the characteristic polynomial of the transfer-matrix $\theta$} \textit{, two of which have multiplicity} 3\textit{, and expressed in the following form:}
				\begin{equation} \label{ny11}
					\lambda_{1,2} = \frac{-h_1 \pm \sqrt{h_1^2 - 4hh_2}}{2h},
				\end{equation}
				\begin{equation} \label{ny12}
					\lambda_{3,4,5} = \frac{-h_3 + \sqrt{h_3^2 - 4hh_4}}{2h},
				\end{equation}
				\begin{equation} \label{ny13}
					\lambda_{6,7,8} = \frac{-h_3 - \sqrt{h_3^2 - 4hh_4}}{2h}.
				\end{equation}
			\end{theorem}
			\newpage
			\section{Exact value of invariant percolation in the triple-chain generalized Ising model with all possible multispin interactions}
			\indent \\
			\indent Spin configurations in the cyclic-closed strip (\ref{1}), in which there exists a side of -1: $\sigma_{0}^m=\sigma_{1}^m=-1$,  $\sigma_{0}^m=\sigma_{2}^m=-1$, $\sigma_{1}^m=\sigma_{2}^m=-1$ at some $m \in \{0, 1, \ldots , L-1\}$, we will call the percolation configurations. \\
			\indent If we exclude from the set of all model configurations the set of percolation configurations, we obtain the set of non-percolation configurations.\\
			\indent Accordingly, the non-percolation probability is the probability that the random state of the model will be from the set of non-percolation configurations. The non-percolation probability of a model is determined by the ratio of two quantities:
			\begin{equation} \label{june2}
				P_L = \frac{Z_{L}'}{Z_{L}},
			\end{equation}
			where $Z_L$ is the partition function (\ref{6}) of the model with Hamiltonian (\ref{5}), $Z_{L}'$ is the analog of the partition function, in which the summation is performed only on the non-percolation  configurations. We will call it the partition function of non-percolation configurations.  \\
			\indent The transfer-matrix of non-percolation configurations $\theta'$ coincides with the transfer-matrix (\ref{12}) of all configurations of the model, but with zero matrix elements corresponding to the percolation configurations:
			\begin{equation} \label{june1}
				\theta' = 
				\begin{pmatrix}
					\textbf{*}&	\textbf{*}&	\textbf{*}&0&	\textbf{*}&0&0&0\\
					\textbf{*}&	\textbf{*}&	\textbf{*}&0&	\textbf{*}&0&0&0\\
					\textbf{*}&	\textbf{*}&	\textbf{*}&0&	\textbf{*}&0&0&0\\
					0&0&0&0&0&0&0&0\\
					\textbf{*}&	\textbf{*}&	\textbf{*}&0&	\textbf{*}&0&0&0\\
					0&0&0&0&0&0&0&0\\
					0&0&0&0&0&0&0&0\\
					0&0&0&0&0&0&0&0
				\end{pmatrix},
			\end{equation}
			\indent To formulate the following theorems, we introduce the matrix:
			\begin{equation} \label{june3}
				\tau' = 
				\begin{pmatrix}
					\theta'_{11}&\theta'_{12}+\theta'_{13}+\theta'_{15}\\
					\theta'_{21}&\theta'_{22}+\theta'_{23}+\theta'_{25}\\
				\end{pmatrix}
			\end{equation}
			\begin{theorem} \label{th8}
				\textbf{The main percolation theorem}\\
				\indent\textit{In the thermodynamic limit for the non-percolation probability $P_L$} (\ref{june2}) \textit{of lattice model with Hamiltonian }(\ref{5}) \textit{there is an equality:}
				\begin{equation} \label{june4}
					\lim_{L\to\infty} \frac{\ln P_L}{L}= \ln\frac{\mu_{max}}{\lambda_{max}},
				\end{equation}
				\textit{where $\mu_{max} = \mu_{max}(H,T)$ is the largest root of the characteristic equation of the matrix $\tau'$ }(\ref{june3})\textit{, and }\textit{$\lambda_{max} = \lambda_{max}(H,T)$ is the largest root} (\ref{n5}) \textit{of the characteristic equation} \textit{ of the matrix $\tau^1$}(\ref{19}).
			\end{theorem}
			\begin{theorem} \label{th9}
				\indent\textit{In a finite closed strip of length $L$ for the non-percolation probability $P_L$} (\ref{june2}) \textit{of the lattice model with Hamiltonian }(\ref{5}) \textit{there is an equality:}
				\begin{equation} \label{june5}
					P_L = \frac{ \sum_{i=1}^{4} {\lambda'_{i}}^L}{\sum_{k=1}^{8} {\lambda_{k}}^L},
				\end{equation}
				\textit{where $\lambda'_{1,2}$ are the roots of the characteristic square equation of the matrix} (\ref{june3}), \textit{and $\lambda'_3, \lambda'_4$, respectively, are equal:}
				\begin{equation} \label{june6}
					\lambda'_3 = \theta'_{52}e^{\frac{2\pi}{3}i} + \theta'_{53}e^{-\frac{2\pi}{3}i} + \theta'_{55},
				\end{equation}
				\begin{equation} \label{june7}
					\lambda'_4 = \theta'_{52}e^{-\frac{2\pi}{3}i} + \theta'_{53}e^{\frac{2\pi}{3}i} + \theta'_{55},
				\end{equation}
				\textit{and $\lambda_k$, $k=1, \ldots, 8$ coincide with the roots }(\ref{ny2})\textit{ —} (\ref{ny4.1}).
			\end{theorem}
			\indent The proofs of the formulated theorems \ref{th1} — \ref{th4}, \ref{th5} — \ref{th9} are given in paragraph 9.
			\newpage
			\section{Planar model with nearest neighbor, next-nearest neighbor and plaquette interactions}
			\indent The triple-chain generalized Ising model with multispin interactions is the minimal possible model on the torus, which unfolding yields a model on the plane.\\
			\indent Consider a planar model with nearest neighbor, next-nearest neighbor, and plaquette interactions, one layer of which contains three plaquettes.\\
			\indent This model is obtained by unfolding the lattice shown in Figure \ref{p1} in the plane with the addition of additional boundary conditions $t_3^m \equiv t_0^m$ and $t_3^{m+1} \equiv t_0^{m+1}$ (Fig. \ref{gonihedric}):
			\begin{figure}[h]
				\centering{\includegraphics[width=1\linewidth]{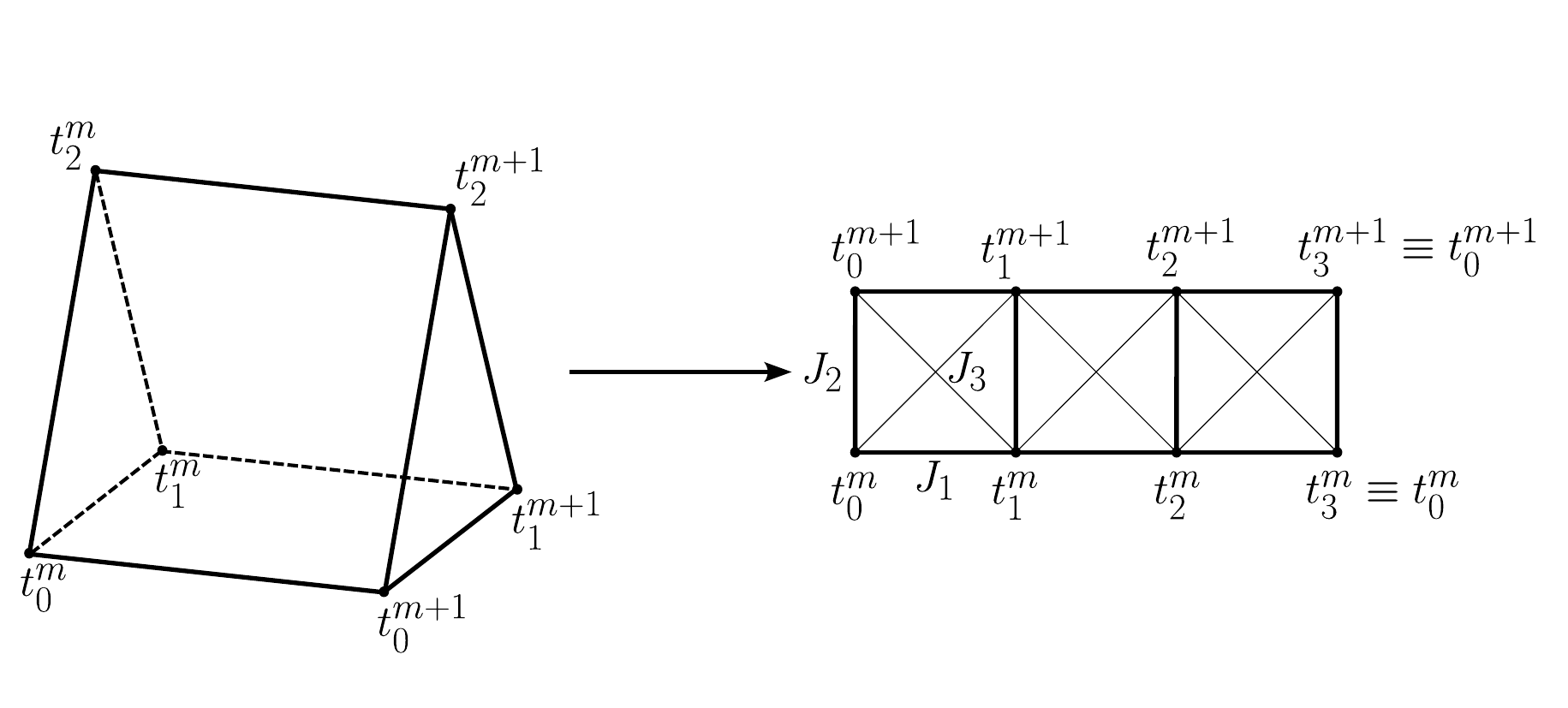}}
				\caption{Planar Ising model with nearest-neighbor, next-nearestneighbor and plaquette interactions obtained by unfolding the triple-chain model on the plane}
				\label{gonihedric}
			\end{figure}
			\indent \\
			\indent Here $J_1$ corresponds to nearest neighbor interactions on the same layer, $J_2$ corresponds to nearest neighbor interactions on adjacent layers, $J_3$ corresponds to next-nearest neighbor interactions, and $J_4$ is not explicitly shown in the figure \ref{gonihedric} but corresponds to plaquette interactions.\\
			\indent The Hamiltonian of the system coincides with (\ref{5}), where the Hamiltonian of one layer of the model is of the form:
			\begin{equation} \label{june8}
				\begin{gathered}
					H^m = -\frac{J_1}{2}\bigg(
					\sigma_{0}^m\sigma_{1}^m + 
					\sigma_{1}^m\sigma_{2}^m + 
					\sigma_{2}^m\sigma_{3}^m + 
					\sigma_{0}^{m+1}\sigma_{1}^{m+1} + 
					\sigma_{1}^{m+1}\sigma_{2}^{m+1} + 
					\sigma_{2}^{m+1}\sigma_{3}^{m+1}
					\bigg) - \\
					-J_2\bigg(
					\frac{1}{2} \sigma_{0}^{m}\sigma_{0}^{m+1} +
					\sigma_{1}^{m}\sigma_{1}^{m+1} +
					\sigma_{2}^{m}\sigma_{2}^{m+1} +
					\frac{1}{2} \sigma_{3}^{m}\sigma_{3}^{m+1}
					\bigg) - \\
					-J_3\bigg(
					\sigma_{0}^{m}\sigma_{1}^{m+1} +
					\sigma_{1}^{m}\sigma_{2}^{m+1} +
					\sigma_{2}^{m}\sigma_{3}^{m+1} +
					\sigma_{1}^{m}\sigma_{0}^{m+1} +
					\sigma_{2}^{m}\sigma_{1}^{m+1} +
					\sigma_{3}^{m}\sigma_{2}^{m+1}
					\bigg) - \\
					-J_4 \bigg(
					\sigma_{0}^{m}\sigma_{1}^{m}\sigma_{0}^{m+1}\sigma_{1}^{m+1} +
					\sigma_{1}^{m}\sigma_{2}^{m}\sigma_{1}^{m+1}\sigma_{2}^{m+1} +
					\sigma_{2}^{m}\sigma_{3}^{m}\sigma_{2}^{m+1}\sigma_{3}^{m+1} 
					\bigg),
				\end{gathered}
			\end{equation}
			and as a transfer-matrix we can use an eighth-order matrix, the structure of which coincides with the transfer-matrix (\ref{29.2}), the elements $a$ — $h$ of which are rewritten according to (\ref{june8}). Then the eigenvalues of the transfer-matrix of this planar model coincide with (\ref{29.22}), and all physical characteristics can be described by the expressions obtained for the triple-chain model with interactions of an even number of spins — the theorems \ref{th3} and \ref{th4} are valid for this model. The expressions for pair correlations formulated in Theorem \ref{th7} are also valid for this model if we assume that all nodes $t_0^m, t_1^m, t_2^m$ are in fact nearest neighbors because of the condition $t_3^m \equiv t_0^m$.\\
			\indent As an example for this planar model, we show plots of free energy, internal energy, specific heat and entropy in the thermodynamic limit in the low-temperature region for four different cases of parameters $J_1, \ldots, J_4$, one of which concerns the so-called planar gonihedric model for which $J_1 = J_2 = k$, $J_3 = -\frac{k}{2}$, $J_4 = \frac{1-k}{2}$ \cite{Bathas}.
			\begin{figure}[H]
				\begin{minipage}[H]{0.495\linewidth}
					\centering{\includegraphics[width=1\linewidth]{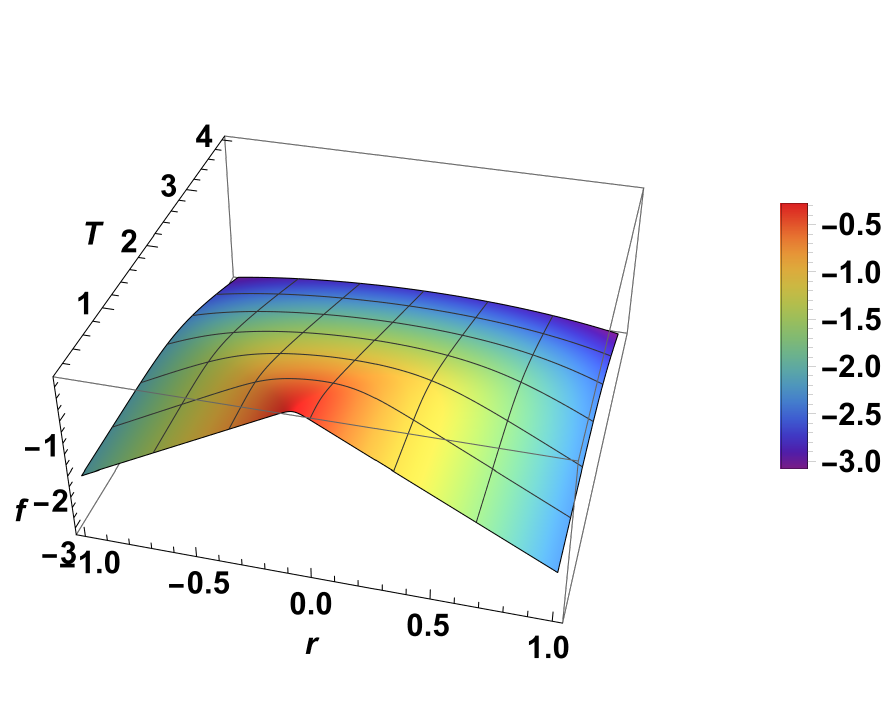}}
				\end{minipage}
				\hfill
				\begin{minipage}[H]{0.495\linewidth}
					\centering{\includegraphics[width=1\linewidth]{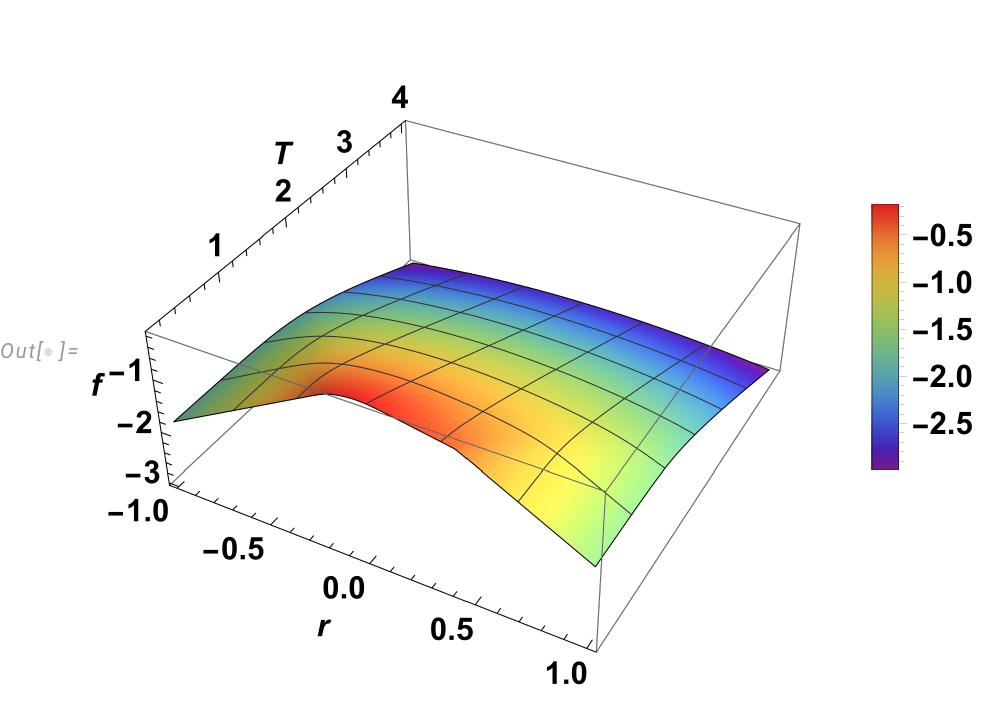}}
				\end{minipage}
				\vfill
				\begin{minipage}[h]{0.495\linewidth}
					\centering{\includegraphics[width=1\linewidth]{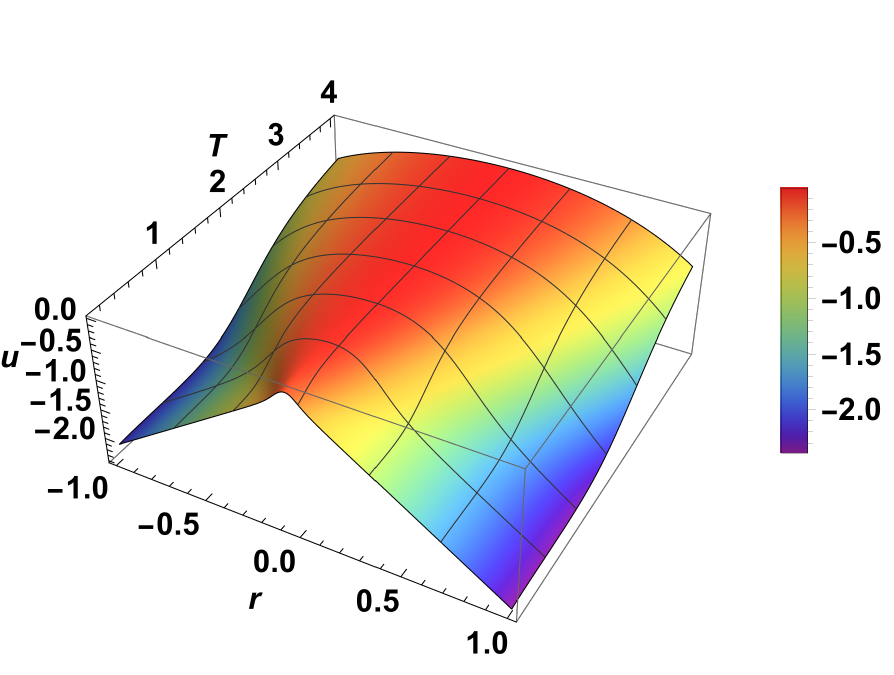}}
				\end{minipage}
				\hfill
				\begin{minipage}[h]{0.495\linewidth}
					\centering{\includegraphics[width=1\linewidth]{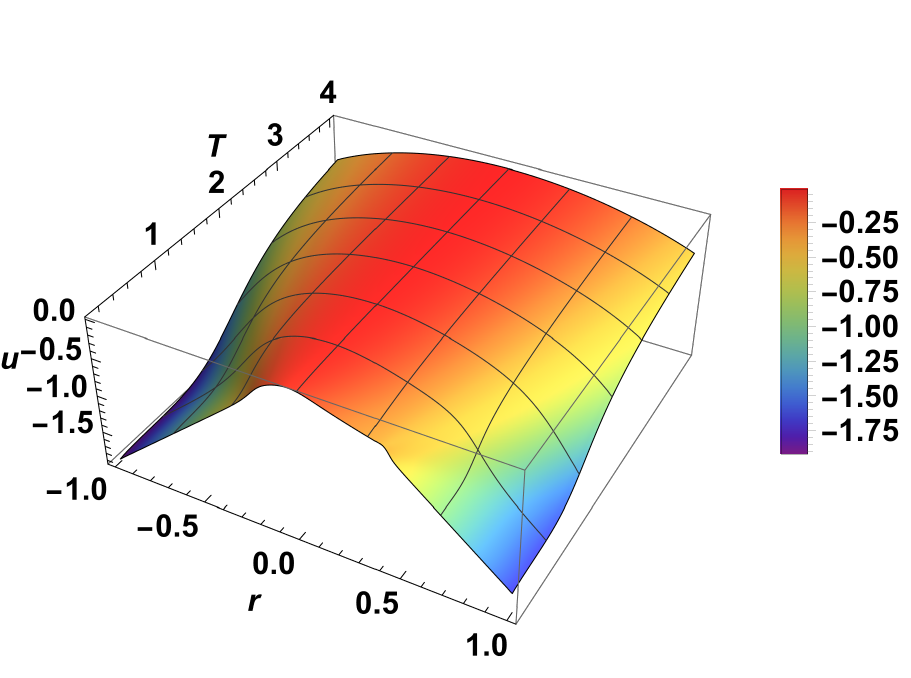}}
				\end{minipage}
				\vfill
				\begin{minipage}[h]{0.495\linewidth}
					\centering{\includegraphics[width=1\linewidth]{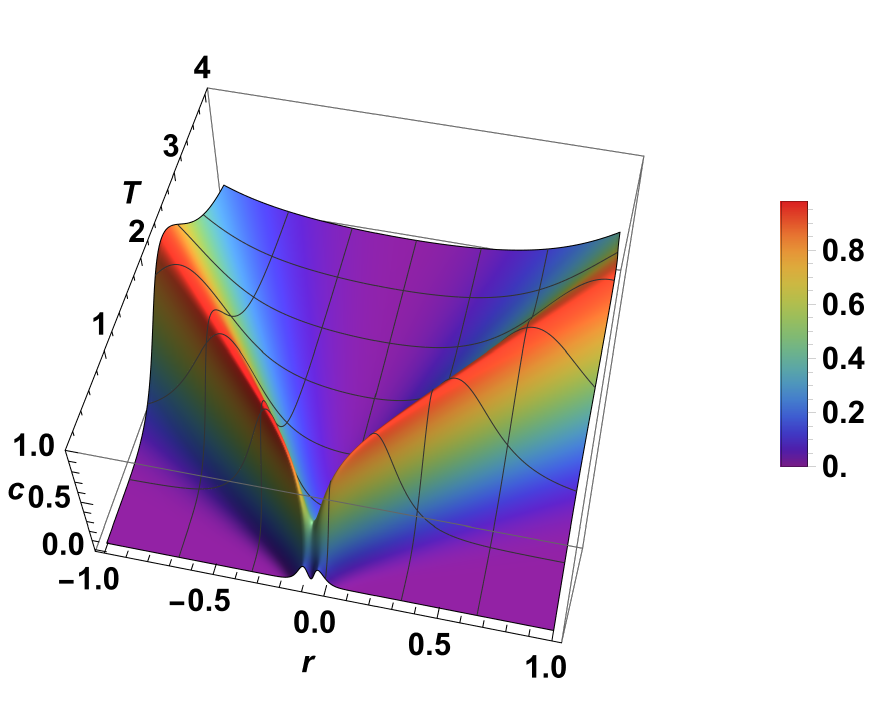}}
				\end{minipage}
				\hfill
				\begin{minipage}[h]{0.495\linewidth}
					\centering{\includegraphics[width=1\linewidth]{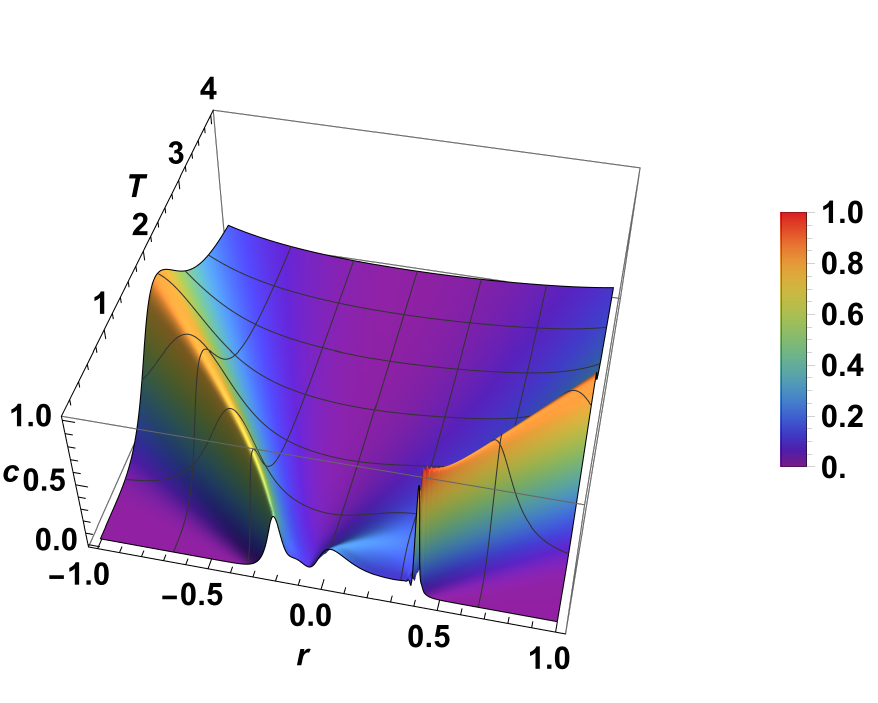}}
				\end{minipage}
				\vfill
				\begin{minipage}[h]{0.495\linewidth}
					\centering{\includegraphics[width=1\linewidth]{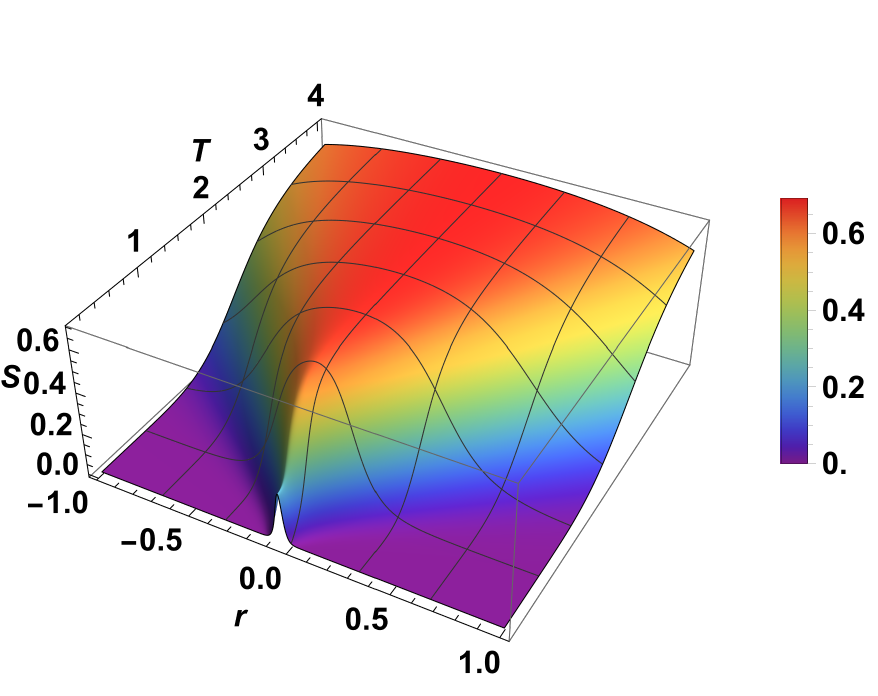}}
				\end{minipage}
				\hfill
				\begin{minipage}[h]{0.495\linewidth}
					\centering{\includegraphics[width=1\linewidth]{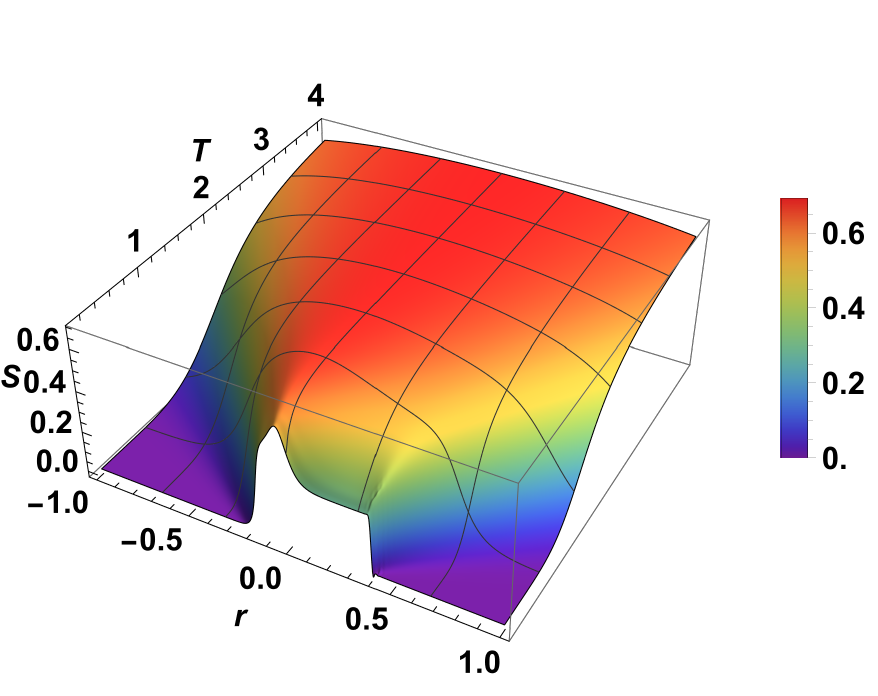}}
				\end{minipage}
				\caption{Plots, respectively, of free energy, internal energy, specific heat and entropy for the cases $J_1 = 0.11, J_2 = 0.12, J_4 = 0.13$ (left) and $J_1=-0.11, J_2=-0.12, J_4=-0.13$ (right). $J_3 = r \in [-1, 1], T\in[0.1, 4]$} 
			\end{figure}
			\begin{figure}[H]
				\begin{minipage}[h]{0.495\linewidth}
					\centering{\includegraphics[width=1\linewidth]{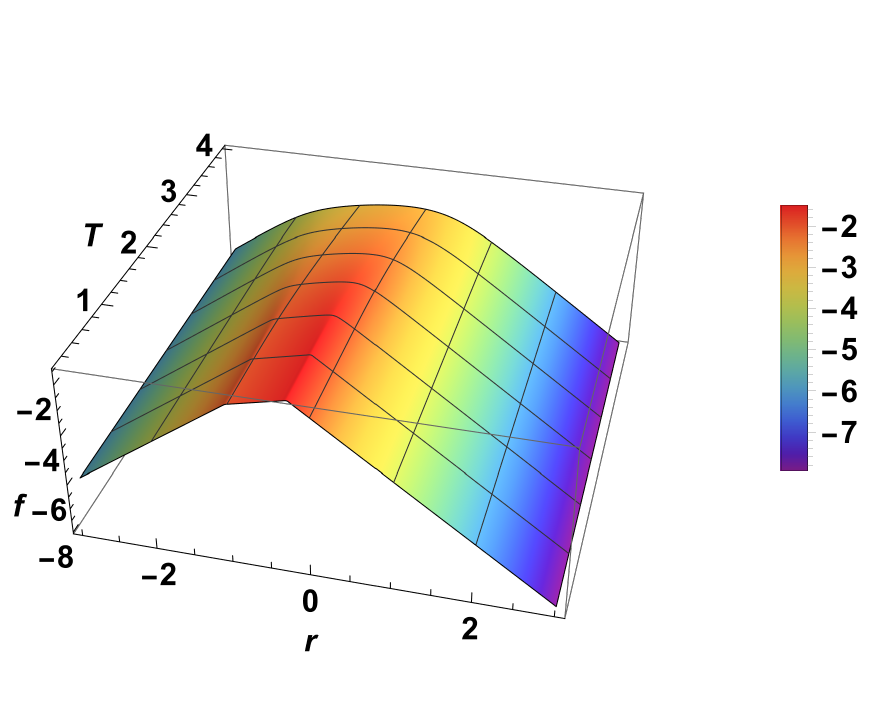}\\ Free energy}
				\end{minipage}
				\hfill
				\begin{minipage}[h]{0.495\linewidth}
					\centering{\includegraphics[width=1\linewidth]{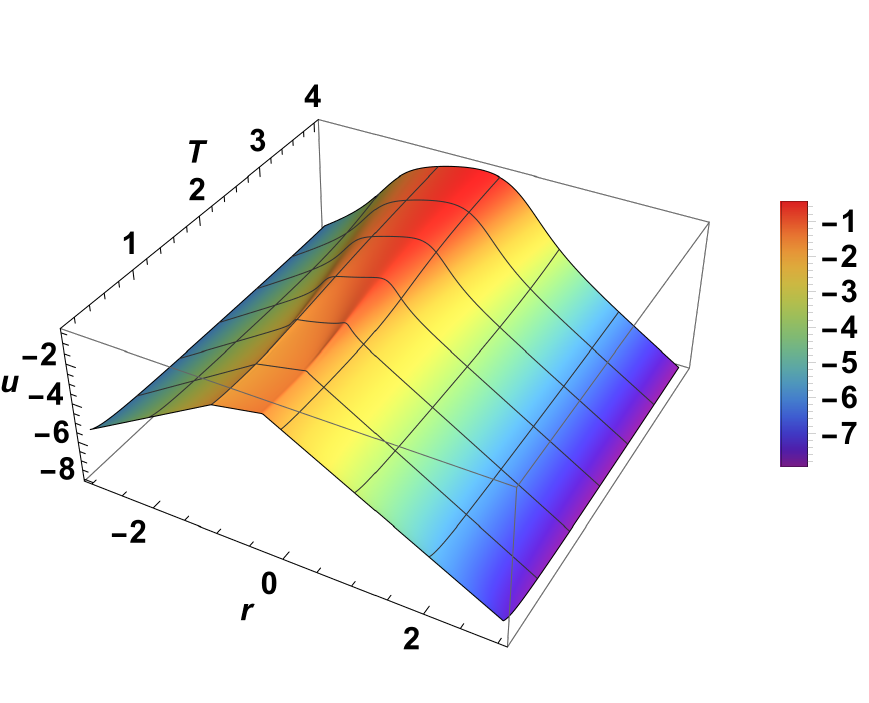} \\ Internal energy}
				\end{minipage}
				\vfill
				\begin{minipage}[h]{0.495\linewidth}
					\centering{\includegraphics[width=1\linewidth]{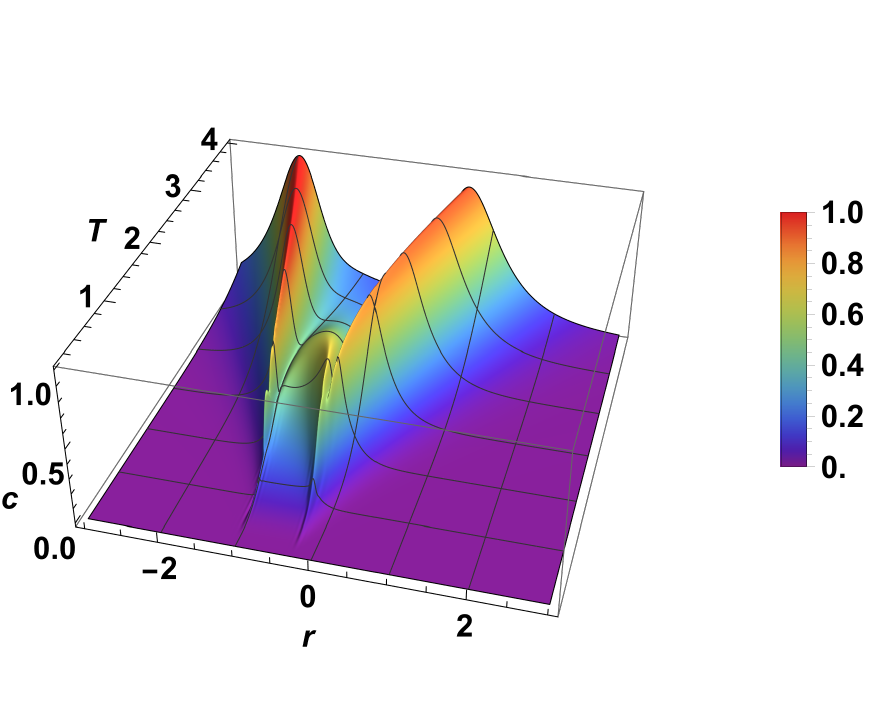} \\ Specific heat}
				\end{minipage}
				\hfill
				\begin{minipage}[h]{0.495\linewidth}
					\centering{\includegraphics[width=1\linewidth]{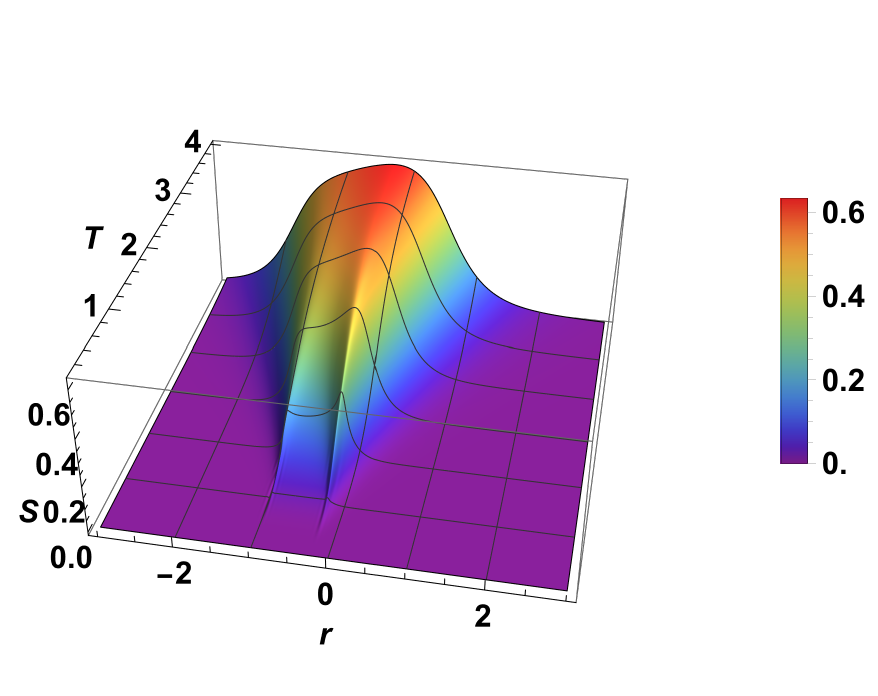} \\ Entropy}
				\end{minipage}
				\caption{Plots of free energy, internal energy, specific heat and entropy for the case $J_1 = \frac{1}{2}, J_2 = 1, J_4 = \frac{1}{2}$ and $J_3 = r \in [-3, 3], T\in[0.1, 4]$} 
			\end{figure}
			\begin{figure}[H]
				\begin{minipage}[h]{0.495\linewidth}
					\centering{\includegraphics[width=1\linewidth]{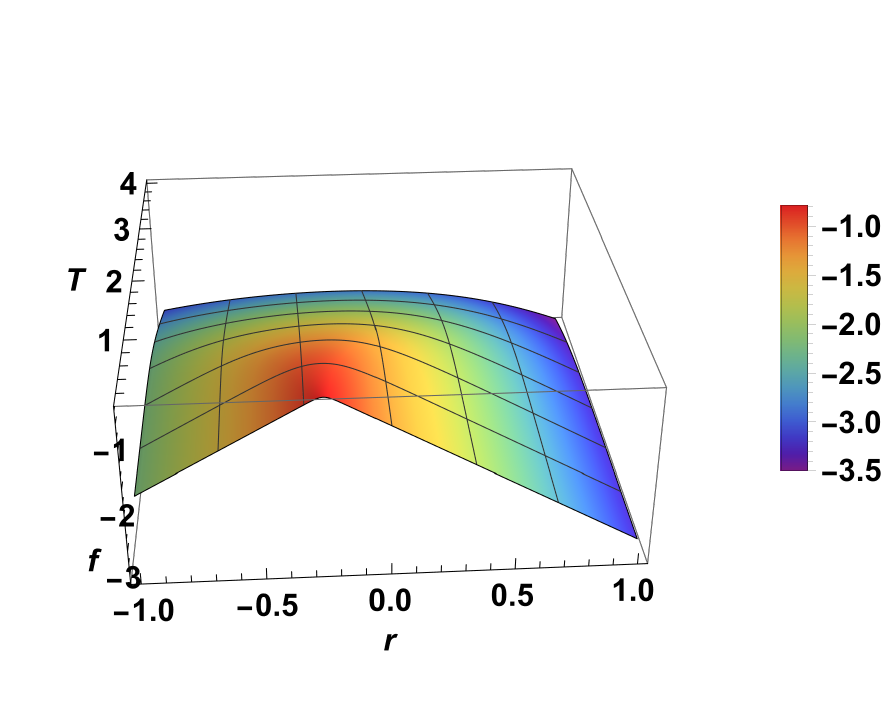}\\ Free energy}
				\end{minipage}
				\hfill
				\begin{minipage}[h]{0.495\linewidth}
					\centering{\includegraphics[width=1\linewidth]{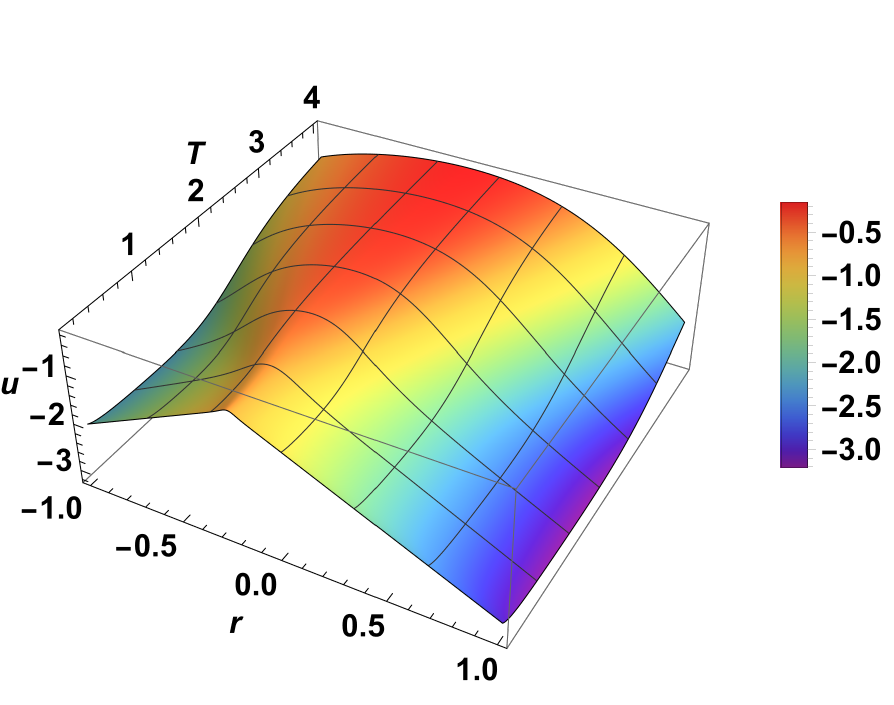} \\ Internal energy}
				\end{minipage}
				\vfill
				\begin{minipage}[h]{0.495\linewidth}
					\centering{\includegraphics[width=1\linewidth]{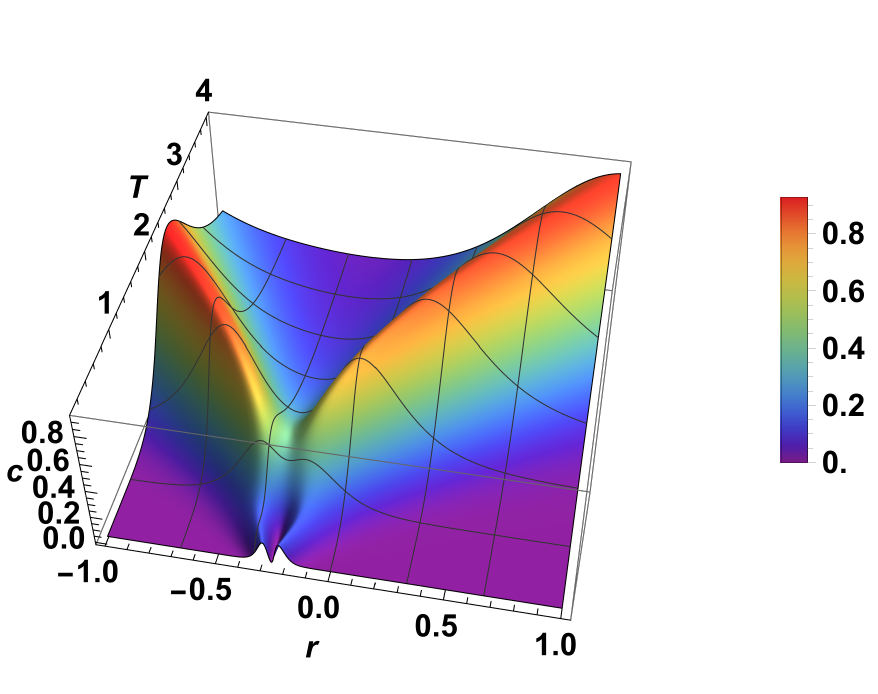} \\ Specific heat}
				\end{minipage}
				\hfill
				\begin{minipage}[h]{0.495\linewidth}
					\centering{\includegraphics[width=1\linewidth]{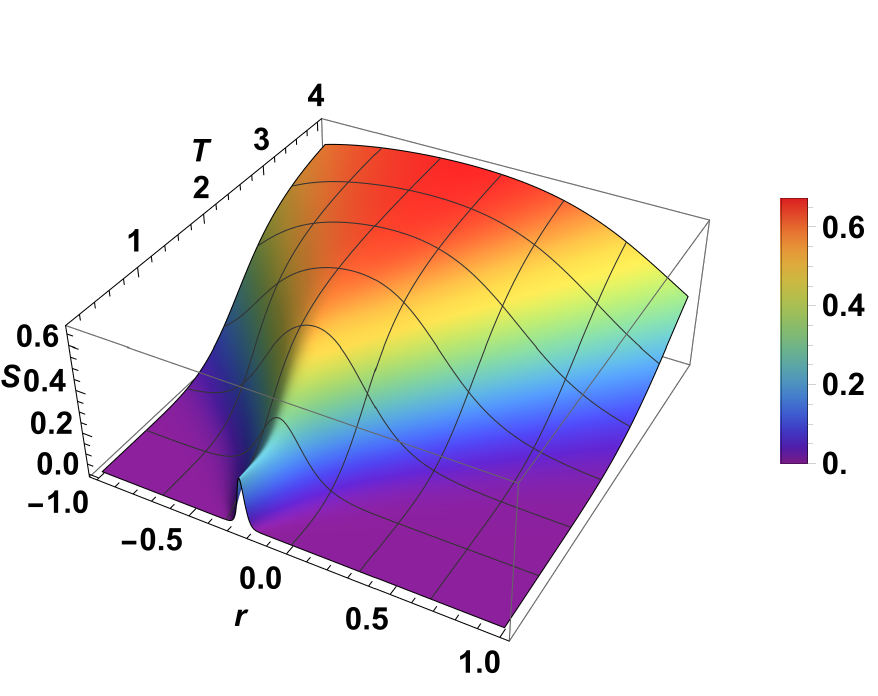} \\ Entropy}
				\end{minipage}
				\caption{Plots of free energy, internal energy, specific heat and entropy for the case $J_1 = J_2 = \frac{1}{2}, J_4 = \frac{1}{4}$ and $J_3 = r \in [-1, 1], T\in[0.1, 4]$} 
				\label{gon15}
			\end{figure}
			\indent The last case (Fig. \ref{gon15}) involves an interaction that yields the so-called gonihedric model (when $J_3=-\frac{1}{4}$). \\
			\indent In \cite{Cirillo}, while constructing phase diagrams for a three-dimensional cubic lattice at variable next-nearest neighbor interaction ($J_3$ in our work), a phase transition in the low-temperature region for two different values of $J_4$ was observed near the value of the parameter $J_3$ corresponding to the gonihedric model (at $\frac{J_3}{J_1} = -\frac{1}{4}$).\\
			\indent In our case, the special behavior of the plots in Figure \ref{gon15} in the low-temperature region is observed at $J_3=-\frac{1}{4}$, which corresponds to the planar gonihedric model ($\frac{J_3}{J_1} = -\frac{1}{2}$). When graphs were plotted on a set of other values of parameters $J_1=J_2$ and $J_4$ (the parameters were taken positive), independent of each other, which were not included, a similar picture was observed: a special behavior of the graphs in the low-temperature region occurred at the point where the relation $\frac{J_3}{J_1} = -\frac{1}{2}$ was observed.
			\subsection{Planar gonihedric model}
\begin{figure}[h]
				\begin{minipage}[h]{0.495\linewidth}
					\centering{\includegraphics[width=0.96\linewidth]{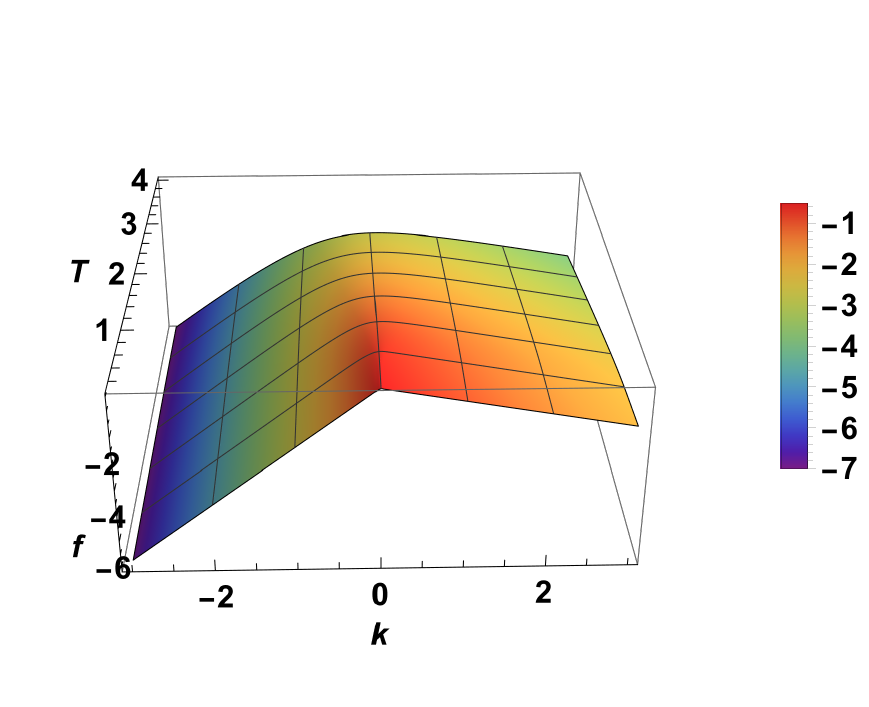}\\ Free energy}
				\end{minipage}
				\hfill
				\begin{minipage}[h]{0.495\linewidth}
					\centering{\includegraphics[width=0.96\linewidth]{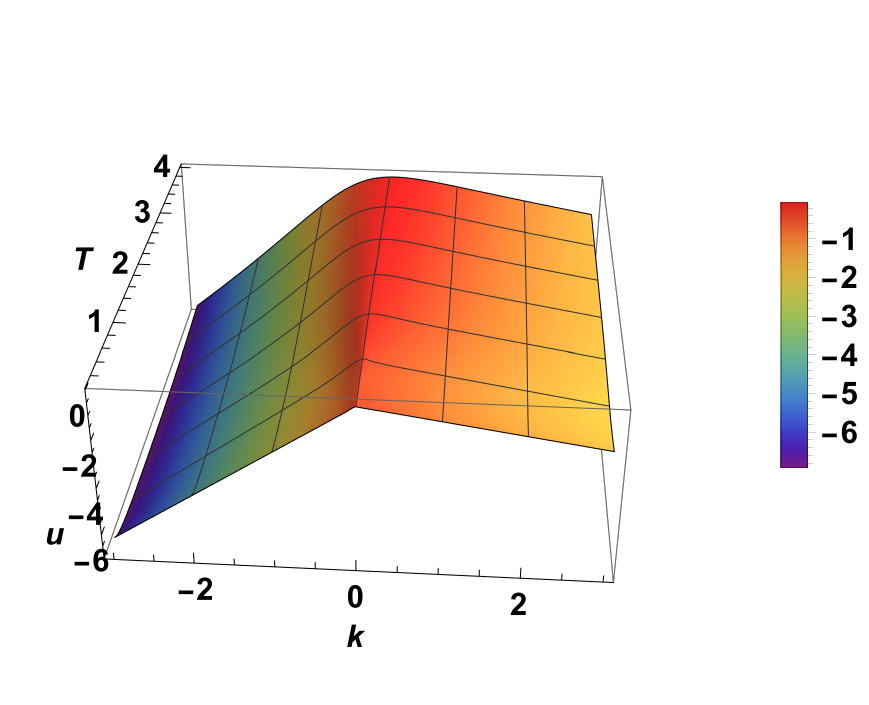} \\ Internal energy}
				\end{minipage}
				\vfill
				\begin{minipage}[h]{0.495\linewidth}
					\centering{\includegraphics[width=0.96\linewidth]{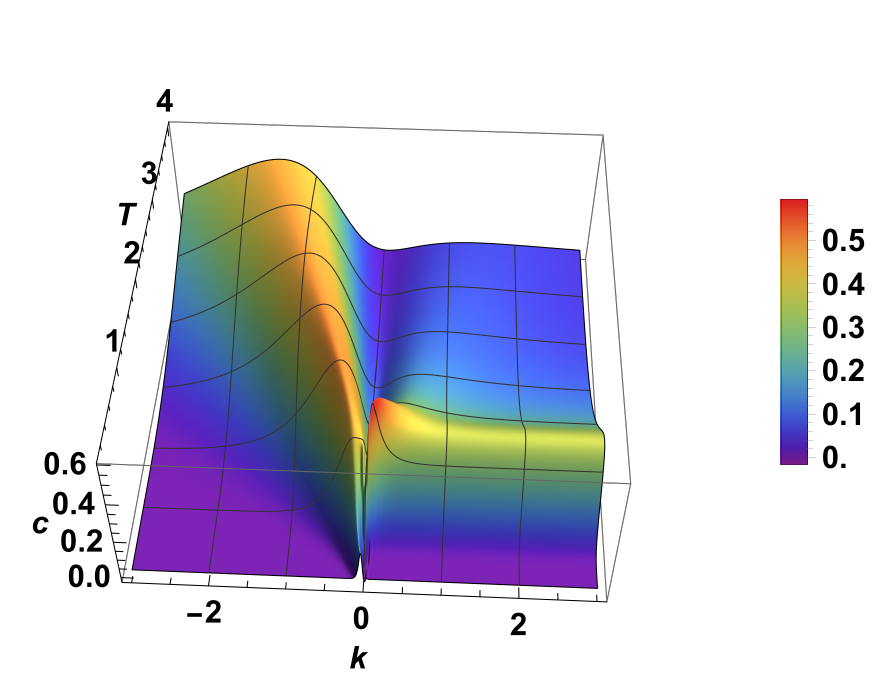} \\ Specific heat}
				\end{minipage}
				\hfill
				\begin{minipage}[h]{0.495\linewidth}
					\centering{\includegraphics[width=0.96\linewidth]{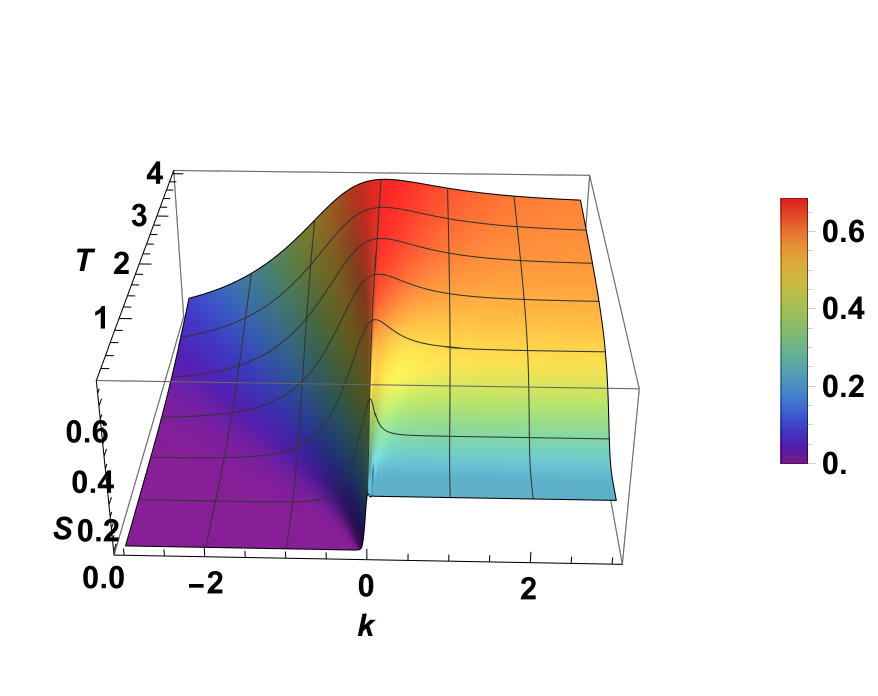} \\ Entropy}
				\end{minipage}
				\caption{Plots of the physical characteristics of the planar gonihedric model at $k \in [-3, 3], T\in[0.1, 4]$} 
				\label{gon16}
			\end{figure}

			\indent \\
			\indent A special case of the planar model with nearest neighbor, next-nearest neighbor, and plaquette interactions discussed above is the so-called planar gonihedric model, which was mentioned when graphing in Figure \ref{gon15}.\\
			\indent The Hamiltonian of one layer of the planar gonihedric model coincides with (\ref{june8}), for which \cite{Bathas}
			\begin{equation} \label{june11}
				\begin{gathered}
					J_1 = J_2 = k, \\
					J_3 = -\frac{k}{2}, \\
					J_4 = \frac{1-k}{2},
				\end{gathered}
			\end{equation}
			i.e. all interspin interactions $J_i$ in the gonihedric model depend on one parameter $k$, and the theorems \ref{th3} — \ref{th7} are valid for it, as well as for the more general case considered above.\\
			\indent As an example, for the case of the planar gonihedric model we have considered, we plot the free energy, internal energy, specific heat, and entropy at the variables temperature $T$ and coefficient $k$ (Fig. \ref{gon16}).
			
			\newpage
			\section{Planar triangular model}
			\indent \\
			\indent Consider a planar triangular model with different nearest neighbor interactions, triple interactions in the same triangle, and interaction with the external field, which, similar to the previous case of the planar model, is obtained by unfolding the lattice shown in Figure \ref{p1} in the plane with the addition of additional boundary conditions $t_3^m \equiv t_0^m$ and $t_3^{m+1} \equiv t_0^{m+1}$ (Fig. \ref{triangle}). For clarity, an additional $m+2$th layer of the model is added to the figure \ref{triangle}:
			\begin{figure}[H]
				\centering{\includegraphics[width=1\linewidth]{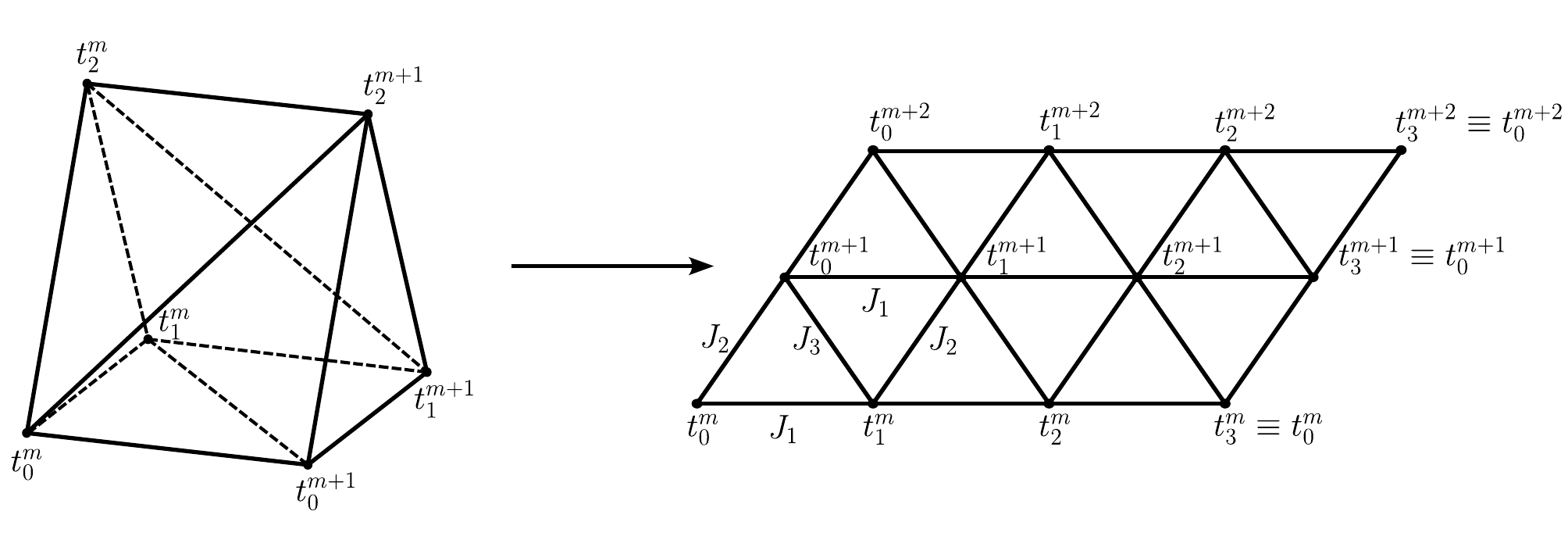}}
				\caption{Planar triangular Ising model with nearest neighbor, next-nearest neighbor and plaquette interactions obtained by unfolding the triple-chain model on the plane}
				\label{triangle}
			\end{figure}
			\indent Here $J_1, J_2, J_3$ correspond to nearest neighbor interactions within a single triangle, $J_4, J_5$ are not shown explicitly but correspond to triple interactions within adjacent triangles, and $H$ corresponds to the interaction with the external field.\\
			\indent The Hamiltonian of the system coincides with (\ref{5}), where the Hamiltonian of one layer of the model is of the form:
			\begin{equation} \label{june9}
				\begin{gathered}
					H^m = -\frac{J_1}{2}\bigg(
					\sigma_{0}^m\sigma_{1}^m +
					\sigma_{0}^m\sigma_{2}^m +
					\sigma_{2}^m\sigma_{3}^m +
					\sigma_{0}^{m+1}\sigma_{1}^{m+1} +
					\sigma_{1}^{m+1}\sigma_{2}^{m+1} +
					\sigma_{2}^{m+1}\sigma_{3}^{m+1} 
					\bigg) - \\
					-J_2 \bigg(
					\frac{1}{2}\sigma_{0}^m\sigma_{0}^{m+1} +
					\sigma_{1}^m\sigma_{1}^{m+1} +
					\sigma_{2}^m\sigma_{2}^{m+1} +
					\frac{1}{2}\sigma_{3}^m\sigma_{3}^{m+1}
					\bigg) - \\
					-J_3 \bigg(
					\sigma_{1}^m\sigma_{0}^{m+1} +
					\sigma_{2}^m\sigma_{1}^{m+1} +
					\sigma_{3}^m\sigma_{2}^{m+1}
					\bigg) - \\
					-J_4 \bigg(
					\sigma_{0}^m\sigma_{1}^m\sigma_{0}^{m+1} +
					\sigma_{1}^m\sigma_{2}^m\sigma_{1}^{m+1} +
					\sigma_{2}^m\sigma_{3}^m\sigma_{2}^{m+1} 
					\bigg) - \\
					-J_5 \bigg(
					\sigma_{1}^m\sigma_{0}^{m+1}\sigma_{1}^{m+1} +
					\sigma_{2}^m\sigma_{1}^{m+1}\sigma_{2}^{m+1} +
					\sigma_{3}^m\sigma_{2}^{m+1}\sigma_{3}^{m+1} 
					\bigg) - \\
					-\frac{H}{2}\bigg(
					\frac{1}{2}\sigma_{0}^m + \sigma_{1}^m + \sigma_{2}^m + \frac{1}{2}\sigma_{3}^m +
					\frac{1}{2}\sigma_{0}^{m+1} + \sigma_{1}^{m+1} + \sigma_{2}^{m+1} + \frac{1}{2}\sigma_{3}^{m+1}
					\bigg),
				\end{gathered}
			\end{equation}
			\indent The transfer-matrix of such a model contains 22 different matrix elements, its eigenvalues coincide with (\ref{ny2}) — (\ref{ny4.1}), and the theorems \ref{th1} and \ref{th2} are valid for it.\\
			\indent As an example for the planar triangular model, we show plots of free energy, internal energy, specific heat, magnetization, magnetic susceptibility, and entropy in the thermodynamic limit in the low-temperature region for two cases of parameters $J_1, \ldots, J_5$ at a constant external field $H$ — for the case when all $J_i > 0$, and when all $J_i < 0$.
			\begin{figure}[H]
				\begin{minipage}[h]{0.495\linewidth}
					\centering{\includegraphics[width=0.9\linewidth]{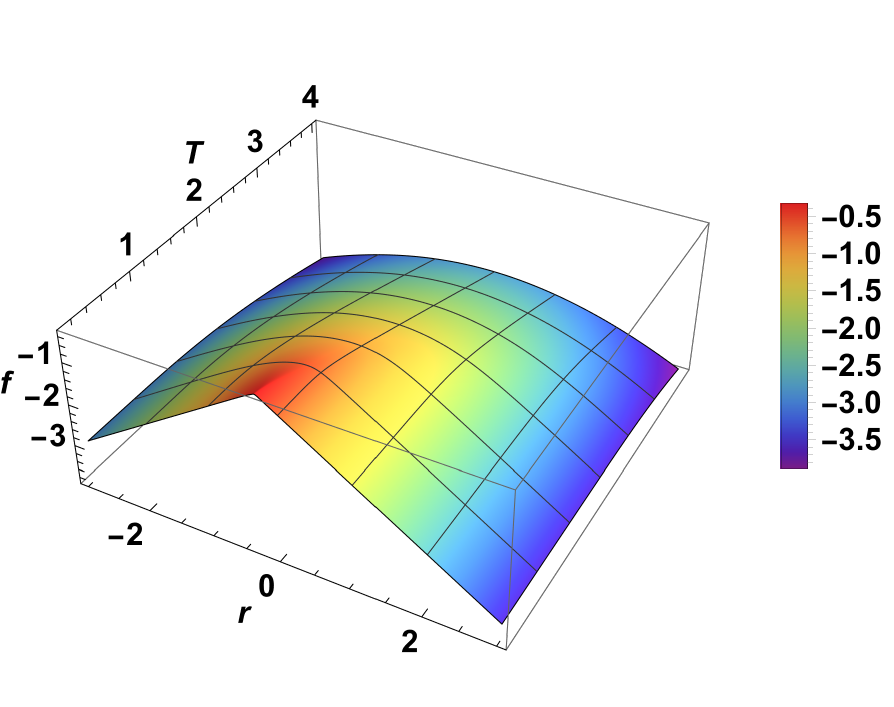}}
				\end{minipage}
				\hfill
				\begin{minipage}[h]{0.495\linewidth}
					\centering{\includegraphics[width=0.9\linewidth]{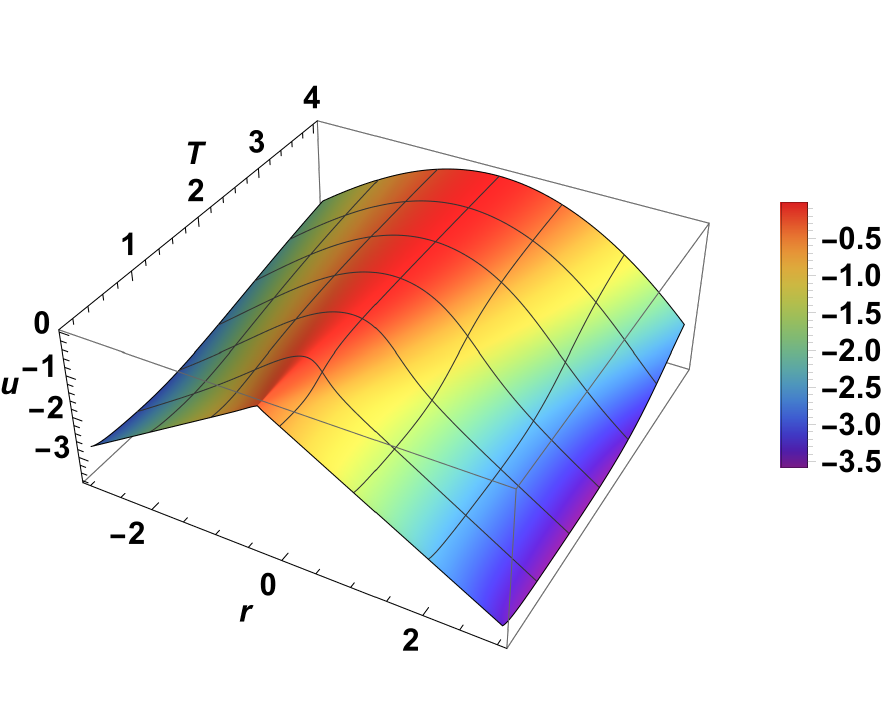}}
				\end{minipage}
				\vfill
				\begin{minipage}[h]{0.495\linewidth}
					\centering{\includegraphics[width=1\linewidth]{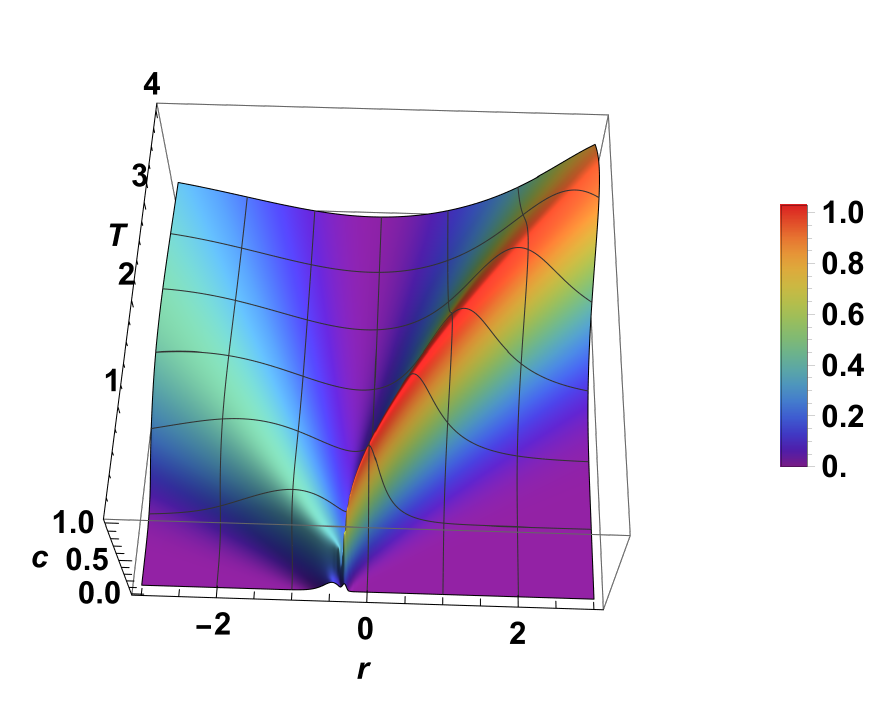}}
				\end{minipage}
				\hfill
				\begin{minipage}[h]{0.495\linewidth}
					\centering{\includegraphics[width=0.9\linewidth]{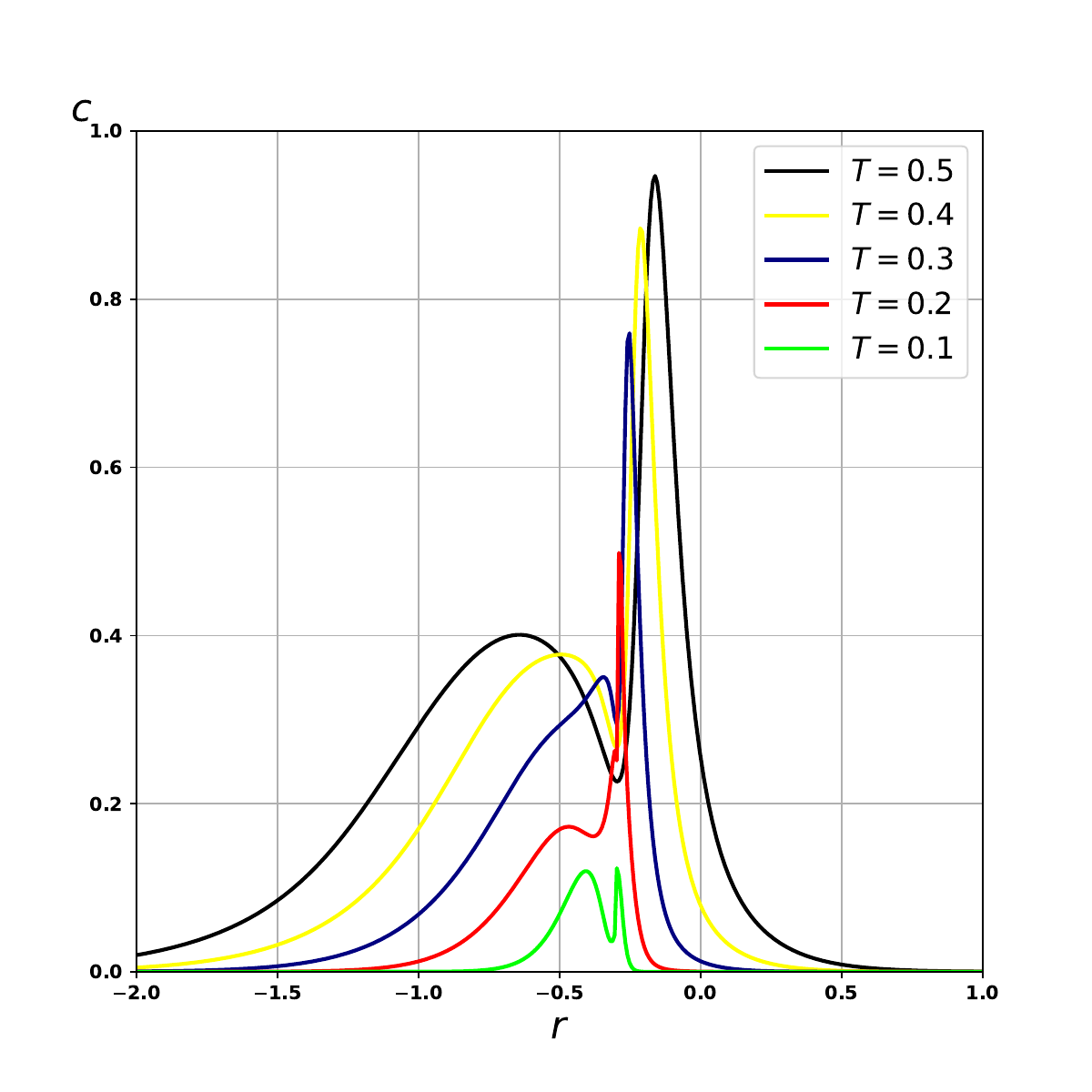}}
				\end{minipage}
				\vfill
				\begin{minipage}[h]{0.495\linewidth}
					\centering{\includegraphics[width=1\linewidth]{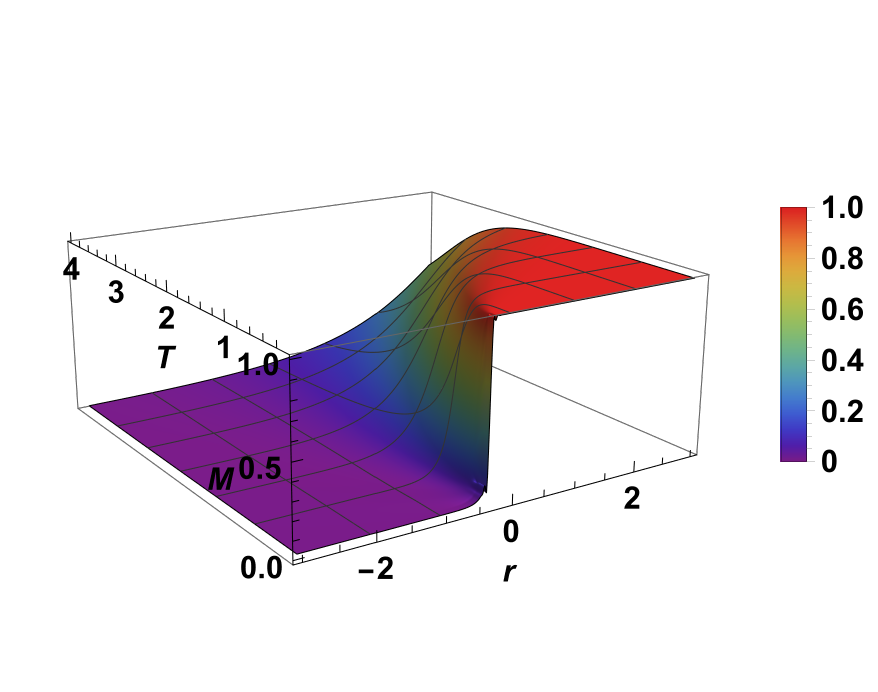}}
				\end{minipage}
				\hfill
				\begin{minipage}[h]{0.495\linewidth}
					\centering{\includegraphics[width=0.9\linewidth]{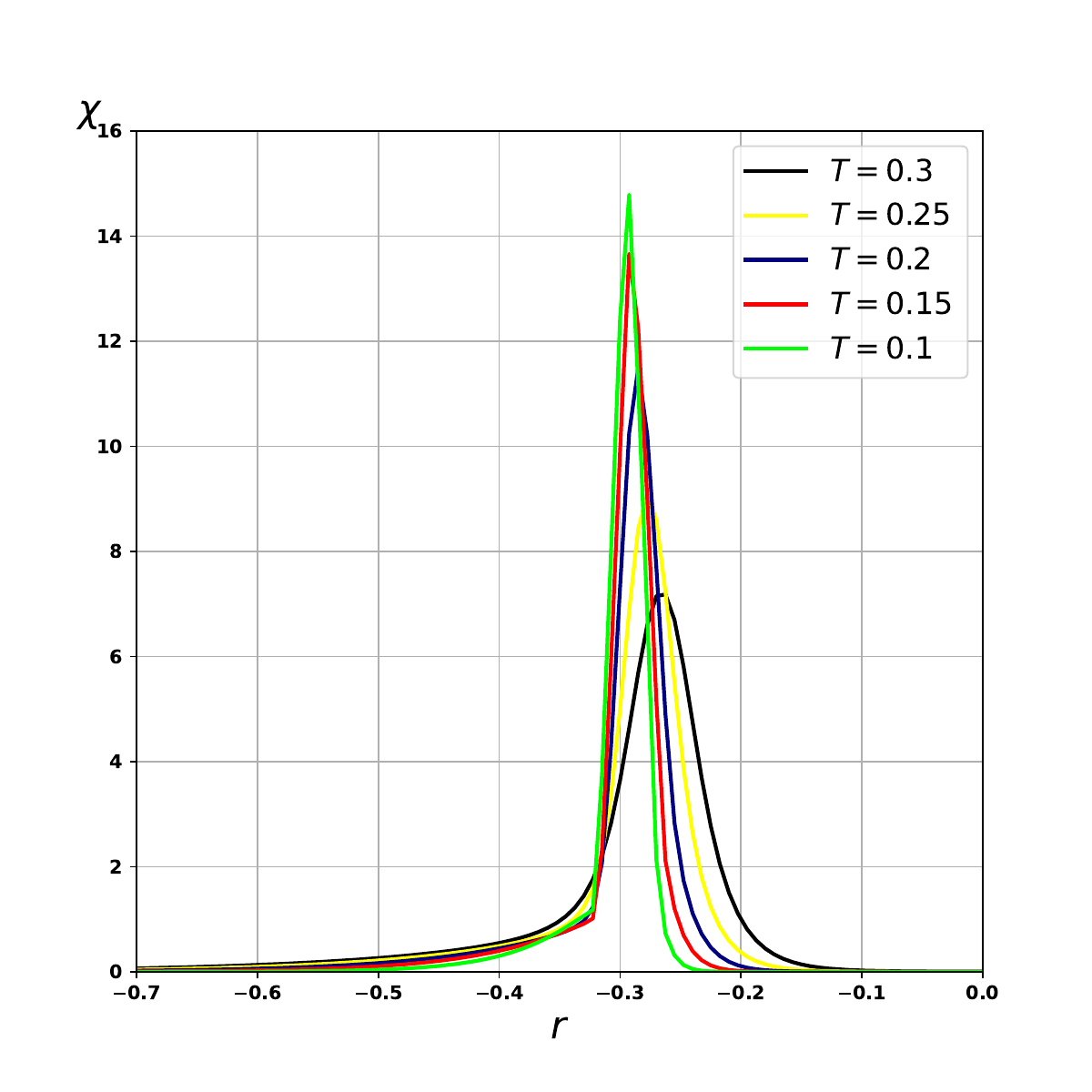}}
				\end{minipage}
				\vfill
				\centering
				\begin{minipage}[h]{0.495\linewidth}
					\centering{\includegraphics[width=1\linewidth]{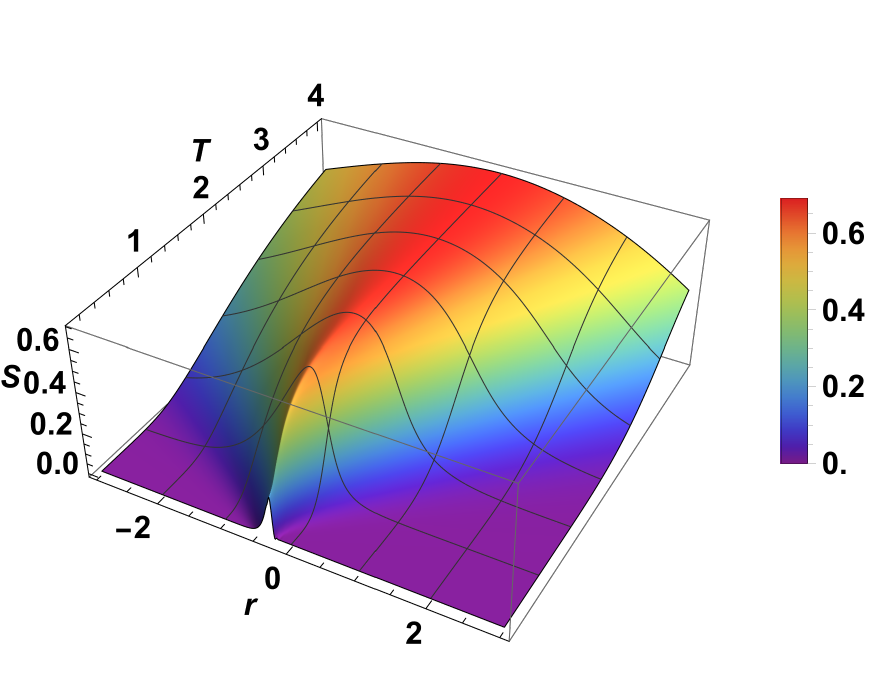}}
				\end{minipage}
				\caption{Plots, respectively, of free energy, internal energy, specific heat and some of its cross sections, magnetization, some susceptibility cross sections and entropy for the case of $J_1 = 0.1, J_2 = 0.11, J_4 = 0.12, J_5 = 0.13, H = 0.15$. $J_3 = r \in [-3, 3], T\in[0.1, 4]$} 
			\end{figure}
			\newpage
			\begin{figure}[H]
				\begin{minipage}[h]{0.495\linewidth}
					\centering{\includegraphics[width=0.9\linewidth]{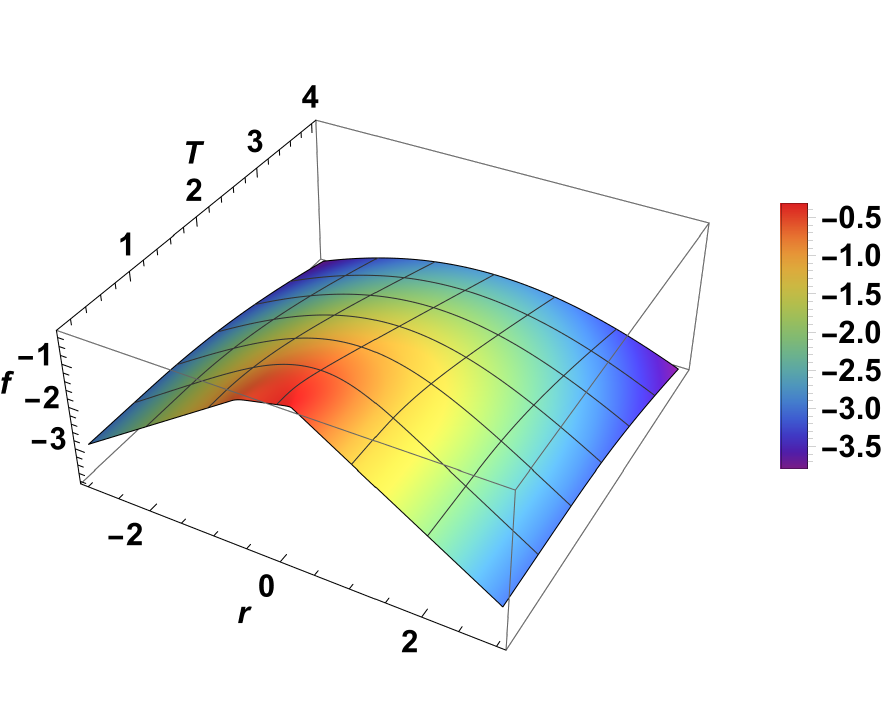}}
				\end{minipage}
				\hfill
				\begin{minipage}[h]{0.495\linewidth}
					\centering{\includegraphics[width=0.9\linewidth]{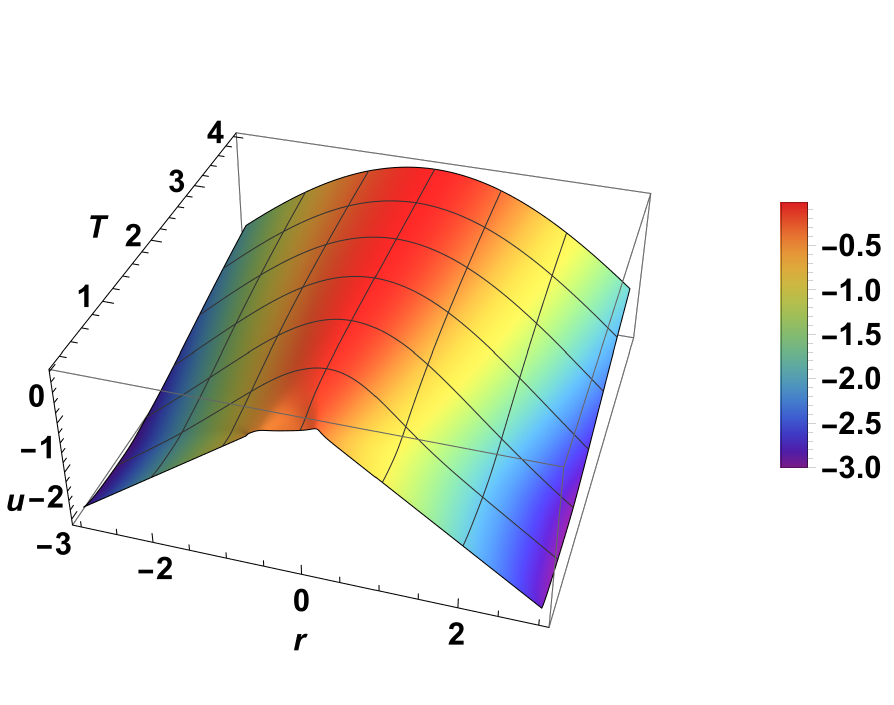}}
				\end{minipage}
				\vfill
				\begin{minipage}[h]{0.495\linewidth}
					\centering{\includegraphics[width=1\linewidth]{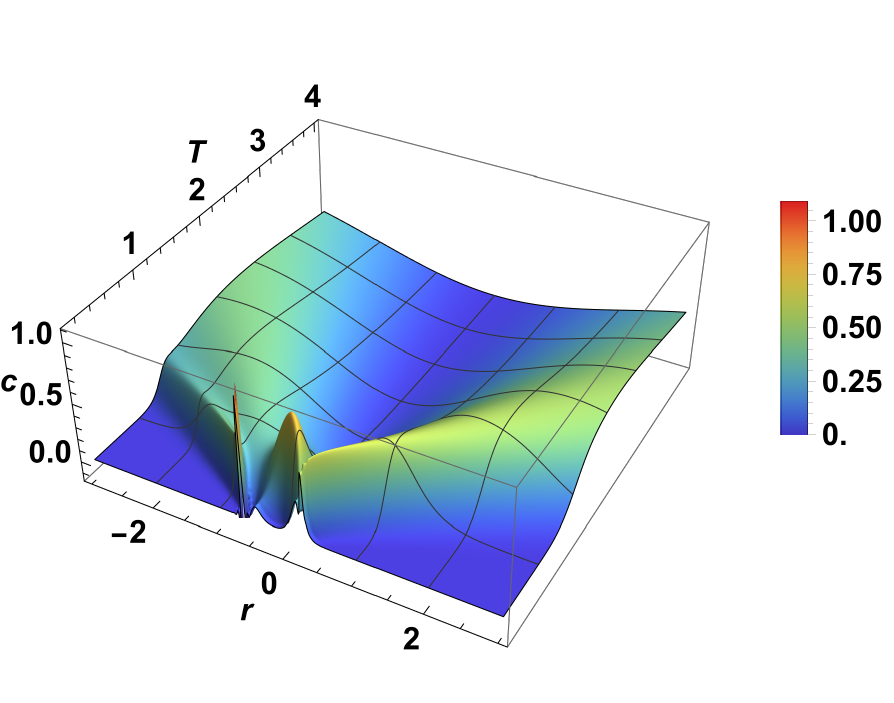}}
				\end{minipage}
				\hfill
				\begin{minipage}[h]{0.495\linewidth}
					\centering{\includegraphics[width=0.9\linewidth]{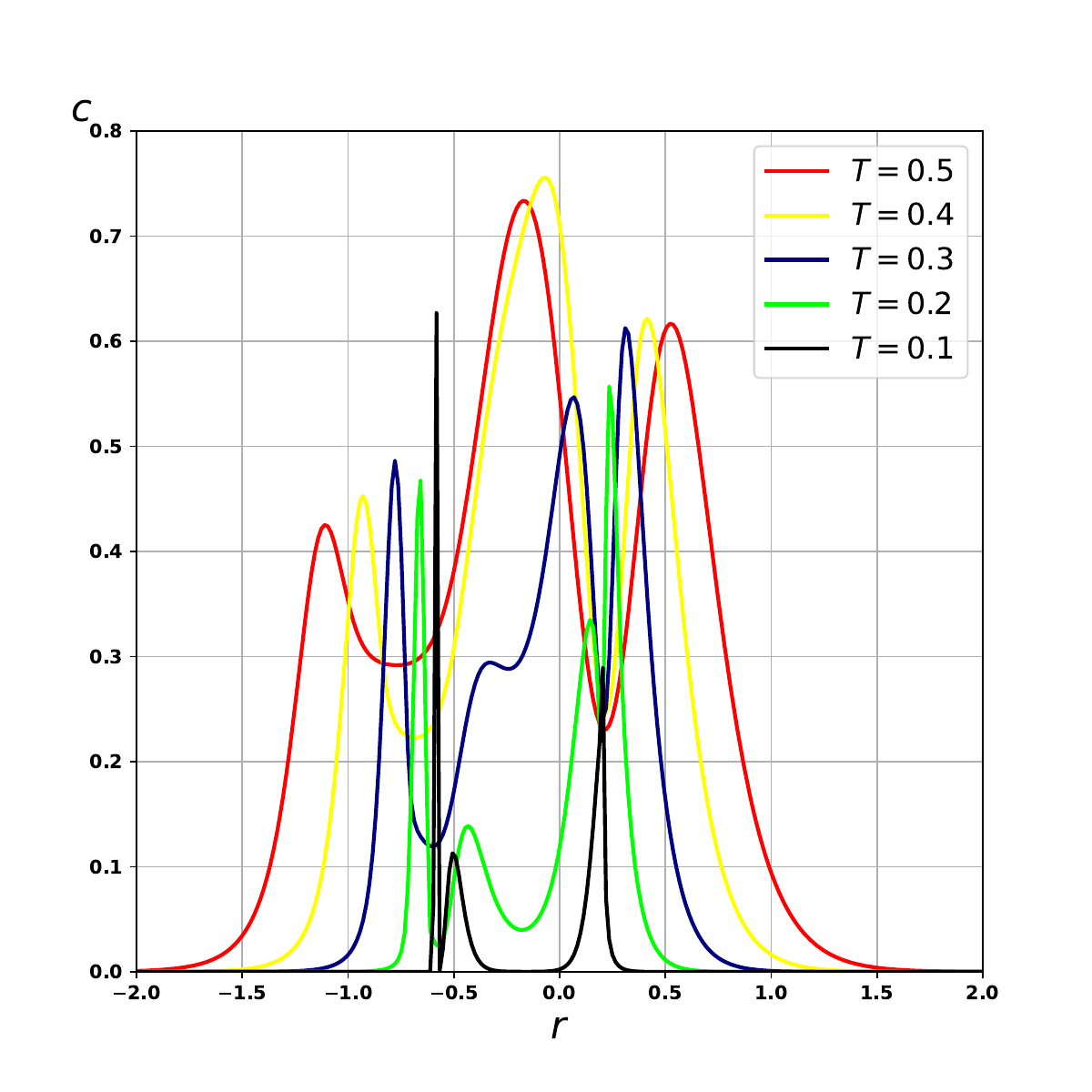}}
				\end{minipage}
				\vfill
				\begin{minipage}[h]{0.495\linewidth}
					\centering{\includegraphics[width=1\linewidth]{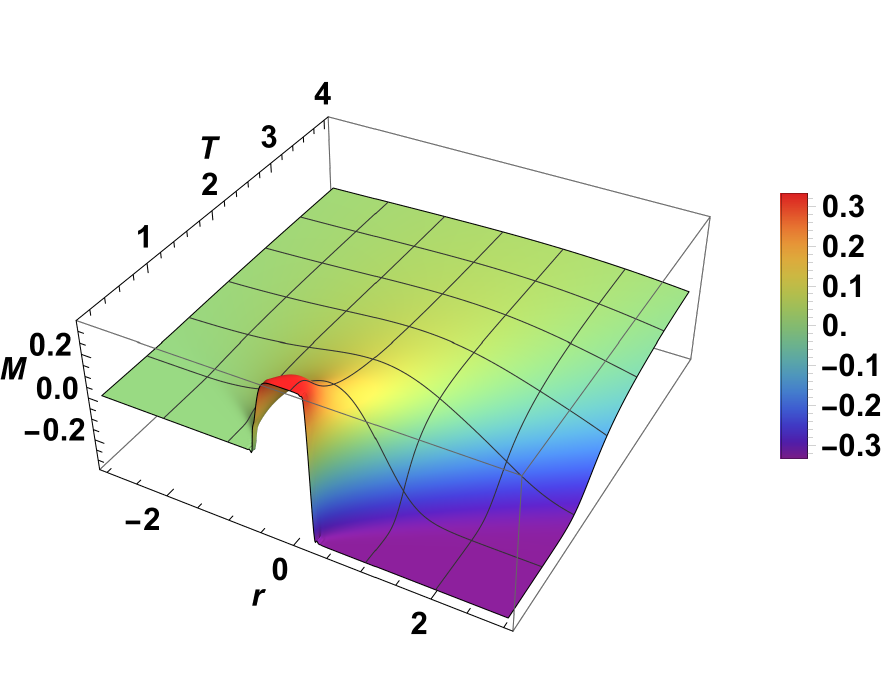}}
				\end{minipage}
				\hfill
				\begin{minipage}[h]{0.495\linewidth}
					\centering{\includegraphics[width=0.9\linewidth]{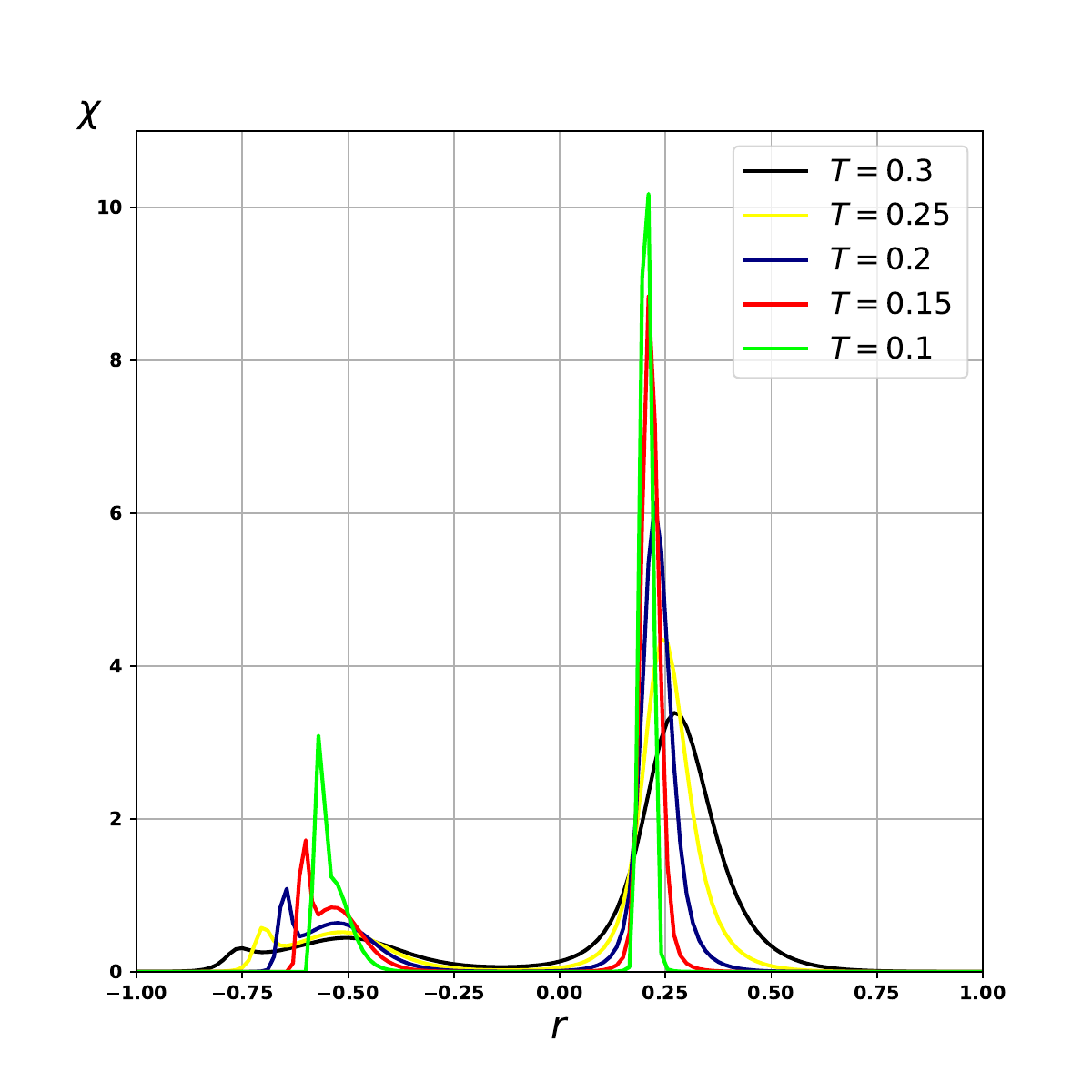}}
				\end{minipage}
				\vfill
				\centering
				\begin{minipage}[h]{0.495\linewidth}
					\centering{\includegraphics[width=1\linewidth]{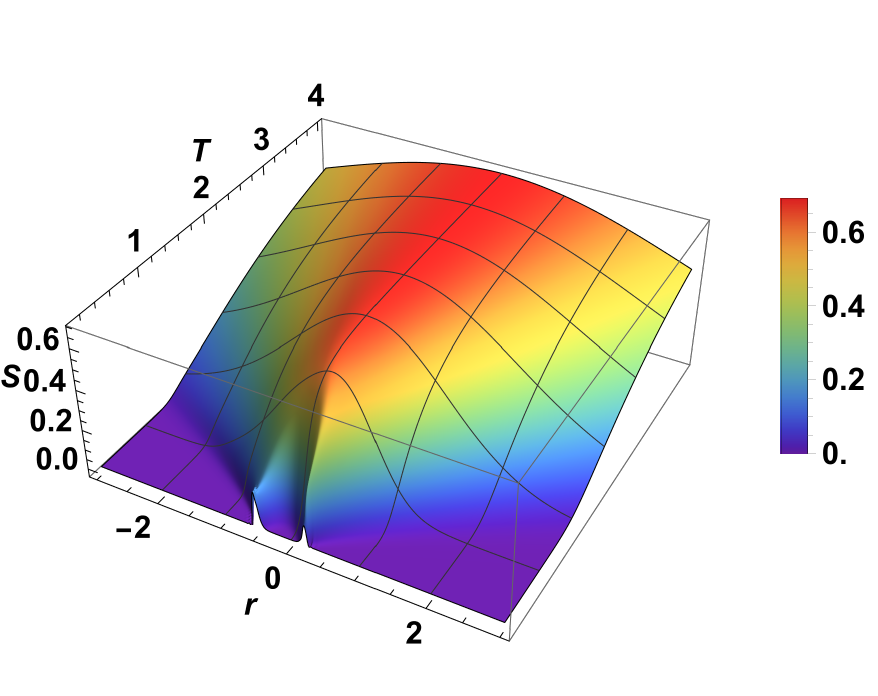}}
				\end{minipage}
				\caption{Plots, respectively, of free energy, internal energy, specific heat and some of its cross sections, magnetization, some susceptibility cross sections and entropy for the case of $J_1 = -0.1, J_2 = -0.11, J_4 = -0.12, J_5 =- 0.13, H = 0.15$. $J_3 = r \in [-3, 3], T\in[0.1, 4]$} 
			\end{figure}
			\newpage
			\section{Examples of physical characteristics in the thermodynamic limit}
\begin{figure}[H]
				\begin{minipage}[h]{0.495\linewidth}
					\centering{\includegraphics[width=1\linewidth]{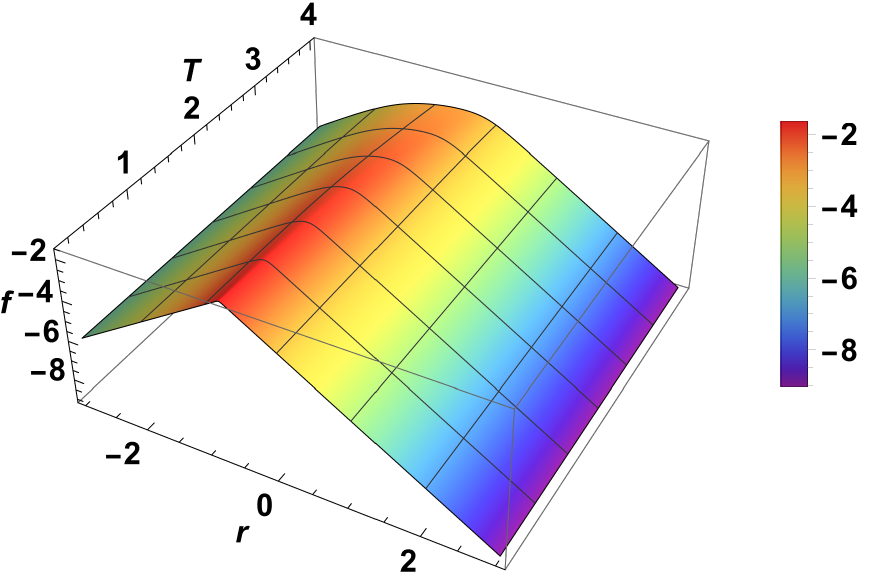} \\ Free energy}
				\end{minipage}
				\hfill
				\begin{minipage}[h]{0.495\linewidth}
					\centering{\includegraphics[width=0.95\linewidth]{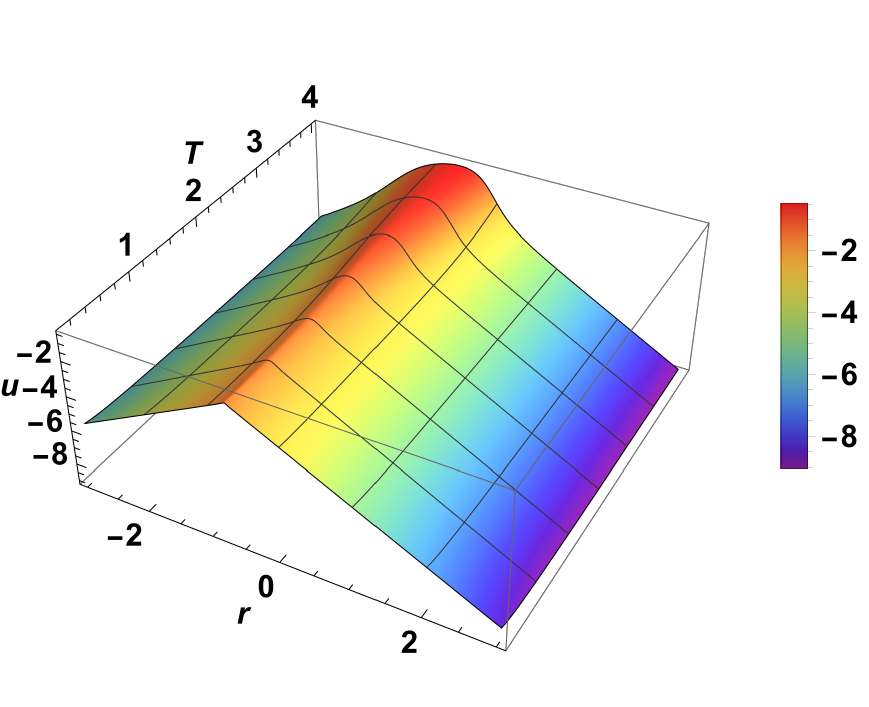} \\ Internal energy}
				\end{minipage}
				\caption{Plots of free energy and internal energy per spin at given interactions $J_1 - J_{19}$ and external field $H$ (table \ref{tab1}) , $r \in [-3,3]$, $T \in [0.1, 4]$} \label{p2}
			\end{figure}

\begin{figure}[h] 
				\centering{\includegraphics[width=0.9\linewidth]{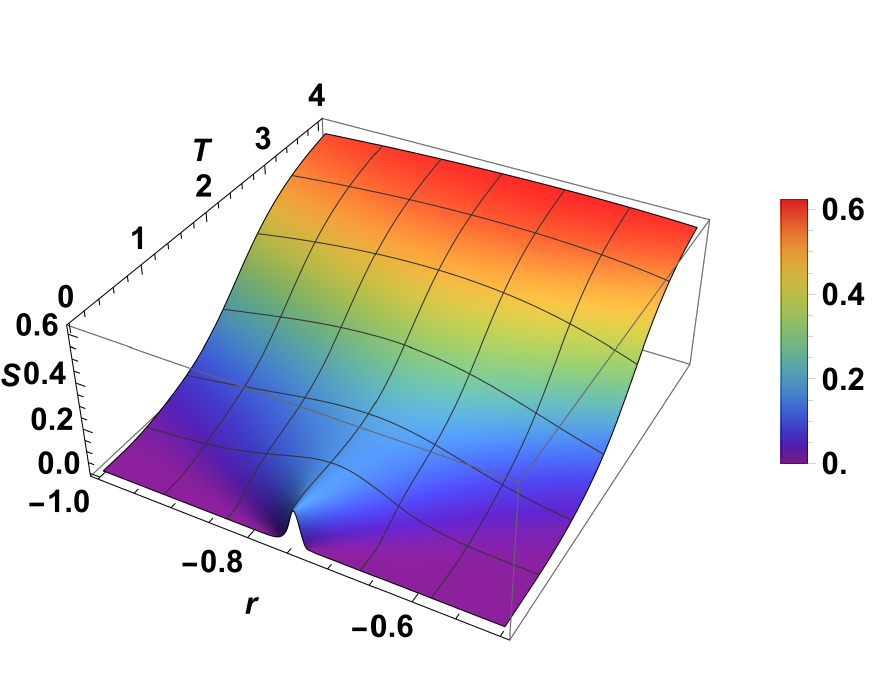}}
				\caption{Entropy graph at given interactions $J_1 - J_{19}$ and external field $H$ (table \ref{tab1}), $r \in [-1,-0.5]$, $T \in [0.05, 4]$.}
				\label{p9}
			\end{figure}

			\indent Let us illustrate the physical characteristics stated in the main theorem \ref{th1}. For this purpose, we will use the analytical expression of the largest eigenvalue $\lambda_{max}$ (\ref{n5}) of the transfer-matrix $\theta$ and apply the formulas (\ref{16}) — (\ref{last4}) to find the thermodynamic characteristics in the thermodynamic limit in the low-temperature region (Fig. \ref{p2} — \ref{p9}) at the following interactions:
			\begin{table} [H]
				\caption{Values of interspin interactions and elementary carriers generating them}
				\label{tab1}
				\begin{center}
					\begin{tabular}{|c|c|c|}
						\hline
						Interaction & Value & Elementary carrier generating interaction\\
						\hline
						${J_1}$ & 0.1 & $\{t_0^m,t_1^m\}$ \\
						\hline
						${J_2}$ & 0.11 & $\{t_0^m, t_0^{m+1}\}$ \\
						\hline
						${J_3}$ & $r$ & $\{t_0^m,t_1^{m+1}\}$ \\
						\hline
						${J_4}$ & $r$ & $\{t_0^m, t_2^{m+1}\}$ \\
						\hline
						${J_5}$ & 0.12 & $\{t_0^m,t_1^m,t_2^m\}$ \\
						\hline
						${J_6}$ & 0.13 & $ \{t_0^m,t_1^m,t_0^{m+1}\}$ \\
						\hline
						${J_7}$ & 0.14 & $\{t_0^m,t_1^m,t_1^{m+1}\}$ \\
						\hline
						${J_8}$ & 0.15 & $ \{t_0^m,t_1^m, t_2^{m+1}\}$ \\
						\hline
						${J_9}$ & 0.16 & $ \{t_0^m,t_0^{m+1}, t_2^{m+1}\}$ \\
						\hline
						${J_{10}}$ & 0.17 & $\{t_0^m,t_0^{m+1},t_1^{m+1}\}$ \\
						\hline
						${J_{11}}$ & 0.18 & $ \{t_0^m,t_1^{m+1},t_2^{m+1}\}$ \\
						\hline
						${J_{12}}$ & 0.19 & $\{t_0^m,t_1^{m},t_2^{m}, t_0^{m+1} \}$ \\
						\hline
						${J_{13}}$ & 0.2 & $\{t_0^m,t_1^{m},t_0^{m + 1}, t_1^{m+1} \}$ \\
						\hline
						${J_{14}}$ & 0.21 & $\{t_0^m,t_1^{m},t_0^{m+1}, t_2^{m+1}\}$ \\
						\hline
						${J_{15}}$ & 0.22 & $\{t_0^m,t_1^{m},t_1^{m+1}, t_2^{m+1} \}$ \\
						\hline
						${J_{16}}$ & 0.23 & $\{t_0^m,t_0^{m+1},t_1^{m + 1}, t_2^{m+1} \}$ \\
						\hline
						${J_{17}}$ & 0.24 & $\{t_0^m,t_1^{m},t_2^{m}, t_0^{m+1},t_1^{m+1}\}$ \\
						\hline
						${J_{18}}$ & 0.25 & $ \{t_0^m,t_1^{m},t_0^{m+1}, t_1^{m+1}, t_2^{m+1}\}$ \\
						\hline
						${J_{19}}$ & 0.26 & $ \{t_0^m,t_1^{m},t_2^{m}, t_0^{m+1}, t_1^{m+1}, t_2^{m+1}\}$ \\
						\hline
						${H}$ & 0.3 & External magnetic field \\
						\hline
					\end{tabular}
				\end{center}
			\end{table}
			\indent \\
			\indent \\
			\indent \\
			
			\begin{figure}[H]
				\begin{minipage}[h]{0.5\linewidth}
					\centering{\includegraphics[width=1\linewidth]{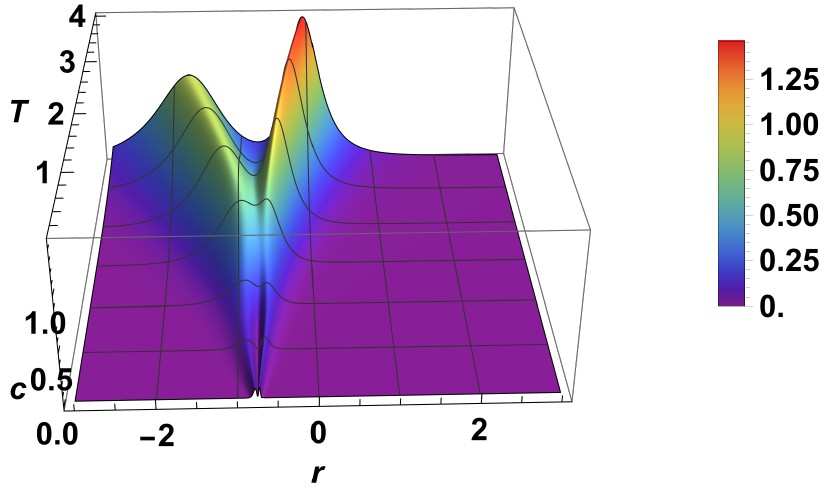}}
				\end{minipage}
				\hfill
				\begin{minipage}[h]{0.49\linewidth}
					\centering{\includegraphics[width=1\linewidth]{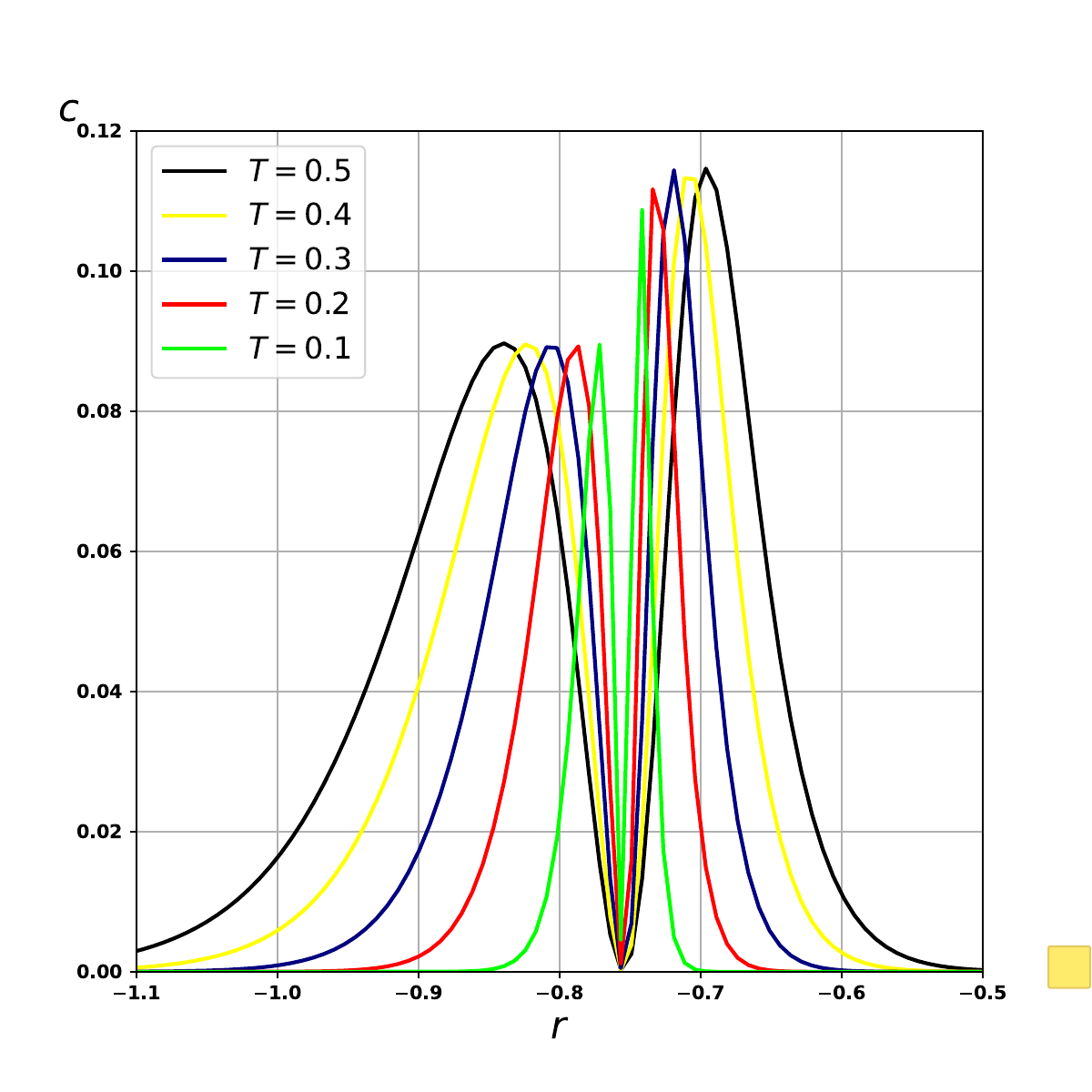}}
				\end{minipage}
				\caption{Plot of specific heat per spin for given interactions $J_1 - J_{19}$ and external field $H$ (table \ref{tab1}) , $r \in [-3,3]$, $T \in [0.1, 4]$ and its cross sections at $T = 0.1$, $T = 0.2$, $T = 0.3$, $T = 0.4$, $T = 0.5$} \label{p4}
			\end{figure}
			\indent \\
			\indent \\
			\indent \\
			\begin{figure}[H]
				\centering{\includegraphics[width=0.9\linewidth]{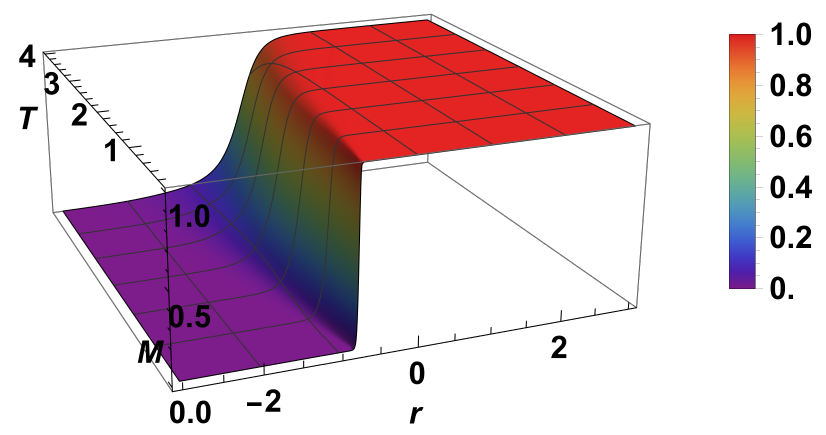}}
				\caption{Plot of magnetization at given interactions $J_1 - J_{19}$ and external field $H$ (table \ref{tab1}), $r \in [-3,3]$, $T \in [0.1, 4]$}
				\label{p6}
			\end{figure}
			\indent The susceptibility cross sections in the low-temperature region are as follows:
			\begin{figure}[H]
				\centering{\includegraphics[width=0.75\linewidth]{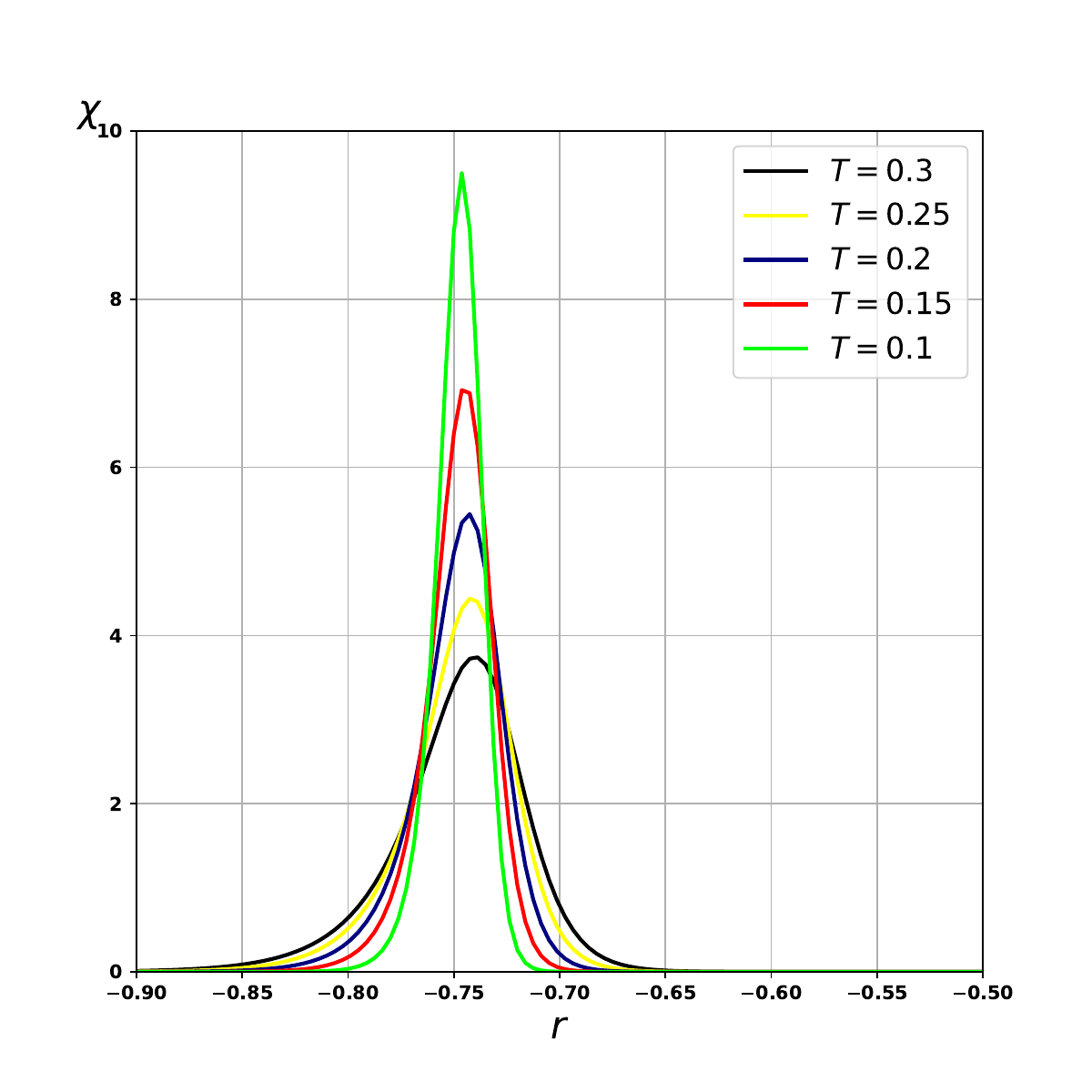}}
				\caption{Plots of susceptibility at given interactions $J_1 - J_{19}$ and external field $H$ (table \ref{tab1}) , $r \in [-0.9,-0.5]$, $T = 0.1$, $T = 0.15$, $T = 0.2$, $T = 0.25$, $T = 0.3$.}
				\label{p8}
			\end{figure}
			\clearpage
			
			\newpage
			\section{Proofs of theorems}
			\subsection{Proof of the theorems \ref{th1}, \ref{th2}}
			\indent \\
			\indent To prove the corresponding theorems, let us find all eigenvalues of the transfer-matrix $\theta$ (\ref{12}).\\
			\indent One of the matrices commuting with $\theta$ is the "rotation" matrix $D_1$ by the angle $\frac{2\pi}{3}$, $D_1:\big\{ \sigma_0^m \rightarrow \sigma_1^{m+1}, \sigma_1^m \rightarrow \sigma_2^{m+1}, \sigma_2^m \rightarrow \sigma_0^{m+1}, m = 0, \ldots, L-1\big\}$:
			\begin{equation} \label{20}
				D_1 = 
				\begin{pmatrix}
					1&0&0&0&0&0&0&0\\
					0&0&1&0&0&0&0&0\\
					0&0&0&0&1&0&0&0\\
					0&0&0&0&0&0&1&0\\
					0&1&0&0&0&0&0&0\\
					0&0&0&1&0&0&0&0\\
					0&0&0&0&0&1&0&0\\
					0&0&0&0&0&0&0&1
				\end{pmatrix},
			\end{equation}
			eigenvalues and basis vectors of corresponding eigenspaces of which are written out in the table \ref{table_vectors_1}:
			\begin{table}[H]
				\centering
				\caption{Eigenvalues and eigenspaces of matrix $D_1$} 		\label{table_vectors_1}
				\begin{tabular}{| c | c |} \hline
					\makecell{Eigenvalue} & \makecell{The basis vectors of the eigenspace}  \\ \hline
					\makecell{1} & \makecell{$
						\begin{gathered}
							\\
							(1, 0, 0, 0, 0, 0, 0, 0)^\intercal \\
							(0, 1, 1, 0, 1, 0, 0, 0)^\intercal \\
							(0, 0, 0, 1, 0, 1, 1, 0)^\intercal \\
							(0, 0, 0, 0, 0, 0, 0, 1)^\intercal \\
						\end{gathered}
						$} \\ \hline
					\makecell{$e^{\frac{2\pi}{3}i}$} & \makecell{$
						\begin{gathered}
							\\
							(0, 0, 0, e^{-\frac{2\pi}{3}i}, 0, e^{\frac{2\pi}{3}i}, 1, 0)^\intercal \\
							(0, e^{\frac{2\pi}{3}i}, e^{-\frac{2\pi}{3}i}, 0, 1, 0, 0, 0)^\intercal
						\end{gathered}
						$} \\ \hline
					\makecell{$e^{-\frac{2\pi}{3}i}$} & \makecell{$
						\begin{gathered}
							\\
							(0, 0, 0, e^{\frac{2\pi}{3}i}, 0, e^{-\frac{2\pi}{3}i}, 1, 0)^\intercal \\
							(0, e^{-\frac{2\pi}{3}i}, e^{\frac{2\pi}{3}i}, 0, 1, 0, 0, 0)^\intercal
						\end{gathered}
						$} \\ \hline
				\end{tabular}
			\end{table}
			
			\indent Using the theorem on the conservation of eigenspaces by commuting matrices \cite{Horn}, and considering that any linear combination of vectors of an eigenspace is also an eigenvector of that eigenspace, let us write down three different kinds of eigenvectors of the transfer-matrix $\theta$:
			\begin{equation} \label{ny16}
				\begin{gathered}
					\overrightarrow{y_1} = (x_1, x_2, x_2, x_3, x_2, x_3, x_3, x_4)^\intercal,\\
					\overrightarrow{y_2} = (0, e^{\frac{2\pi}{3}i}x_1, e^{-\frac{2\pi}{3}i}x_1, e^{-\frac{2\pi}{3}i}x_2, x_1,e^{\frac{2\pi}{3}i}x_2, x_2, 0)^\intercal,\\
					\overrightarrow{y_3} = (0, e^{-\frac{2\pi}{3}i}x_1, e^{\frac{2\pi}{3}i}x_1, e^{\frac{2\pi}{3}i}x_2, x_1, e^{-\frac{2\pi}{3}i}x_2, x_2, 0)^\intercal.
				\end{gathered}
			\end{equation}
			\indent As a result, all 8 eigenvectors of the matrix $\theta$ will be represented by these types of vectors, namely: four eigenvectors of the form $\overrightarrow{y_1}$, two eigenvectors of the form $\overrightarrow{y_2}$ and two eigenvectors of the form $\overrightarrow{y_3}$. \\
			\indent From the structure of the eigenvectors (\ref{ny16}) and the structure of the transfer-matrix $\theta$ (\ref{12}), four eigenvalues of the matrix $\theta$ will coincide with the eigenvalues of the matrix $\tau^1$ (\ref{19}), two will coincide with the eigenvalues of the matrix $\tau^2$ (\ref{n10}), and two more will coincide with the eigenvalues of the matrix $\tau^3$ (\ref{n14}).\\
			\indent The characteristic equation of the matrix $\tau^1$ (\ref{19}) is an equation of degree four (\ref{ny1}) whose coefficients $a,b,c,d$ are given in the appendix. In the general case it is solved by the method of Ferrari, also described in the appendix, according to which and taking into account the substitutions (\ref{ny1.1}), four of the eight eigenvalues of the transfer-matrix $\theta$ have explicit expressions (\ref{ny2}), (\ref{ny3}). \\
			\indent The characteristic equations of the matrices $\tau^2$ (\ref{n10}), $\tau^3$ (\ref{n14}) are quadratic, respectively:
			\[
			\begin{gathered}
				\lambda^2 + b_2\lambda + c_2 = 0,\\
				\lambda^2 + b_3\lambda + c_3 = 0,
			\end{gathered}
			\]
			whose coefficients $b_2, c_2, b_3, c_3$ respectively coincide with the expressions (\ref{ny1.2}) and, consequently, their eigenvalues are expressed as (\ref{ny4}), (\ref{ny4.1}).\\
			\indent The transfer-matrix $\theta$ is real, with positive elements. Then, by the Perron-Frobenius theorem, it has a simple largest eigenvalue corresponding to an eigenvector with all positive components: comparing the obtained expressions (\ref{ny2}) — (\ref{ny4.1}), we have that the largest eigenvalue is $\lambda_1$ coinciding with (\ref{n5}).\\
			\indent The largest eigenvalue $\lambda_{max}(H,T)$ of the transfer-matrix satisfies the equation (\ref{ny1}). Differentiating and doubly differentiating it by the external field $H$, we obtain the expressions (note that $d'_H = 0$ according to the form of the matrix (\ref{19})):
			\[
			\begin{gathered}
				4\lambda_{max}^3\frac{\partial \lambda_{max}}{\partial H}  + a'_H\lambda_{max}^3 + 3a\lambda_{max}^2\frac{\partial \lambda_{max}}{\partial H}  + b'_H\lambda_{max}^2 +  2b\lambda_{max}\frac{\partial \lambda_{max}}{\partial H}  + \\
				+ c'_H\lambda_{max} + c\frac{\partial \lambda_{max}}{\partial H} = 0,
			\end{gathered}
			\]
			\[
			\begin{gathered}
				12\lambda_{max}^2\biggl(\frac{\partial \lambda_{max}}{\partial H}\biggr)^2 + 4 \lambda_{max}^3 \frac{\partial^2 \lambda_{max}}{\partial H^2} + a''_{HH}\lambda_{max}^3  + 6a'_H\lambda_{max}^2\frac{\partial \lambda_{max}}{\partial H} + \\
				+ 6a\lambda_{max}\biggl(\frac{\partial \lambda_{max}}{\partial H}\biggr)^2 
				+ 3a\lambda_{max}^2\frac{\partial^2 \lambda_{max}}{\partial H^2} + b''_{HH}\lambda_{max}^2 + 4b'_H\lambda_{max}\frac{\partial \lambda_{max}}{\partial H} + \\
				+ 2b\biggl(\frac{\partial \lambda_{max}}{\partial H}\biggr)^2 +  2b\lambda_{max}\frac{\partial^2 \lambda_{max}}{\partial H^2} 
				+ c''_{HH}\lambda_{max} + 2c'_H\frac{\partial \lambda_{max}}{\partial H} + c \frac{\partial^2 \lambda_{max}}{\partial H^2} = 0,
			\end{gathered}
			\]
			from which the expressions (\ref{last5}), (\ref{last6}) are obtained.\\
			\indent Similarly, differentiating equation (\ref{ny1}) by temperature $T$, we obtain the expression:
			\[
			\begin{gathered}
				4\lambda_{max}^3\frac{\partial \lambda_{max}}{\partial T}  + a'_T\lambda_{max}^3 + 3a\lambda_{max}^2\frac{\partial \lambda_{max}}{\partial T}  + b'_T\lambda_{max}^2 +  2b\lambda_{max}\frac{\partial \lambda_{max}}{\partial T}  + \\
				+ c'_T\lambda_{max} + c\frac{\partial \lambda_{max}}{\partial T} + d'_T = 0,
			\end{gathered}
			\]
			from which we get (\ref{last11}).\\
			\indent Expressions for the partial derivatives of $a'_H,b'_H,c'_H, a''_{HH}, b''_{HH}, c''_{HH}$ are given in the appendix.
			\subsection{Proof of theorems \ref{th3}, \ref{th4}}
			\indent \\
			\indent In the first special case, in addition to commuting with the matrix $D_1$ (\ref{20}), the transfer-matrix $\theta$ commutes with the matrix $D_2$ due to central symmetry:
			\begin{equation} \label{46}
				D_2 = 
				\begin{pmatrix}
					0&0&0&0&0&0&0&1\\
					0&0&0&0&0&0&1&0\\
					0&0&0&0&0&1&0&0\\
					0&0&0&0&1&0&0&0\\
					0&0&0&1&0&0&0&0\\
					0&0&1&0&0&0&0&0\\
					0&1&0&0&0&0&0&0\\
					1&0&0&0&0&0&0&0
				\end{pmatrix},
			\end{equation}
			eigenvalues and basis vectors of corresponding eigenspaces of which are written out in the table \ref{table_vectors_2}:
			\begin{table}[H]
				\centering
				\caption{Eigenvalues and eigenspaces of the matrix $D_2$} 		\label{table_vectors_2}
				\begin{tabular}{| c | c |} \hline
					\makecell{Eigenvalue} & \makecell{The basis vectors of the eigenspace}  \\ \hline
					\makecell{$1$} & \makecell{$
						\begin{gathered}
							\\
							(1, 0, 0, 0, 0, 0, 0, 1)^\intercal \\
							(0, 1, 0, 0, 0, 0, 1, 0)^\intercal\\
							(0, 0, 1, 0, 0, 1, 0, 0)^\intercal \\
							(0, 0, 0, 1, 1, 0, 0, 0)^\intercal
						\end{gathered}
						$} \\ \hline
					\makecell{$-1$} & \makecell{$
						\begin{gathered}
							\\
							(-1, 0, 0, 0, 0, 0, 0, 1)^\intercal \\
							(0, -1, 0, 0, 0, 0, 1, 0)^\intercal \\
							(0, 0, -1, 0, 0, 1, 0, 0)^\intercal \\
							(0, 0, 0, -1, 1, 0, 0, 0)^\intercal
						\end{gathered}
						$} \\ \hline
				\end{tabular}
			\end{table}
			\indent Using the theorem on the conservation of eigenspaces by commuting matrices \cite{Horn}, and assuming that all eigenvectors of matrix $D_2$ are either centrally symmetric or centrally antisymmetric, and that the transfer-matrix is centrally symmetric for this case, let us write down the first two possible kinds of eigenvectors of the transfer-matrix, which will be either centrally symmetric or centrally antisymmetric \cite{Andrew}:
			\begin{equation} \label{49}
				\begin{gathered}
					\overrightarrow{y_4} = (x_1, x_2, x_2, x_2, x_2, x_2, x_2, x_1)^\intercal, \\
					\overrightarrow{y_5} = (x_1, x_2, x_2, -x_2, x_2, -x_2, -x_2, -x_1)^\intercal.
				\end{gathered}
			\end{equation}
			\indent Moreover, using the commutation $\theta$ with the matrix $D_1$, applying to which the condition of central symmetric or central antisymmetric eigenvector, we find linear combinations of vectors from eigenspaces (table \ref{table_vectors_1}) that are centrally symmetric or centrally antisymmetric, which are defined uniquely (without taking into account some multiplier $x$):
			\begin{equation} \label{ny19}
				\begin{gathered}
					\overrightarrow{y_6} = (0, e^{\frac{2\pi}{3}i}x, e^{-\frac{2\pi}{3}i}x, x, x, e^{-\frac{2\pi}{3}i}x, e^{\frac{2\pi}{3}i}x, 0)^\intercal, \\
					\overrightarrow{y_7} = (0, -e^{\frac{2\pi}{3}i}x, -e^{-\frac{2\pi}{3}i}x, x, -x, e^{-\frac{2\pi}{3}i}x, e^{\frac{2\pi}{3}i}x, 0)^\intercal,\\
					\overrightarrow{y_8} = (0, x, e^{-\frac{2\pi}{3}i}x, e^{\frac{2\pi}{3}i}x, e^{\frac{2\pi}{3}i}x, e^{-\frac{2\pi}{3}i}x, x, 0)^\intercal, \\
					\overrightarrow{y_9} = 	(0, -x, -e^{-\frac{2\pi}{3}i}x, e^{\frac{2\pi}{3}i}x, -e^{\frac{2\pi}{3}i}x, e^{-\frac{2\pi}{3}i}x, x, 0)^\intercal.
				\end{gathered}
			\end{equation}
			\indent As a result, all 8 eigenvectors of the matrix $\theta$ will be represented by these types of vectors, namely: two eigenvectors of the form $\overrightarrow{y_4}$, two eigenvectors of the form $\overrightarrow{y_5}$, and one eigenvector of the form $\overrightarrow{y_6}$ — $\overrightarrow{y_9}$ each.\\
			\indent From the structure of the eigenvectors (\ref{49}), (\ref{ny19}), two eigenvalues of the matrix $\theta$ will coincide with the eigenvalues of the matrix $\tau^4$ (\ref{45}), two will coincide with the eigenvalues of the matrix $\tau^5$ (\ref{54}), and four will coincide with the expressions (\ref{61}) — (\ref{64}).\\
			\indent Similar to the previous proof, the expressions for the eigenvalues of matrices $\tau^4, \tau^5$ of size $2 \times 2$ coincide with the expressions (\ref{ny6}), (\ref{ny7}), and the largest eigenvalue of matrix $\theta$ is $\lambda_1$, the expression for which coincides with (\ref{ny5}).
			
			\subsection{Proof of theorems \ref{th5}, \ref{th6}}
			\indent \\
			\indent In the second special case, in addition to commuting with the matrix $D_1$ (\ref{20}), the transfer-matrix $\theta$ commutes with the matrix $D_3$:
			\begin{equation} \label{ny20}
				D_3 = 
				\begin{pmatrix}
					0&0&0&1&0&0&0&0\\
					0&0&1&0&0&0&0&0\\
					0&1&0&0&0&0&0&0\\
					1&0&0&0&0&0&0&0\\
					0&0&0&0&0&0&0&1\\
					0&0&0&0&0&0&1&0\\
					0&0&0&0&0&1&0&0\\
					0&0&0&0&1&0&0&0
				\end{pmatrix},
			\end{equation}
			eigenvalues and basis vectors of corresponding eigenspaces of which are written out in the table \ref{table_vectors_3}:
			\begin{table}[H]
				\centering
				\caption{Eigenvalues and eigenspaces of the matrix $D_3$} 		\label{table_vectors_3}
				\begin{tabular}{| c | c |} \hline
					\makecell{Eigenvalue} & \makecell{The basis vectors of the eigenspace}  \\ \hline
					\makecell{$1$} & \makecell{$
						\begin{gathered}
							\\
							(0, 0, 0, 0, 1, 0, 0, 1)^\intercal \\
							(0, 0, 0, 0, 0, 1, 1, 0)^\intercal \\
							(1, 0, 0, 1, 0, 0, 0, 0)^\intercal \\
							(0, 1, 1, 0, 0, 0, 0, 0)^\intercal
						\end{gathered}
						$} \\ \hline
					\makecell{$-1$} & \makecell{$
						\begin{gathered}
							\\
							(0, 0, 0, 0, -1, 0, 0, 1)^\intercal \\
							(0, 0, 0, 0, 0, -1, 1, 0)^\intercal \\
							(-1, 0, 0, 1, 0, 0, 0, 0)^\intercal \\
							(0, -1, 1, 0, 0, 0, 0, 0)^\intercal
						\end{gathered}
						$} \\ \hline
				\end{tabular}
			\end{table}
			\indent Similarly to the previous proofs, using the eigenspace conservation theorem for commuting matrices and considering that linear combinations of vectors from eigenspaces are also eigenvectors of this subspace, let us write down one of the possible kinds of eigenvectors of the matrix $\theta$, taking as a basis the eigenspace of the matrix $D_3$ corresponding to the eigenvalue 1:
			\begin{equation} \label{ny23}
				\begin{gathered}
					\overrightarrow{y_{10}} = (x_1, x_2, x_2, x_3, x_2, x_3, x_3, x_4)^\intercal
				\end{gathered}
			\end{equation}
			\indent Comparing the types of vectors $\overrightarrow{y_{1}}$ и $	\overrightarrow{y_{10}}$, уточним вид собственного вектора трансфер-матрицы $\theta$: 
			\begin{equation} \label{ny24}
				\begin{gathered}
					\overrightarrow{y_{10_1}}= (x_1, x_2, x_2, x_1, x_2, x_1, x_1, x_2)^\intercal
				\end{gathered}
			\end{equation}
			\indent From the form of the eigenvector (\ref{ny24}), we obtain that two eigenvalues of the matrix $\theta$ will coincide with the eigenvalues of the matrix $\tau^6$:
			\begin{equation} \label{ny25}
				\begin{gathered}
					\tau^6 = 
					\begin{pmatrix}
						\theta_{11} + \theta_{14} + \theta_{16} + \theta_{17}& \theta_{12} + \theta_{13} + \theta_{15} + \theta_{18}\\
						\theta_{21} + \theta_{24} + \theta_{26} + \theta_{27}& \theta_{22} + \theta_{23} + \theta_{25} + \theta_{28}\\
					\end{pmatrix} = \\
					= \begin{pmatrix}
						\frac{q_1r\big(3+p^4q_2^4q_3^4u^4\big)}{puq_2q_3}& \frac{3pq_2}{q_3ru} + \frac{q_3^3u^3}{p^3q_2^3r}\\
						\frac{3pq_2}{q_3ru} + \frac{q_3^3u^3}{p^3q_2^3r}& \frac{r\big(3q_2^4q_3^4+p^4u^4\big)}{pq_1q_2^3q_3^3u}\\
					\end{pmatrix},
				\end{gathered}
			\end{equation}
			expressions for which coincide with (\ref{ny11}). \\
			\indent To find the remaining six eigenvalues of the matrix $\theta$, we divide its characteristic polynomial by the characteristic polynomial of the matrix $\tau^6$. As a result, we have a polynomial which is factorized into three identical square trinomials:
			\begin{equation} \label{ny26}
				\begin{gathered}
					\bigg(p^6q_1q_2^4q_3^4r^2u^2\lambda^2 + p^5q_2q_3r^3u\bigg[q_2^2q_3^2\big(q_1^2+q_2^2q_3^2\big) - p^4u^4\big(1+q_1^2q_2^6q_3^6\big)\bigg]\lambda + \\
					+q_1\bigg[p^4q_2^4q_3^4\big(r^4-p^4\big) +p^4u^4\big(q_2^8+q_3^8\big) - p^8r^4u^4\big(1+q_2^8q_3^8\big) + q_2^4q_3^4u^8\big(p^{12}r^4-1\big) \bigg]\bigg)^3 = \\
					= \bigg(h\lambda^2 + h_3\lambda + h_4\bigg)^3,
				\end{gathered}
			\end{equation}
			from which we obtain two eigenvalues of the matrix $\theta$, each of which has multiplicity 3 and whose expressions coincide with (\ref{ny12}), (\ref{ny13}).\\
			\indent Comparing the expressions (\ref{ny11}) — (\ref{ny13}) for the eigenvalues of the matrix $\theta$, we obtain that the largest one is $\lambda_1$, coinciding with (\ref{ny10}).
			\indent \\
			\subsection{Proof of theorems \ref{th8}, \ref{th9}}
			\indent \\
			\indent  The proof is based on commuting the matrix $\theta'$ (\ref{june1}) with the matrix $D_1$ (\ref{20}). \\
			\indent Proceeding from the structure of the transfer-matrix $\theta'$ and the type of eigenvectors (\ref{ny16}) obtained from the eigen subspaces of the matrix $D_1$, let us write down three possible types of eigenvectors of the matrix $\theta'$:
			\begin{equation} \label{june10}
				\begin{gathered}
					\overrightarrow{y_{1}}'= (x_1, x_2, x_2, 0, x_2, 0, 0, 0)^\intercal, \\
					\overrightarrow{y_{2}}'= (0, e^{\frac{2\pi}{3}i}x_1, e^{-\frac{2\pi}{3}i}x_1, 0, x_1, 0, 0, 0)^\intercal, \\
					\overrightarrow{y_{3}}'= (0, e^{-\frac{2\pi}{3}i}x_1, e^{\frac{2\pi}{3}i}x_1, 0, x_1, 0, 0, 0)^\intercal.
				\end{gathered}
			\end{equation}
			\indent As a result, four of the eight eigenvectors of the matrix $\theta'$ will be represented by these types of vectors, namely: two eigenvectors of the form $\overrightarrow{y_{1}}'$, one eigenvector of the form $\overrightarrow{y_{2}}'$ and one eigenvector of the form $\overrightarrow{y_{3}}'$.\\
			\indent The remaining four eigenvectors of the matrix $\theta'$ are zero.\\
			\indent The structure of the eigenvectors (\ref{june10}) shows that two of the eight eigenvalues of the matrix $\theta'$ coincide with the eigenvalues of the matrix $\tau'$ (\ref{june3}). Two more, respectively, with the expressions (\ref{june6}), (\ref{june7}), and the remaining four eigenvalues are zero.\\
			\indent Moreover, the largest eigenvalue of the transfer-matrix $\theta'$ coincides with the largest eigenvalue of the matrix $\tau'$, and the analog of the partition function $Z'_{L}$ is equal to the sum of $L$th powers of the eigenvalues of the transfer-matrix $\theta'$.\\
			\indent Thus, in the thermodynamic limit, the equality (\ref{june4}) is true, and in the finite-length strip, the expression for the probability of non-percolation coincides with (\ref{june5}).\\
			\indent \\
			\section{Conclusion}\label{conclusion}
			\indent In this paper, for the considered triple-chain rotation invariant generalized Ising model with all possible multispin interactions for a closed spin chain, the exact values of the partition function, free energy, internal energy, specific heat, magnetization, susceptibility, and entropy in the thermodynamic limit at $L\to\infty$, as well as the exact expression for the partition function in a finite band of length $L$ are found by the transfer-matrix method. \\
			\indent The main results of the paper are formulated and proved in the form of two theorems for the general case. In the first theorem, the expressions of physical characteristics of the model in the thermodynamic limit are calculated using the exact expression for the largest eigenvalue of the transfer-matrix $\theta$. In the second theorem, an expression for finding the partition function in a finite strip of length $L$ is obtained using all 8 eigenvalues of the transfer-matrix $\theta$, the expressions for which were obtained explicitly. Exact expressions for all 8 eigenvalues of the constructed transfer-matrix of size $8 \times 8$ are found: 4 eigenvalues are expressed through the roots of the fourth degree equation, which is solved by the Ferrari method described in the appendix, and the other four eigenvalues are expressed through the roots of two quadratic trinomials. The structure of the eigenvectors of the transfer-matrix is also shown. \\
			\indent With the use of analytical expressions of the largest eigenvalue of the transfer-matrix, an example illustrating the mentioned physical characteristics in the thermodynamic limit is given: at a variable temperature and two variable spin interactions corresponding to the interaction of next-nearest neighbors and equal to each other, three-dimensional plots of free energy, internal energy, specific heat, magnetization, susceptibility and entropy and cross sections of some of them at fixed temperatures are constructed. \\
			\indent The theorems analogous to the general case are formulated and proved for two special cases: the first of which is a model with multispin interactions of even number of spins, and the second is a model with some interactions of two, three, four and six spins. \\
			\indent For both special cases additional matrices (\ref{46}, \ref{ny20}) different from the matrix for the general case (\ref{20}), commuting with the transfer-matrix, were found, which allowed to clarify the type of eigenvectors and simplify the results obtained. \\
			\indent For the main special case, exact analytical expressions for all eigenvalues are given using the transfer-matrix method, four of which, including the largest one, are found as roots of two quadratic equations, and another four as roots of four linear equations. We obtain exact analytic expressions for the partition function in a strip of finite length $L$, as well as exact expressions for the free energy, internal energy, specific heat, and entropy of the model in the thermodynamic limit at $L\to\infty$, which, due to the simple form of the largest eigenvalue of the transfer-matrix, also take a simplified form. In addition, the structure of the eigenvectors of the transfer-matrix is also shown. \\
			\indent An exact analytical expression of pair correlations in the thermodynamic limit was also found for this case, the structure of ground states was shown, for each of which an example of the corresponding configuration of the model was given, and it was also shown that the model has zero magnetization in the absence of an external field. With the help of two special projections of the ground states, the possibility of a special behavior of the correlation length at the boundaries of the ground states was illustrated. \\
			\indent In the second special case, by means of the theorem on the conservation of eigenspaces by commuting matrices, the exact expressions for all eigenvalues of the transfer-matrix are found as two roots of multiplicity one of one square trinomial, and two roots of multiplicity three of the second square trinomial. \\
			\indent Two theorems similar to those for the general case are also formulated and proved for this case, and expressions for the partition function in a finite strip of length $L$ and expressions for the free energy, internal energy, specific heat and entropy in the thermodynamic limit at $L\to\infty$ are explicitly derived.\\
			\indent We formulate and prove two theorems on the probability of non-percolation, invariant with respect to rotation by an angle $2 \pi/3$, in a closed strip of finite length $L $, and in the thermodynamic limit.  \\
			\indent As an important special case of the model with multispin interactions of even number of spins, the model with nearest-neighbor, next-nearest neighbor and plaquette interactions is considered separately. \\
			\indent Results for the flat gonihedric model are formulated separately. \\  
			\indent The large parameter selection possibilities allowed us to obtain accurate characterizations of physical quantities for the planar triangular model as well.
			\section{Appendix}
			\textit{\textbf{Ferrari method for solving the fourth degree equation}}\\
			\indent To find the roots of the fourth degree equation (\ref{ny1}), we will use the Ferrari method \cite{Cardano}.\\
			\indent Suppose there is an equation of degree four:
			\begin{equation} \label{ny27}
				\lambda^4+a\lambda^3+b\lambda^2+c\lambda + d = 0,
			\end{equation}
			for which, in the context of our work.
			\begin{equation} \label{last8}
				\begin{gathered}
					a = - \tau_{11}^1- \tau_{22}^1 - \tau_{33}^1 - \tau_{44}^1, \\
					\begin{gathered}
						b =
						\tau_{11}^1\tau_{22}^1+\tau_{11}^1\tau_{33}^1+\tau_{11}^1\tau_{44}^1+\tau_{22}^1\tau_{33}^1
						+\tau_{22}^1\tau_{44}^1+\tau_{33}^1\tau_{44}^1-\\
						-\tau_{24}^1\tau_{42}^1-\tau_{34}^1\tau_{43}^1-\tau_{23}^1\tau_{32}^1-\tau_{12}^1\tau_{21}^1-\tau_{13}^1\tau_{31}^1-\tau_{14}^1\tau_{41}^1,
					\end{gathered}		\\
					\begin{gathered}
						c = 
						-\tau_{11}^1\tau_{22}^1\tau_{33}^1-\tau_{11}^1\tau_{22}^1\tau_{44}^1-\tau_{11}^1\tau_{33}^1\tau_{44}^1+\tau_{11}^1\tau_{24}^1\tau_{42}^1+\tau_{11}^1\tau_{34}^1\tau_{43}^1+\tau_{11}^1\tau_{23}^1\tau_{32}^1-\\
						-\tau_{22}^1\tau_{33}^1\tau_{44}^1-\tau_{24}^1\tau_{43}^1\tau_{32}^1-\tau_{23}^1\tau_{34}^1\tau_{42}^1+\tau_{24}^1\tau_{42}^1\tau_{33}^1+\tau_{43}^1\tau_{34}^1\tau_{22}^1+\tau_{32}^1\tau_{23}^1\tau_{44}^1+\\
						+\tau_{12}^1\tau_{21}^1\tau_{33}^1+\tau_{12}^1\tau_{21}^1\tau_{44}^1-\tau_{12}^1\tau_{24}^1\tau_{41}^1-\tau_{12}^1\tau_{23}^1\tau_{31}^1-\tau_{13}^1\tau_{32}^1\tau_{21}^1-\tau_{13}^1\tau_{34}^1\tau_{41}^1+\\
						+\tau_{13}^1\tau_{31}^1\tau_{22}^1+\tau_{13}^1\tau_{31}^1\tau_{44}^1+\tau_{14}^1\tau_{41}^1\tau_{22}^1+\tau_{14}^1\tau_{41}^1\tau_{33}^1-\tau_{14}^1\tau_{21}^1\tau_{42}^1-\tau_{14}^1\tau_{43}^1\tau_{31}^1,\\
					\end{gathered} \\
					\begin{gathered}
						d = 
						\tau_{11}^1\tau_{22}^1\tau_{33}^1\tau_{44}^1+\tau_{11}^1\tau_{24}^1\tau_{43}^1\tau_{32}^1+\tau_{11}^1\tau_{23}^1\tau_{34}^1\tau_{42}^1-\tau_{11}^1\tau_{24}^1\tau_{42}^1\tau_{33}^1-\tau_{11}^1\tau_{43}^1\tau_{34}^1\tau_{22}^1-\tau_{11}^1\tau_{32}^1\tau_{23}^1\tau_{44}^1-\\
						-\tau_{12}^1\tau_{21}^1\tau_{33}^1\tau_{44}^1-\tau_{12}^1\tau_{24}^1\tau_{43}^1\tau_{31}^1-\tau_{12}^1\tau_{23}^1\tau_{34}^1\tau_{41}^1+\tau_{12}^1\tau_{24}^1\tau_{41}^1\tau_{33}^1+\tau_{12}^1\tau_{21}^1\tau_{34}^1\tau_{43}^1+\tau_{12}^1\tau_{31}^1\tau_{23}^1\tau_{44}^1+\\
						+\tau_{13}^1\tau_{21}^1\tau_{32}^1\tau_{44}^1+\tau_{13}^1\tau_{31}^1\tau_{42}^1\tau_{24}^1+\tau_{13}^1\tau_{34}^1\tau_{41}^1\tau_{22}^1-\tau_{13}^1\tau_{24}^1\tau_{32}^1\tau_{41}^1-\tau_{13}^1\tau_{34}^1\tau_{42}^1\tau_{21}^1-\tau_{13}^1\tau_{31}^1\tau_{22}^1\tau_{44}^1-\\
						-\tau_{14}^1\tau_{21}^1\tau_{32}^1\tau_{43}^1-\tau_{14}^1\tau_{31}^1\tau_{42}^1\tau_{23}^1-\tau_{14}^1\tau_{41}^1\tau_{22}^1\tau_{33}^1+\tau_{14}^1\tau_{23}^1\tau_{32}^1\tau_{41}^1+\tau_{14}^1\tau_{21}^1\tau_{42}^1\tau_{33}^1+\tau_{14}^1\tau_{31}^1\tau_{43}^1\tau_{22}^1.
					\end{gathered}
				\end{gathered}
			\end{equation}
			\indent By Ferrari method, if $y_1$ is an arbitrary valid positive solution of an auxiliary equation of degree three:
			\begin{equation} \label{n1}
				y^3 + Ay^2 + By + C = 0,
			\end{equation}
			for which
			\[
			\begin{gathered}
				A = - b, \\
				B = ac - 4d, \\
				C = -a^2d + 4bd - c^2,
			\end{gathered}
			\]
			then the four roots of the original equation of degree four (\ref{ny27}) are found as roots of two quadratic equations
			\begin{equation} \label{n2}
				\lambda^2 + \frac{a}{2}\lambda + \frac{y_1}{2} = \pm \sqrt{\biggl(\frac{a^2}{4} - b + y_1\biggr)\lambda^2 + \biggl(\frac{a}{2}y_1 - c\biggr)\lambda + \biggl(\frac{y_1^2}{4} - d\biggr)},
			\end{equation}
			where the root expression in the right side is a complete square trinomial.\\
			\indent Using the Cardano-Vieta formula \cite{Tabachnikov}, we write one of the roots of the equation (\ref{n1}) in the form:
			\begin{equation} \label{n3}
				y_1 = \frac{1}{3}\biggl(-A + 2\sqrt{A^2 - 3B} \sin \biggl[\frac{1}{3} \biggl(\operatorname{arcsin}\biggl[\frac{2A^3 - 9 AB + 27C}{2\sqrt{(A^2 - 3B)^3}}\biggr] + 2\pi\biggr) \biggr]\biggr),
			\end{equation}
			then the equation (\ref{n2}) can be represented as two quadratic equations:
			\begin{equation} \label{n4}
				\lambda^2 + \frac{a}{2}\lambda + \frac{y_1}{2} = \pm \biggl(\sqrt{\frac{a^2}{4} - b + y_1}\lambda + \operatorname{sign}\biggl(\frac{a}{2}y_1 - c\biggr)\sqrt{\frac{y_1^2}{4} - d}\biggr),
			\end{equation}
			solving which, we obtain four roots of the original equation of degree four.\\
			\\
			\indent \textit{\textbf{Computation of partial derivatives of the coefficients of equation (\ref{ny1}) by the external field $H$.}}\\ 
			\indent For clarity, let us write out a matrix consisting of partial derivatives on the outer field $H$ matrix elements of matrix (\ref{19}):
			\begin{equation} \label{last7}
				\begin{pmatrix}
					\frac{3}{T}\tau_{11}^1& \frac{2}{T}\tau_{12}^1&\frac{1}{T} \tau_{13}^1& 0\\
					\frac{2}{T}\tau_{21}^1& \frac{1}{T}\tau_{22}^1& 0& -\frac{1}{T}\tau_{24}^1\\
					\frac{1}{T}\tau_{31}^1&0& -\frac{1}{T}\tau_{33}^1& -\frac{2}{T} \tau_{34}^1\\
					0& -\frac{1}{T} \tau_{42}^1& -\frac{2}{T}\tau_{43}^1& - \frac{3}{T}\tau_{44}^1
				\end{pmatrix} 
			\end{equation}
			\indent From the expressions for the coefficients (\ref{last8}) we obtain expressions for partial derivatives:
			\[
			\begin{gathered}
				a'_H = \frac{1}{T}\bigg(-3 \tau_{11}^1- \tau_{22}^1 +\tau_{33}^1 + 3\tau_{44}^1\bigg), \\
				\begin{gathered}
					b'_H =
					\frac{1}{T}\bigg(-4\tau_{11}^1\tau_{22}^1+
					2\tau_{11}^1\tau_{33}^1
					-2\tau_{22}^1\tau_{44}^1-
					4\tau_{33}^1\tau_{44}^1
					+2\tau_{24}^1\tau_{42}^1+
					4\tau_{34}^1\tau_{43}^1-
					4\tau_{12}^1\tau_{21}^1-
					2\tau_{13}^1\tau_{31}^1\bigg),
				\end{gathered}		\\
				\begin{gathered}
					c'_H = 
					\frac{1}{T}\bigg(-3\tau_{11}^1\tau_{22}^1\tau_{33}^1-
					\tau_{11}^1\tau_{22}^1\tau_{44}^1+
					\tau_{11}^1\tau_{33}^1\tau_{44}^1+
					\tau_{11}^1\tau_{24}^1\tau_{42}^1-
					\tau_{11}^1\tau_{34}^1\tau_{43}^1+
					3\tau_{11}^1\tau_{23}^1\tau_{32}^1+\\
					+3\tau_{22}^1\tau_{33}^1\tau_{44}^1+
					3\tau_{24}^1\tau_{43}^1\tau_{32}^1+
					3\tau_{23}^1\tau_{34}^1\tau_{42}^1-
					3\tau_{24}^1\tau_{42}^1\tau_{33}^1-
					3\tau_{43}^1\tau_{34}^1\tau_{22}^1-
					3 \tau_{32}^1\tau_{23}^1\tau_{44}^1+\\
					+3\tau_{12}^1\tau_{21}^1\tau_{33}^1+
					\tau_{12}^1\tau_{21}^1\tau_{44}^1-
					\tau_{12}^1\tau_{24}^1\tau_{41}^1-
					3\tau_{12}^1\tau_{23}^1\tau_{31}^1-
					3\tau_{13}^1\tau_{32}^1\tau_{21}^1+
					\tau_{13}^1\tau_{34}^1\tau_{41}^1+\\
					+3\tau_{13}^1\tau_{31}^1\tau_{22}^1-
					\tau_{13}^1\tau_{31}^1\tau_{44}^1+
					\tau_{14}^1\tau_{41}^1\tau_{22}^1-
					\tau_{14}^1\tau_{41}^1\tau_{33}^1-
					\tau_{14}^1\tau_{21}^1\tau_{42}^1+
					\tau_{14}^1\tau_{43}^1\tau_{31}^1\bigg),\\
				\end{gathered} 
			\end{gathered}
			\]

			\[
			\begin{gathered}
				a''_{HH} = \frac{1}{T^2}\bigg(- 9\tau_{11}^1- \tau_{22}^1 - \tau_{33}^1 - 9\tau_{44}^1\bigg), \\
				\begin{gathered}
					b''_{HH} =
					\frac{1}{T^2}\bigg(16\tau_{11}^1\tau_{22}^1+
					4\tau_{11}^1\tau_{33}^1+
					4\tau_{22}^1\tau_{44}^1+
					16\tau_{33}^1\tau_{44}^1
					-4\tau_{24}^1\tau_{42}^1-
					16\tau_{34}^1\tau_{43}^1-
					16\tau_{12}^1\tau_{21}^1-
					4\tau_{13}^1\tau_{31}^1\bigg),
				\end{gathered}		\\
				\begin{gathered}
					c''_{HH} = 
					\frac{1}{T^2}\bigg(-9\tau_{11}^1\tau_{22}^1\tau_{33}^1-
					\tau_{11}^1\tau_{22}^1\tau_{44}^1-
					\tau_{11}^1\tau_{33}^1\tau_{44}^1+
					\tau_{11}^1\tau_{24}^1\tau_{42}^1+
					\tau_{11}^1\tau_{34}^1\tau_{43}^1+
					9\tau_{11}^1\tau_{23}^1\tau_{32}^1-\\
					-9\tau_{22}^1\tau_{33}^1\tau_{44}^1-
					9\tau_{24}^1\tau_{43}^1\tau_{32}^1-
					9\tau_{23}^1\tau_{34}^1\tau_{42}^1+
					9\tau_{24}^1\tau_{42}^1\tau_{33}^1+
					9\tau_{43}^1\tau_{34}^1\tau_{22}^1+
					9\tau_{32}^1\tau_{23}^1\tau_{44}^1+\\
					+9\tau_{12}^1\tau_{21}^1\tau_{33}^1+
					\tau_{12}^1\tau_{21}^1\tau_{44}^1-
					\tau_{12}^1\tau_{24}^1\tau_{41}^1-
					9\tau_{12}^1\tau_{23}^1\tau_{31}^1-
					9\tau_{13}^1\tau_{32}^1\tau_{21}^1-
					\tau_{13}^1\tau_{34}^1\tau_{41}^1+\\
					+9\tau_{13}^1\tau_{31}^1\tau_{22}^1+
					\tau_{13}^1\tau_{31}^1\tau_{44}^1+
					\tau_{14}^1\tau_{41}^1\tau_{22}^1+
					\tau_{14}^1\tau_{41}^1\tau_{33}^1-
					\tau_{14}^1\tau_{21}^1\tau_{42}^1-
					\tau_{14}^1\tau_{43}^1\tau_{31}^1\bigg).\\
				\end{gathered} \\
			\end{gathered}
			\]

		\end{document}